\documentclass[12pt,preprintnumbers,superscriptaddress,amsmath,amssymb,nofootinbib]{revtex4}

\usepackage{graphicx}
\usepackage{dcolumn}
\usepackage{bm}
\usepackage{amssymb}
\usepackage{amsmath}
\usepackage{epsfig}    
\usepackage{color}
\usepackage{slashed}
\usepackage[hypertex]{hyperref}
\usepackage{multirow}
\begin{document} 

\title{Extracting the mass scale of a second Higgs boson \\ from a deviation in 
$h(125)$ couplings}

\author{Simone Blasi}
\email{simone.blasi.91@gmail.com}
\affiliation{INFN, Sezione di Firenze, and Department of Physics and Astronomy, University of Florence, Via
G. Sansone 1, 50019 Sesto Fiorentino, Italy}

\author{Stefania De Curtis}
\email{decurtis@fi.infn.it}
\affiliation{INFN, Sezione di Firenze, and Department of Physics and Astronomy, University of Florence, Via
G. Sansone 1, 50019 Sesto Fiorentino, Italy}

\author{Kei Yagyu}
\email{yagyu@fi.infn.it}
\affiliation{INFN, Sezione di Firenze, and Department of Physics and Astronomy, University of Florence, Via
G. Sansone 1, 50019 Sesto Fiorentino, Italy}

\begin{abstract}

\noindent
We investigate the correlation between a possible deviation in the discovered Higgs boson $h(125)$ couplings
from the Standard Model prediction and the 
mass scale ($M_{\text{2nd}}$) of the next-to-lightest Higgs boson in models with non-minimal Higgs sectors. 
In particular, we comprehensively study a class of next-to-minimal Higgs sectors which satisfy the electroweak
$\rho$ parameter to be one at tree level. 
We derive an upper limit on $M_{\text{2nd}}$ 
by imposing bounds from perturbative unitarity, vacuum stability, triviality and electroweak precision data 
as functions of the deviation in the $hVV$ ($V=W,Z$) couplings. 
Furthermore, we discuss the complementarity between these bounds and the current LHC data, e.g.,
by considering direct searches for additional Higgs bosons 
and indirect constraints arising from the measured $h(125)$ signal strengths. 

\end{abstract}
\maketitle

\newpage
\section{Introduction}

The existence of at least one isospin doublet scalar field is strongly suggested by the
discovery of the Higgs boson $h(125)$ at the LHC and the measurement of
its properties, which are consistent with those of the Standard Model (SM) Higgs boson~\cite{LHC1}.  
This experimental fact brings us to the natural question whether the observed Higgs boson is unique or
it corresponds to one of the resonances of a wider structure.
The latter possibility requires the Higgs sector
to be extended from the minimal form. 
On the other hand, a non-minimal shape of the Higgs sector is expected by
New Physics (NP) paradigms (e.g., 
composite Higgs models and supersymmetry) embedded in physics beyond the SM (BSM).
Therefore, the detection of a second Higgs boson would be a clear 
evidence of NP. 

The mass of a Higgs boson is one of the most critical parameters for its direct detection at collider
experiments. 
In this paper, we aim to systematically derive the limits
on the mass of a second Higgs boson 
in next-to-minimal renormalizable Higgs sectors, 
i.e., those composed of one isospin doublet plus an
extra Higgs field with a non-vanishing Vacuum Expectation Value (VEV). 
If we require the VEV of the extra Higgs field not to spoil the relation
for the electroweak parameter $\rho = 1$ at tree level,
the simplest three choices are: the Higgs Singlet Model (HSM), the 2 Higgs Doublet Model (2HDM)~\cite{review_thdm} 
and the Georgi-Machacek (GM) model~\cite{GM1,GM2}\footnote{In fact, although 
the Higgs sector of the GM model is composed of one iso-doublet and two iso-triplet scalar fields, 
the latter can be packaged as one $SU(2)_L\times SU(2)_R$ bi-triplet. 
We thus regard the GM model as the next-to-minimal Higgs sector containing triplets. }. 
These extensions, which are commonly understood as low energy
descriptions of underlying NP scenarios, have been largely
considered in the literature because of their connection with several 
open questions of the SM. It is known that the HSM can
provide a candidate for dark matter \cite{HSM-RGE},
while models with Higgs triplets are involved in generating
neutrino masses through the so-called type II seesaw mechanism \cite{type2}.
Conversely, the interest in 2HDMs is mainly motivated by
supersymmetric extensions of the SM.
Furthermore, the reason to consider models with an extended Higgs sector is phenomenological. 
As mentioned above, the discovery of a second Higgs boson
would necessarily require to build-up a non-minimal structure for the Higgs sector.
In addition, non-minimal Higgs sectors can be indirectly probed by precise measurements
of the $h$ couplings to 
SM particles, because the $h$ state is there generally realized through a
non-zero mixing among all the other scalars with the same quantum numbers,
resulting deviations in the $h$ couplings from the SM prediction.
Regarding, e.g., the coupling to a vector boson pair, 
the current $1\sigma$ uncertainty at the LHC is indeed
about 10\% \cite{LHC1}, so that there is still room for a sensible NP contribution.
Such uncertainty is expected to
be reduced to $\sim$ 5\% by the forthcoming LHC High Luminosity option \cite{HLLHC_ATLAS,HLLHC_CMS}
and even to 0.5\% at future $e^+e^-$ colliders \cite{ILC}.

Basically, the masses of extra Higgs bosons within a given model
are free parameters. However, it is possible
to extract their order of magnitude by taking into account theoretical issues within particular
BSM scenarios. For example, it is known that perturbative unitarity 
constrains the size of dimensionless 
quartic couplings in the Higgs potential which actually enter the expression of the physical Higgs boson masses.
Originally, this method was applied 
to set an upper limit on the SM Higgs boson mass by Lee, Quigg and Thacker~\cite{LQT}.
Afterwards, the same technique was carried out in
various extended Higgs sectors such as 2HDMs~\cite{thdm_PU1,thdm_PU2,thdm_PU3,thdm_PU4}. 
Besides that, the reliability of a perturbative approach requires
the scalar potential to be bounded from below in any direction of the field space.  
Such requirement is usually referred to as vacuum stability
and it provides further constraints on the parameter space of non-minimal Higgs sectors.
Furthermore, one requires the absence of Landau poles up to a certain cutoff of a given model: this is 
the so-called triviality constraint. 

Bounds on the mass of extra Higgs bosons can be extracted by considering experimental issues as well.
We here take into account the constraints coming from the 
Electroweak Precision Tests (EWPTs) and the currently available LHC data, which are
based both on the null excess of signatures from 
direct searches for extra Higgs bosons and on the analysis of the $h(125)$ signal strengths.
Combining both theoretical and experimental requirements, 
we restrict the possible allowed values of the extra Higgs masses, depending on the model and its 
parameter configurations. 

It is useful for this analysis to discuss the decoupling and the alignment limits of 
non-minimal Higgs sectors. 
The decoupling limit is defined in such a way that all the masses of the extra Higgs bosons are 
taken to infinity, and eventually
only the $h$ state remains light. 
In this limit, the extended Higgs sectors effectively reduce to
the minimal one and all the observables relevant to $h$, such as the couplings to SM particles, 
do not deviate from the SM prediction. 
On the other hand, the alignment limit is defined such that 
the $h$ state and the Nambu-Goldstone (NG) bosons emerging from the electroweak
spontaneous symmetry breaking fill the same doublet field, which carries the whole VEV $v$, 
fixed by $v = (\sqrt{2}G_F)^{-1/2}$, with $G_F$ being the Fermi constant. 
In this limit, all the $h$ couplings to SM particles such as $hVV$ ($V=W^\pm,Z$) and $hf\bar{f}$ 
align to be the same as their SM values at tree level. 
Notice that the decoupling limit can only be taken if the alignment 
limit is realized.
Therefore, a deviation from the alignment limit, which gives a non-zero deviation 
in $h$ couplings from the SM prediction, 
results in an upper bound on the extra Higgs boson masses. 
In particular, we investigate the correlation between
a deviation in the $h$ coupling
to a vector boson pair and the bounds on the extra Higgs boson masses by taking into account the
aforementioned constraints for each extended Higgs sector. 

This paper is organized as follows. 
In Sec.~\ref{sec:model}, we define the three models with non-minimal Higgs sectors, i.e., 
the HSM, the 2HDM and the GM model. 
In Sec.~\ref{sec:bound}, we numerically extract, for each model, the upper limit on the mass of the
next-to-lightest Higgs boson by imposing
the theoretical constraints (unitarity, vacuum stability, triviality) and those coming from the EWPTs.
In Sec.~\ref{sec:compl}, the complementarity between the bounds from the LHC data and those of Sec.~\ref{sec:bound}
is studied.
Our conclusions are summarized in Sec.~\ref{sec:conc}. 

\section{Extended Higgs sectors \label{sec:model}}

\begin{table}[t]
\begin{center}
\begin{tabular}{l||l}\hline\hline 
Model    & Scalar fields $\sim$ $(SU(2)_L,U(1)_Y)$ \\\hline
HSM      & $\Phi({\bf 2},1/2)$ and $S({\bf 1},0)$\\
2HDM     & $\Phi_1({\bf 2},1/2)$ and $\Phi_2({\bf 2},1/2)$\\
GM model & $\Phi({\bf 2},1/2)$, $\xi({\bf 3},0)$ and $\chi({\bf 3},1)$\\
\hline\hline
\end{tabular} 
\end{center}
\caption{Electroweak quantum numbers of the scalar fields in the non-minimal Higgs sectors
considered. } \label{fields}
\end{table}

We briefly review the three extended Higgs models
considered in this paper, namely, the HSM, the 2HDM and the GM model. 
The scalar field content is summarized in Table~\ref{fields}. 
Throughout the paper, we use the shorthand notation: $s_X^{}= \sin X$, $c_X^{}= \cos X$ and $t_X^{} = \tan X$ for an arbitrary angle parameter $X$. 
In each model, the
symbol $h$ is used to denote the discovered Higgs boson at the LHC with a mass of 125 GeV ($m_h=125$ GeV).  

\subsection{Higgs Singlet Model}

The most general scalar potential in the HSM has the following form:
\begin{align}
V(\Phi,S)_{\text{HSM}} =& \mu^2|\Phi|^2+\lambda |\Phi|^4 
+\mu_{\Phi S}|\Phi|^2 S+ \lambda_{\Phi S} |\Phi|^2 S^2
+t_S^{}S +m^2_SS^2+ \mu_SS^3+ \lambda_SS^4, 
\end{align} 
where the doublet and the singlet fields are respectively parameterized as 
\begin{align}
\Phi = 
\begin{pmatrix}
G^+\\
\frac{\phi+v+iG^0}{\sqrt{2}}
\end{pmatrix},~
S=v_S+s. \label{hsm1}
\end{align}
In Eq.~(\ref{hsm1}), $G^\pm$ and $G^0$ are the NG bosons which are absorbed by the longitudinal components of the $W^\pm$ and $Z$ bosons, respectively.  
The VEV of the singlet field $v_S^{}$ contributes neither to the electroweak symmetry breaking nor to the fermion mass generation.  
As a consequence, we can set $v_S^{} = 0$ without loss of generality because of the shift symmetry of the singlet VEV~\cite{Dawson,HSM}, and 
we will adopt it throughout this paper. 

From the tadpole conditions
\begin{align}\label{eq:tadpole}
\frac{\partial V}{\partial \phi}\Bigg|_0  = 
\frac{\partial V}{\partial s}\Bigg|_0 = 0, 
\end{align}
we can eliminate $t_S^{}$ and $\mu^2$ parameters. In Eq.~(\ref{eq:tadpole}) and in the following, the symbol $|_0$ denotes 
that all the scalar fields are taken to be zero after the derivative. 
The mass eigenstates for the neutral Higgs bosons can be defined as
\begin{align}
\left(
\begin{array}{c}
s\\
\phi
\end{array}\right)=
R(\alpha)
\left(
\begin{array}{c}
H_S\\
h
\end{array}\right),~~\text{with}~~
R(\theta) = \begin{pmatrix}
c_\theta & -s_\theta  \\
s_\theta & c_\theta  
\end{pmatrix}.   \label{hsm2}
\end{align}
The squared matrix elements $M_{ij}^2$ $(i,j = 1,2)$ in the basis ($s,\phi$) 
are given in terms of the parameters in the potential as
\begin{align}
M^2_{11}= 2m_S^2+ v^2\lambda_{\Phi S}  ,\quad
M^2_{22}=2\lambda v^2,\quad M^2_{12}=v\mu_{\Phi S}.  \label{hsm3}
\end{align}
The mass eigenvalues for $H_S$ $(m_{H_S}^{})$ and $h$ $(m_h)$, and the mixing angle $\alpha$
are easily obtained:
\begin{align}
&m_{H_S}^2=M_{11}^2c^2_\alpha +M_{22}^2s^2_\alpha +M_{12}^2s_{2\alpha} ,~
m_h^2=M_{11}^2s^2_\alpha +M_{22}^2c^2_\alpha -M_{12}^2s_{2\alpha},~
t_{2\alpha} =\frac{2M_{12}^2}{M_{11}^2-M_{22}^2}. \label{hsm-alpha}
\end{align}
These relations can be inverted to express each element of the squared mass matrix in terms of the mass eigenvalues:
\begin{align}
M_{11}^2 = m_{H_S}^2c_\alpha^2 + m_h^2 s_\alpha^2,\quad
M_{22}^2 = m_{H_S}^2s_\alpha^2 + m_h^2 c_\alpha^2,\quad 
M_{12}^2 = \frac{1}{2}s_{2\alpha}(m_{H_S}^2-m_h^2). \label{lam}
\end{align}
From Eqs.~(\ref{hsm3}) and (\ref{hsm-alpha}), we see that 
the decoupling limit is defined by $m_S^2 \to \infty$, where 
$m_{H^{}_S}$ and $\alpha$ become infinity and zero, respectively. 
On the other hand, the alignment limit is obtained by taking $\mu_{\Phi S}^{} \to 0$,
in which the mixing angle $\alpha$ vanishes. 
In this limit, $m_{H^{}_S}$ is not necessarily large. 
Conversely, from the second equations of (\ref{hsm3}) and (\ref{lam})
we find that a large value of $m_{H^{}_S}$ with non-zero $\alpha$ is realized only by taking a large value of
the quartic coupling $\lambda$. 
Therefore, in the case $\alpha \neq 0 $, there must be an upper limit on $m_{H^{}_S}$ 
by imposing perturbative unitarity, vacuum stability and triviality bounds (see App.~\ref{app1}). 
We will quantitatively derive such limit in the next section. 
Following the above discussion, the HSM can be described by 5 independent parameters after fixing the VEV $v$ and $m_h$:
\begin{align}
m_{H^{}_S},~~s_\alpha,~~\lambda_S,~~\lambda_{\Phi S},~~\mu_{S}. 
\end{align}
The kinetic and Yukawa terms (here and in the following we explicitly write only those 
for the third generation fermions) are given by  
\begin{align}
{\cal L} = (D_\mu \Phi)^\dagger (D^\mu \Phi) + \frac{1}{2}(\partial_\mu S)(\partial^\mu S) 
- \left(y_t\bar{Q}_L^3 \Phi^c t_R^{} + y_b\bar{Q}_L^3 \Phi b_R^{} + y_\tau\bar{L}_L^3\Phi \tau_R^{} + \text{h.c.} \right), 
\end{align}
where $Q_L^3 =(t,b)_L$ and $L_L^3 =(\nu_\tau,\tau)_L$ and $\Phi^c = i\tau_2 \Phi^*$. 
The covariant derivative for $\Phi$ reads
\begin{align}
D_\mu \Phi =\partial_\mu\Phi -ig_2\frac{\tau^a}{2}W_\mu^a\Phi - \frac{i}{2}g_1B_\mu \Phi, \label{cov_phi}
\end{align}
where $g_2$ and $g_1$ are the $SU(2)_L$ and $U(1)_Y$ gauge couplings, respectively. 
We see that the singlet field $S$ does not couple to the SM fermions and gauge bosons, so that 
interaction terms for $H_S$ with SM fields are only generated by the mixing. 
In terms of the mass eigenstates of the Higgs bosons, we thus obtain:
\begin{align}
{\cal L}_{\text{int}} = \left(\frac{h}{v}c_\alpha + \frac{H_S}{v}s_\alpha \right)(2m_W^2 W_\mu^+ W^{-\mu} + m_Z^2 Z_\mu Z^\mu -m_f \bar{f}f),~~(f = t,b,\tau). 
\end{align}

\subsection{2-Higgs Doublet Model}

In order to avoid tree level flavour changing neutral currents in the 2HDM, 
we impose a discrete $Z_2$ symmetry~\cite{GW}, which can be softly broken. 
The $Z_2$ charge assignment for the two doublets is: $(\Phi_1,\Phi_2) \to (+\Phi_1, -\Phi_2)$. 
The Higgs potential is given by  
\begin{align}
V(\Phi_1,\Phi_2)_{\text{2HDM}}&=\mu_1^2|\Phi_1|^2+\mu_2^2|\Phi_2|^2-\mu_{12}^2(\Phi_1^\dagger \Phi_2 +\text{h.c.})
+\frac{1}{2}\lambda_1|\Phi_1|^4+\frac{1}{2}\lambda_2|\Phi_2|^4\notag\\
& +\lambda_3|\Phi_1|^2|\Phi_2|^2+\lambda_4|\Phi_1^\dagger\Phi_2|^2
+\frac{1}{2}\lambda_5[(\Phi_1^\dagger\Phi_2)^2+\text{h.c.}], 
\end{align}
where $\mu_{12}^2$ is the soft-breaking term of the $Z_2$ symmetry. 
In general, the $\mu_{12}^2$ and $\lambda_5$ parameters can be complex, but we assume them to be real for simplicity. 

A convenient basis for the Higgs fields is the so-called Higgs basis $(\Phi,\Psi)$~\cite{HBasis} which is related to 
the original one $(\Phi_1,\Phi_2)$ by  
\begin{align}
\begin{pmatrix}
\Phi_1\\
\Phi_2
\end{pmatrix}
=R(\beta)
\begin{pmatrix}
\Phi\\
\Psi
\end{pmatrix}, \label{hb1}
\end{align}
where $t_\beta = v_2/v_1$ with $v_{i}$ the VEV of $\Phi_i$. In this basis, the VEV $v(=\sqrt{v_1^2 + v_2^2})$ 
and the NG bosons ($G^\pm$ and $G^0$) belong to the same doublet. Namely, 
\begin{align}
\Phi=\begin{pmatrix}
G^+\\
\frac{h_1'+v+ iG^0}{\sqrt{2}}
\end{pmatrix},\quad 
\Psi=\begin{pmatrix}
H^+\\
\frac{h_2'+ iA}{\sqrt{2}}
\end{pmatrix}. \label{hb2}
\end{align} 
In Eq.~(\ref{hb2}), $H^\pm$ and $A$ are the physical singly-charged and CP-odd Higgs bosons, respectively. 
The two CP-even Higgs states $h_1'$ and $h_2'$ can mix each other. 
Their mass eigenstates are defined by 
\begin{align}
\begin{pmatrix}
h_1'\\
h_2'
\end{pmatrix} = R(\alpha-\beta)
\begin{pmatrix}
H\\
h
\end{pmatrix}. \label{even}
\end{align}
By imposing the tadpole conditions
\begin{align}
\frac{\partial V}{\partial h_1}\Bigg|_0  = 
\frac{\partial V}{\partial h_2}\Bigg|_0 = 0, 
\end{align}
we can eliminate $\mu_1^2$ and $\mu_2^2$. 
The masses for $A$ ($m_A^{}$) and $H^\pm$ ($m_{H^\pm}^{}$) are then given by 
\begin{align}
m_{H^\pm}^2&=M^2-\frac{v^2}{2}(\lambda_4+\lambda_5),\quad m_A^2=M^2-v^2\lambda_5, \label{ma}
\end{align}
where $M^2 = \mu_{12}^2/(s_\beta c_\beta)$. 
The relation between the CP-even Higgs boson masses and the 
matrix elements $M_{ij}^2$ in the ($h_1',h_2'$) basis
is given in Eqs.~(\ref{hsm-alpha}) and (\ref{lam}) after the replacement of $(\alpha, H_S) \to (\alpha -\beta, H)$. 
The squared matrix elements are given in terms of the potential parameters by the following relations:
\begin{align}
M_{11}^2&=v^2(\lambda_1 c^4_\beta+\lambda_2 s^4_\beta)+\frac{v^2}{2}\lambda_{345}s^2_{2\beta},\quad
M_{22}^2=M^2+v^2s^2_\beta c^2_\beta(\lambda_1+\lambda_2-2\lambda_{345}),\notag \\
M_{12}^2&=v^2s_\beta c_\beta\left(\lambda_2 s^2_\beta-\lambda_1c^2_\beta + c_{2\beta}\lambda_{345}\right),  \label{lam_thdm}
\end{align}
where $\lambda_{345}=\lambda_3+\lambda_4+\lambda_5$. 
From the discussion above, the scalar potential can be fully described
by
the following set of independent parameters (after fixing $v$ and $m_h$):
\begin{align}
&m_A^{},~~m_{H^\pm},~~m_{H},~~c_{\beta-\alpha},~~t_\beta,~~M^2, 
\end{align}
with $0 < \beta -\alpha < \pi$. 

Let us discuss the decoupling and the alignment limits in the 2HDM. 
The decoupling limit is given by $M^2 \to \infty$, by which 
all the masses of $H^\pm$, $A$ and $H$ become infinity, and $t_{2(\beta-\alpha)}$ becomes zero 
(equivalently $s_{\beta-\alpha} \to 1$).
On the other hand, the alignment limit is defined by taking $s_{\beta-\alpha} \to 1$, 
so that the $h_1'$ state in Eq.~(\ref{hb2}) corresponds to the $h$ state. 
Similarly to the HSM, if we take $s_{\beta - \alpha}\neq 1$
the decoupling limit cannot be reached, because
a large value for $m_H$ is only realized by 
a large value of the scalar quartic couplings (which are disfavored, e.g., by perturbative unitarity).
This is clear from the relation:
\begin{align}
M^2_{11} = m_H^2 c_{\beta-\alpha}^2 + m_h^2s_{\beta-\alpha}^2 = v^2\left(\lambda_1 c^4_\beta+\lambda_2 s^4_\beta + \frac{\lambda_{345}}{2}s^2_{2\beta}\right).  \label{thdm2}
\end{align}

Regarding the kinetic and Yukawa interaction terms, they
are given by  
\begin{align}
{\cal L} = \sum_{i=1,2}(D_\mu \Phi_i)^\dagger (D^\mu \Phi_i)
- \left(y_t\bar{Q}_L^3 \Phi^c_t t_R^{} + y_b\bar{Q}_L^3 \Phi_b b_R^{} + y_\tau\bar{L}_L^3\Phi_\tau \tau_R^{} + \text{h.c.} \right), 
\end{align}
where the covariant derivative $D_\mu$ is the same as
Eq.~(\ref{cov_phi}). The $\Phi_f$ ($f=t,b,\tau$) fields
are $\Phi_1$ or $\Phi_2$ depending on the $Z_2$-charge assignment for the right handed fermions. 
For the latter, there are 4 independent choices which lead to 4 different types of Yukawa
interactions \cite{Grossman,Barger}, referred to as Type-I, Type-II, Type-X and Type-Y
\cite{typeX}. 
The combination of $(\Phi_t,\Phi_b,\Phi_\tau)$ is determined in each type
as follows:
\begin{align}
\begin{split}
&(\Phi_t,\Phi_b,\Phi_\tau) = (\Phi_2,\Phi_2,\Phi_2)~~~\text{for Type-I}, \\
&(\Phi_t,\Phi_b,\Phi_\tau) = (\Phi_2,\Phi_1,\Phi_1)~~~\text{for Type-II}, \\
&(\Phi_t,\Phi_b,\Phi_\tau) = (\Phi_2,\Phi_2,\Phi_1)~~~\text{for Type-X}, \\
&(\Phi_t,\Phi_b,\Phi_\tau) = (\Phi_2,\Phi_1,\Phi_2)~~~\text{for Type-Y}. 
\end{split}\label{type}
\end{align}
For example, Type-II is realized by setting the charge
assignments as $(t_R,b_R,\tau_R) \to (-t_R,+b_R,+\tau_R)$ and taking
all the left-handed fermions to be $Z_2$-even. 
In terms of the Higgs boson mass eigenstates, we obtain the following interaction terms:
\begin{align}
{\cal L}_{\text{int}} &= \left(\frac{h}{v}s_{\beta-\alpha} + \frac{H}{v}c_{\beta-\alpha} \right)(2m_W^2 W_\mu^+ W^{-\mu} + m_Z^2 Z_\mu Z^\mu )\notag\\
&-\sum_{f=t,b,\tau} \frac{m_f}{v}\left[(s_{\beta-\alpha} + \xi_f c_{\beta-\alpha})\bar{f}f h+(c_{\beta-\alpha} - \xi_f s_{\beta-\alpha})\bar{f}f H  
-2iI_f \xi_f \bar{f}\gamma_5f A \right]\notag\\
&- \frac{\sqrt{2}}{v} \left[\bar{t}(m_b \xi_b P_R - m_t \xi_t P_L) bH^+ + \bar{\nu}_\tau \, m_\tau \xi_\tau P_R \,\tau H^+  +\text{h.c.}\right], 
\end{align}
where $I_f=+1/2 (-1/2)$ for $f = t(b,\tau)$ and
$P_{L,R}$ are the projection operators for left-/right-handed fermions. The mixing factors
$\xi_f$ are given by:
\begin{align}
 \begin{split}
 & \xi_t = \text{cot} \, \beta \ \text{for all Types} \\
 & \xi_b = \text{cot} \, \beta \ \text{for Type-I,\,-X}, \ - \text{tan} \, \beta \ \text{for Type-II,\,-Y} \\
 & \xi_\tau = \text{cot} \, \beta \ \text{for Type-I,\,-Y}, \ - \text{tan} \, \beta \ \text{for Type-II,\,-X}. \\
 \end{split}
\end{align}
It is straightforward to check that, in the alignment limit $s_{\beta-\alpha} \to 1$, 
all the $h$ couplings coincide with those of the SM Higgs boson at tree level. 

\subsection{Georgi-Machacek Model \label{sec:gm}}

The Higgs potential in the GM model can be constructed in terms of 
the $SU(2)_L \times SU(2)_R$ bi-doublet $\mathbf{\Phi}$ and bi-triplet $\Delta$ fields:
\begin{align}
\mathbf{\Phi}\equiv (\Phi^c,\Phi) = \left(
\begin{array}{cc}
\phi^{0*} & \phi^+ \\
-\phi^- & \phi^0
\end{array}\right),\quad 
\Delta\equiv(\chi^c,\xi,\chi)= \left(
\begin{array}{ccc}
\chi^{0*} & \xi^+ & \chi^{++} \\
-\chi^- & \xi^0 & \chi^{+} \\
\chi^{--} & -\xi^- & \chi^{0} 
\end{array}\right), \label{eq:Higgs_matrices}
\end{align}
where $\chi^c = C_3\chi^*$ is the charge conjugated $\chi$ field with 
\begin{align}
C_3 = \begin{pmatrix}
0 &0&1\\
0&-1&0\\
1&0&0
\end{pmatrix}. 
\end{align}
The neutral component fields are expressed by 
\begin{align}
\phi^0&=\frac{1}{\sqrt{2}}(\phi_r+v_\phi+i\phi_i), \quad 
\chi^0=\frac{1}{\sqrt{2}}(\chi_r+i\chi_i)+v_\chi,\quad \xi^0=\xi_r+v_\xi, \label{eq:neutral}
\end{align}
where $v_\phi$, $v_\chi$ and $v_\xi$ being the VEV's for $\phi^0$, $\chi^0$ and $\xi^0$, respectively. 

Assuming the vacuum alignment: $v_\chi = v_\xi (\equiv v_\Delta^{})$, 
the VEV $\langle \Delta \rangle$ is proportional to the $3\times 3$ identity matrix. 
In this configuration, the global $SU(2)_L \times SU(2)_R$ symmetry is spontaneously broken down to the 
$SU(2)_V$ symmetry, i.e., the so-called custodial symmetry. 

The most general scalar potential for the $\Phi$ and $\Delta$ fields is given by 
\begin{align}
V(\mathbf{\Phi},\Delta)_{\text{GM}}&=\mu_\Phi^2\text{tr}(\mathbf{\Phi}^\dagger\mathbf{\Phi})+\mu_\Delta^2\text{tr}(\Delta^\dagger\Delta)
+\lambda_1[\text{tr}(\mathbf{\Phi}^\dagger\mathbf{\Phi})]^2+\lambda_2[\text{tr}(\Delta^\dagger\Delta)]^2
+\lambda_3\text{tr}[(\Delta^\dagger\Delta)^2]\notag\\
&+\lambda_4\text{tr}(\mathbf{\Phi}^\dagger\mathbf{\Phi})\text{tr}(\Delta^\dagger\Delta)
+\lambda_5\text{tr}\left(\mathbf{\Phi}^\dagger\frac{\tau^a}{2}\mathbf{\Phi}\frac{\tau^b}{2}\right)
\text{tr}(\Delta^\dagger t^a\Delta t^b)\notag\\
&+\mu_1\text{tr}\left(\mathbf{\Phi}^\dagger \frac{\tau^a}{2}\mathbf{\Phi}\frac{\tau^b}{2}\right)(P^\dagger \Delta P)^{ab}
+\mu_2\text{tr}\left(\Delta^\dagger t^a\Delta t^b\right)(P^\dagger \Delta P)^{ab}, \label{eq:pot}
\end{align}
where $\tau^a$ and $t^a$ ($a=1$--$3$) are the $2\times 2$ and $3\times 3$ matrix representations of the $SU(2)$ generators, respectively. 
The unitary matrix $P$ enters in the similarity transformation $P (-i\epsilon^a)P^\dagger = t^a$ with $\epsilon^a$ being 
the adjoint representation of the $SU(2)$ generators. The explicit form of $P$ is given as
\begin{align}
P=\left(
\begin{array}{ccc}
-1/\sqrt{2} & i/\sqrt{2} & 0 \\
0 & 0 & 1 \\
1/\sqrt{2} & i/\sqrt{2} & 0
\end{array}\right). 
\end{align}
We note that 
the custodial symmetry in the Higgs potential is broken by the $U(1)_Y$ hypercharge gauge interaction at one-loop level. 
The effects of the loop induced custodial-breaking terms, which cannot be 
described in terms of $\mathbf{\Phi}$ and $\Delta$, were discussed in Ref.~\cite{Simone}. 

The mass eigenstates of the physical Higgs bosons can be classified in terms of the $SU(2)_V$ multiplets, namely, 
the 5-plet $(H_5^{\pm\pm},H_5^\pm,H_5^0)$, the two 3-plets $(H_3^\pm,H_3^0)$ and $(G^\pm,G^0)$, 
and the two singlets $H_1$ and $h$. 
The Higgs bosons belonging to the same $SU(2)_V$ multiplet are degenerate in mass. 
These mass eigenstates are related to the original states given in Eq.~(\ref{eq:Higgs_matrices}) by the following transformations:
\begin{align}\label{eq:GMmasseigen}
\begin{pmatrix}
\chi_i\\
\phi_i
\end{pmatrix}
&=R(\beta)
\begin{pmatrix}
G^0\\
H_3^0
\end{pmatrix}, \notag \\
\begin{pmatrix}
\phi^\pm \\
\xi^\pm \\
\chi^\pm
\end{pmatrix}
&=
R_5^\pm
\begin{pmatrix}
\phi^\pm \\
\xi^{\prime \pm}\\
H_5^\pm
\end{pmatrix}
= R_5^\pm R_{\beta}
\begin{pmatrix}
G^\pm\\
H_3^\pm\\
H_5^\pm
\end{pmatrix},~
\begin{pmatrix}
\xi_r\\
\phi_r\\
\chi_r
\end{pmatrix}=R_5^r\begin{pmatrix}
\xi_r'\\
\phi_r\\
H_5^0
\end{pmatrix} = R_5^r R_{\alpha}
\begin{pmatrix}
H_1\\
h\\
H_5^0
\end{pmatrix}, 
\end{align}
where $R_5^\pm$ and $R_5^r$ are the orthogonal $3\times 3$ matrices which separate the 5-plet Higgs bosons from the other singly-charged 
and CP-even scalar states, respectively. 
Their explicit forms are given by
\begin{align}
R_5^\pm =
\begin{pmatrix}
1&0&0 \\
0 & \frac{1}{\sqrt{2}} & -\frac{1}{\sqrt{2}} \\
0 & \frac{1}{\sqrt{2}} & \frac{1}{\sqrt{2}} 
\end{pmatrix},\quad
R_5^r=
\left(
\begin{array}{ccc}
\frac{1}{\sqrt{3}} &0& -\sqrt{\frac{2}{3}}\\
0 & 1 &0\\
\sqrt{\frac{2}{3}} & 0 & \frac{1}{\sqrt{3}}
\end{array}\right). 
\end{align}
The other two matrices involved in Eq.~(\ref{eq:GMmasseigen}) are
\begin{align}
R_\beta = 
\left(
\begin{array}{ccc}
s_\beta & c_\beta & 0\\
c_\beta & -s_\beta & 0\\
0&0&1
\end{array}
\right),\quad 
R_\alpha = 
\left(
\begin{array}{ccc}
 c_\alpha & -s_\alpha &0\\
 s_\alpha & c_\alpha &0\\
0 & 0 & 1 
\end{array}\right), 
\end{align}
where $t_\beta = v_\phi/(2\sqrt{2}v_\Delta^{})$ and $\alpha$ describes the mixing of the CP-even singlet states. 
By using the two independent tadpole conditions
\begin{align}
\frac{\partial V}{\partial \phi_r}\Bigg|_0  = 
\frac{\partial V}{\partial \chi_r}\Bigg|_0 = 0, 
\end{align}
we can  eliminate the $\mu_\Phi^2$ and $\mu_\Delta^2$ parameters. 
The mass eigenvalues of the $SU(2)_V$ 5-plet and 3-plet Higgs bosons $(m_{H_5}$ and $m_{H_3}$) are then given by 
\begin{align}
m_{H_5}^2 &= v^2\left(\lambda_3 c_\beta^2 -\frac{3}{2}\lambda_5 s_\beta^2 \right) - \frac{v\mu_1}{\sqrt{2}}t_\beta s_\beta -3\sqrt{2} v\mu_2 c_\beta ,\quad
m_{H_3}^2 = -\frac{v^2}{2}\lambda_5 -\frac{v\mu_1}{\sqrt{2}c_\beta}. \label{m3m5}
\end{align}
The relation between the masses of the CP-even Higgs bosons and the
matrix elements $M_{ij}^2$ in the basis of $(\xi_r',\phi_r)$
can be obtained by using Eqs.~(\ref{hsm-alpha}) and (\ref{lam}) after the replacement of $H_S \to H_1$. 
The matrix elements have the following expressions in terms of the potential parameters:
\begin{align}
M^2_{11}&= v^2(3\lambda_2+\lambda_3)c_\beta^2 -\frac{v\mu_1}{\sqrt{2}} t_\beta s_\beta + \frac{3v\mu_2}{\sqrt{2}}c_\beta,\notag\\
M^2_{22}&=8v^2\lambda_1s_\beta^2, \notag\\
M^2_{12}&=\sqrt{\frac{3}{2}} \left[v^2(2\lambda_4+\lambda_5)c_{\beta} +\frac{v\mu_1 }{\sqrt{2}}\right]s_{\beta}.
\label{eq:Meven_ele}
\end{align}
Summarizing, the GM model can be described by 7 independent quantities, after fixing $v$ and $m_h$:
\begin{align}
m_{H_5}^{},~~m_{H_3}^{},~~m_{H_1}^{},~~t_{\beta},~~s_\alpha,~~\mu_1,~~\mu_2. 
\end{align}

Let us now discuss the decoupling and the alignment limits. 
The decoupling limit is realized by taking $t_\beta \to \infty$ with a negative value of $\mu_1$. 
In this limit, $m_{H_3}^{}$ and $m_{H_5}^{}$ become infinity as seen by Eq.~(\ref{m3m5}). 
In addition, among the matrix elements given in Eq.~(\ref{eq:Meven_ele}), only $M_{11}^2$ becomes infinity, so that 
we obtain $m_{H_1}^{} \to \infty$ and $\alpha \to 0$. 
Differently from the HSM and the 2HDM, the alignment limit cannot be taken without the decoupling limit, because 
$t_\beta \to \infty$ is necessarily required in order to have the doublet field $\mathbf{\Phi}$ carrying the 
whole VEV $v$.

Finally, let us explicitly write the kinetic and Yukawa Lagrangian terms of the GM model: 
\begin{align}
\mathcal{L}&=\frac{1}{2}\text{tr}(D_\mu \mathbf{\Phi})^\dagger (D^\mu \mathbf{\Phi})
+\frac{1}{2}\text{tr}(D_\mu \Delta)^\dagger (D^\mu \Delta) 
-\left(y_t\bar{Q}_L^3 \Phi^c\, t_R^{} +y_b\bar{Q}_L^3 \,\Phi\, b_R^{} +y_\tau\bar{L}_L^3 \,\Phi\, \tau_R^{} +\text{h.c.}\right), \label{yuk}
\end{align}
where the covariant derivatives are expressed as 
\begin{align}
D_\mu \mathbf{\Phi} =\partial_\mu\mathbf{\Phi} -ig_2\frac{\tau^a}{2}W_\mu^a\mathbf{\Phi} + ig_1B_\mu \mathbf{\Phi}\frac{\tau^3}{2},\\
D_\mu \Delta =\partial_\mu\Delta -ig_2t^aW_\mu^a\Delta + ig_1B_\mu \Delta t^3. \label{cov}
\end{align}
The trilinear interaction terms of the physical Higgs bosons with SM fermions and
gauge bosons are given by:
\begin{align}\label{eq:int_gm}
\begin{split}
{\cal L}_{\text{int}} = \, & \ \ \ \frac{2m_W^2}{v}\left(c_{hVV}^{}h + c_{H_1VV}H_1 -\frac{c_\beta}{\sqrt{3}} H_5^0 \right)W_\mu^+ W^{-\mu}
 \\ & + \frac{m_Z^2}{v}\left(  c_{hVV}h +  c_{H_1VV} H_1 +\frac{2c_\beta}{\sqrt{3}}H_5^0 \right)Z_\mu Z^\mu  \\
& -\sum_{f=t,b,\tau} \frac{m_f}{v}\left(\frac{c_\alpha}{s_\beta}\bar{f}f h + \frac{s_\alpha}{s_\beta}\bar{f}f H_1  
-2iI_f \cot\beta \bar{f}\gamma_5f H_3^0  \right) \\
&- \frac{\sqrt{2}}{v}\cot\beta \left[\bar{t}(m_b P_R - m_t P_L) bH_3^+ + \bar{\nu}_\tau \, m_\tau  P_R \,\tau H_3^+  +\text{h.c.}\right], 
\end{split}
\end{align}
where
\begin{align}
c_{hVV} = s_\beta c_\alpha - \frac{2\sqrt{6}}{3}\,c_\beta s_\alpha,~~c_{H_1VV} =s_\beta s_\alpha + \frac{2\sqrt{6}}{3}\,c_\beta c_\alpha. \label{kappas}
\end{align}
There are several remarkable points to be stressed about the interaction terms given in Eq.~(\ref{eq:int_gm}). 
First, the 5-plet Higgs bosons have a fermiophobic nature, i.e., they do not couple to fermions, 
but they couple to the SM gauge bosons. 
Second, the coupling structure of the 3-plet Higgs bosons $H_3^0$ and $H_3^\pm$
is the same as that for $A$ and $H^\pm$ in the Type-I 2HDM. 
Finally, the SM-like Higgs boson $h$ coupling to the gauge bosons can be
larger than the SM prediction, because of the factor of $2\sqrt{6}/3$ as shown in Eq.~(\ref{kappas}). 
This is a peculiarity of the GM model, because $\kappa_V \leq 1$ in the HSM and the 2HDMs. 
\begin{table}[t]
\begin{center}
\begin{tabular}{l|l|l|l|l}\hline\hline 
           & Decoupling limit     & Alignment limit & $\kappa_V \equiv g_{hVV}^{}/g_{hVV}^{\text{SM}}$ & $\kappa_f \equiv g_{hff}^{}/g_{hff}^{\text{SM}}$ \\\hline
HSM        & $m_S^2 \to \infty$   & $s_\alpha \to 0$ & $c_\alpha$ & $c_\alpha$ \\\hline
2HDM       & $M^2\to \infty$      & $s_{\beta-\alpha} \to 1$ & $s_{\beta-\alpha}$ & $s_{\beta-\alpha} + \xi_f c_{\beta-\alpha}$\\\hline
GM model   & $t_\beta \to \infty $ & $t_\beta \to \infty$  & $s_\beta c_\alpha - 2\sqrt{2/3}\,c_\beta s_\alpha$ & $c_\alpha/s_\beta$ \\
\hline\hline
\end{tabular} 
\end{center}
\caption{Summary of the relevant properties of the HSM, the 2HDM and the GM model. } 
\label{prop2}
\end{table}

In Table~\ref{prop2}, we summarize the relevant properties of the three extended Higgs sectors here considered.

\section{Upper bound on the extra Higgs boson masses \label{sec:bound}}

As discussed in the previous section,
if the alignment limit is not realized, 
there must be an upper bound on the extra Higgs boson masses. 
In this section, we numerically evaluate such upper bound by taking into account perturbative unitarity,
vacuum stability and triviality constraints. 
In addition, we will concern about the compatibility with the experimental values of
the $S$ and $T$ parameters introduced by Peskin and Takeuchi~\cite{Peskin} which describe the oblique
corrections to the electroweak processes. 
In particular, we require 
the predictions of $\Delta S \equiv S_{\text{NP}}-S_{\text{SM}}$ and $\Delta T \equiv T_{\text{NP}}-T_{\text{SM}}$
to be within the 95\% CL region of the $\Delta \chi^2$ fit (corresponding to $\Delta \chi^2 \leq 5.99$), 
where $S_{\text{NP}}~(T_{\text{NP}})$ and $S_{\text{SM}}~(T_{\text{SM}})$ are  
the New Physics and the SM prediction for the $S$ ($T$) parameters, respectively. 
The observed values of $\Delta S$ and $\Delta T$ are~\cite{st} 
\begin{align}
\Delta S_{\text{exp}} = 0.05\pm 0.09,\quad \Delta T_{\text{exp}} = 0.08\pm 0.07, 
\end{align}
with the correlation factor $\rho_{\text{ST}}$ being 0.91. 
The analytic formulae for $\Delta S$ and $\Delta T$ are given in Appendix~\ref{appST} 
for each of the extended Higgs models.

We introduce $M_{\text{2nd}}$ as the mass of the next-to-lightest Higgs boson, assuming the lightest one to be the discovered Higgs boson $h$.
By definition, $M_{\text{2nd}}$ corresponds to $m_{H_S}^{}$ in the HSM, while
in the 2HDM and the GM model, it is defined by: 
\begin{equation}
\begin{split}
M_{\text{2nd}} = \text{Min}(m_{H^\pm}^{},m_A^{},m_H^{}) &~~\text{in the 2HDM},~~\\
M_{\text{2nd}} = \text{Min}(m_{H_5}^{},m_{H_3}^{},m_{H_1}^{}) &~~\text{in the GM model}. 
\end{split}
\end{equation}
In the following analysis, we enforce the bounds in two steps denoted by ``Bound A'' and ``Bound B''. 
First, we impose the tree level perturbative unitarity, the vacuum stability and the constraint from the $S$ and $T$ parameters
(Bound A). 
In addition to the constraints of Bound A, we then impose 
the improved vacuum stability and the triviality constraints (Bound B) evaluated
by using the running coupling constants arising from the
one-loop renormalization group equations (RGEs) (see Appendix~\ref{sec:rge}). 
Thus, Bound A does not depend on the energy scale, while Bound B does. 
In particular, Bound B depends on the theory cutoff connected with the appearance of
Landau poles or vacuum instability, both requiring further contributions from a more
fundamental theory.
\begin{figure}[t]
\begin{center}
\includegraphics[scale=0.32]{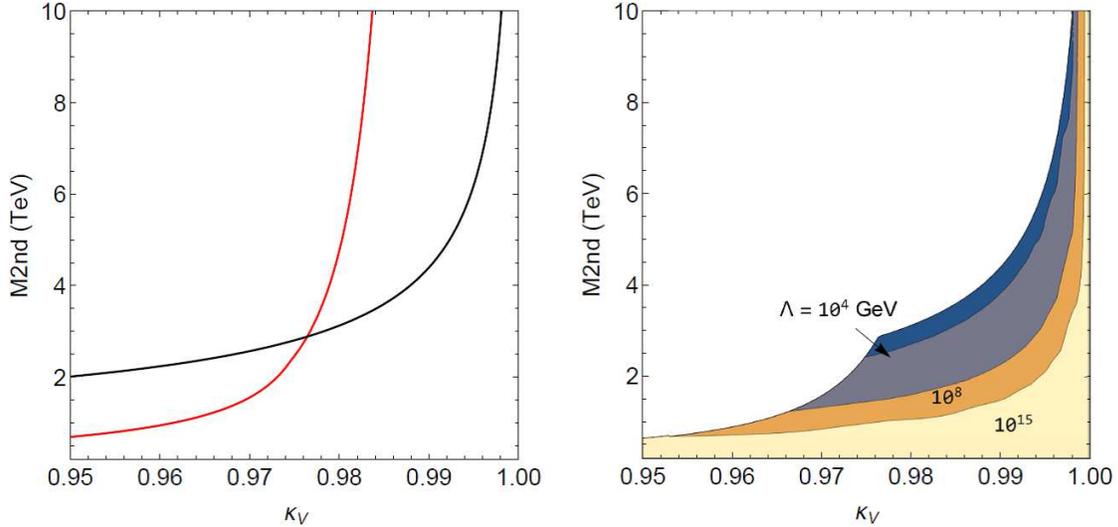}
\caption{Bounds on $M_{\text{2nd}} = m_{H_S}^{}$ in the HSM as functions of $\kappa_V = g_{\text{hVV}}/
g_{\text{hvv}}^{\,\text{SM}}$.
On the left panel, we show the upper limit on $M_{\text{2nd}}$ given by the tree level perturbative unitarity and the vacuum stability constraints (black line) 
and that given by the electroweak $S$ and $T$ parameters (red line).
On the right panel, the upper limit on $M_{\text{2nd}}$ is shown by imposing Bound B, where the
cutoff of the theory is shown as contours on this plane.}
\label{fig1}
\end{center}
\end{figure}
Let us discuss the results for the HSM.
We scan over the parameters $\lambda_{\Phi S}$, $\lambda_S$ and $\mu_S^{}$ of the potential and,
in the left panel of Fig.~\ref{fig1}, 
we show the upper bound on $M_{\text{2nd}}= m_{H_S}^{}$ as a function of $\kappa_V = g_{\text{hVV}}/
g_{\text{hvv}}^{\,\text{SM}}$ under Bound A. 
We separately take into account
the perturbative unitarity and vacuum stability constraints
(black curve) and the constraints from the $S$ and $T$ parameters (red curve). 
We can see that, at $\kappa_V^{} \simeq 0.975$, the constraint providing the most stringent bound on $M_{\text{2nd}}$
is interchanged, namely, for $\kappa_V \lesssim 0.975$ it is the one from the $S$ and $T$ parameters,
while for $\kappa_V \gtrsim 0.975$, the perturbative unitarity and the vacuum stability bounds give the stronger bound. 
When $\kappa_V^{}$ is getting close to 1, the bound on $M_{\text{2nd}}$ becomes milder 
and it disappears in the alignment limit $\kappa_V^{} \to 1$. 

On the right panel, we show the allowed values of
$M_{\text{2nd}}$ under Bound B as a function of $\kappa_V^{}$.
The cutoff scale $\Lambda$ of the theory is indicated by the contours in this plot.
We conclude that, if the deviation in the $hVV$ coupling from the SM prediction is measured 
to be, e.g., larger than 1\% ($\kappa_V^{}\leq 0.99$),
the extra Higgs boson mass in the HSM is expected to be below $\sim 4 \, \text{TeV}$. 
Furthermore, if we require the theory cutoff to be larger than e.g., $10^8$ ($10^{15}$) GeV, then 
the extra Higgs boson mass is expected to be below $\sim 2~(1)$ TeV, assuming $\kappa_V^{} \leq 0.99$.
In other words, if there is no other NP contribution modifying the running of the coupling
constants of the HSM up to $10^{15}$ GeV, then a $1\%$ deviation in the $hVV$ coupling implies
a bound on the extra Higgs boson mass around 1 TeV.
As shown in Sec.~\ref{sec:compl}, these values are still allowed by direct searches and
by the signal strength analysis at the LHC.
\begin{figure}[t]
\begin{center}
\includegraphics[scale=0.82]{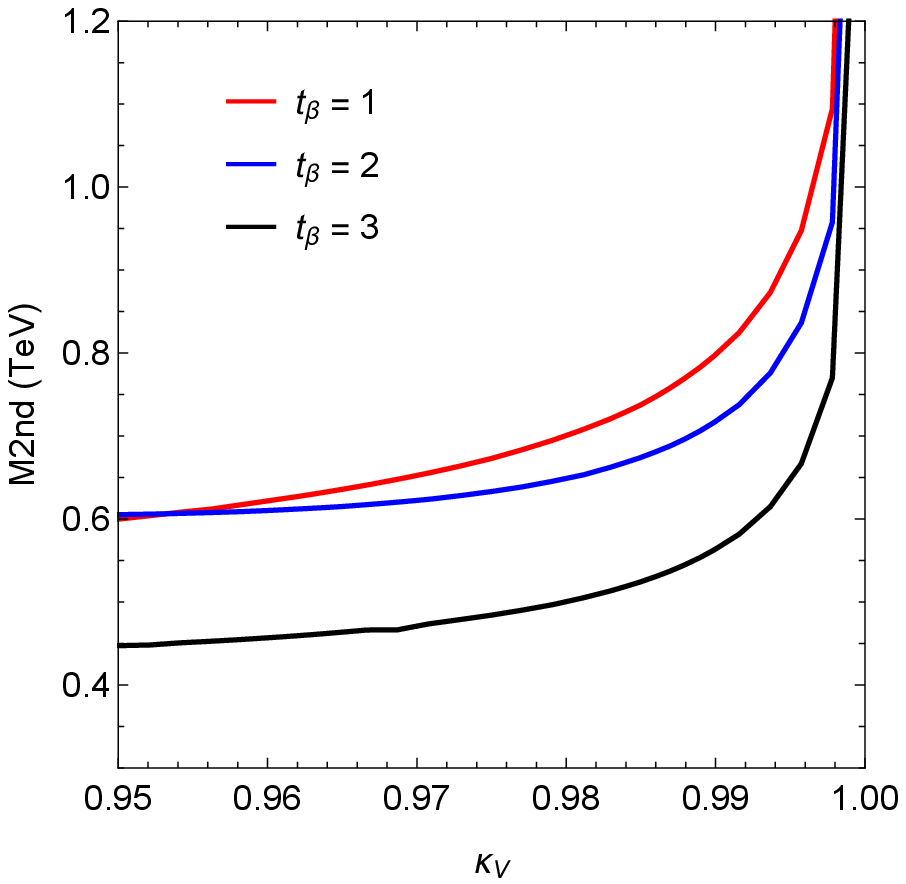}
\includegraphics[scale=0.82]{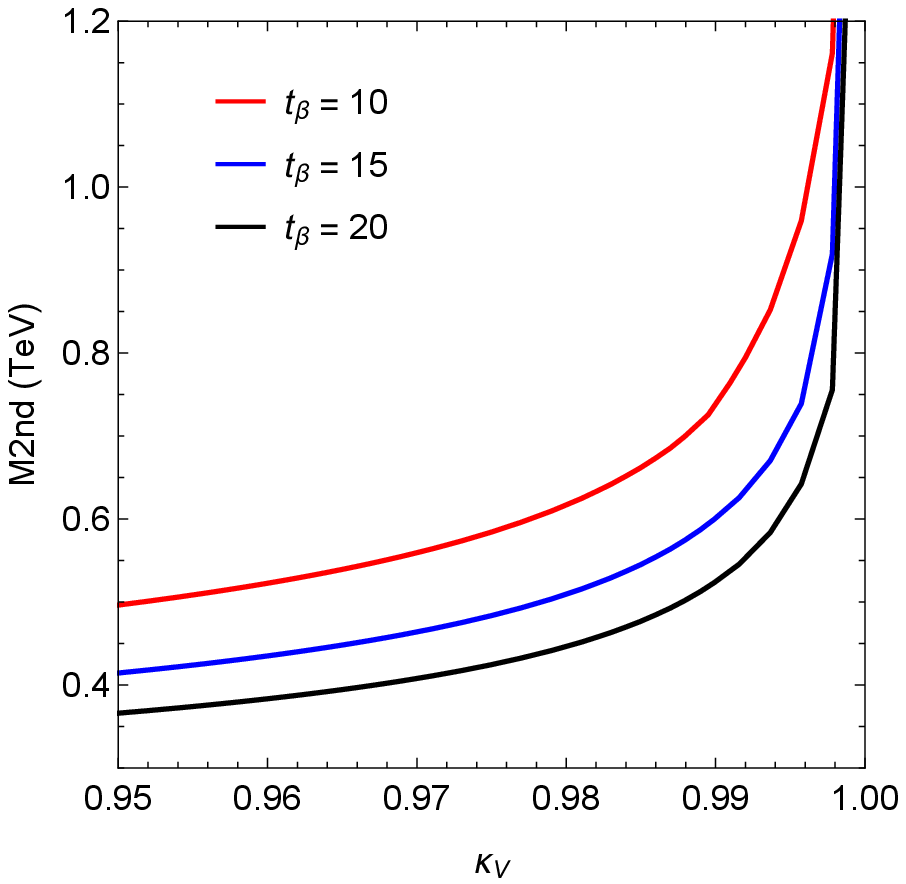}
\caption{Upper limit on $M_{\text{2nd}} = \text{Min}(m_{H^\pm}^{},m_A^{},m_H^{})$ obtained by imposing Bound A as a function of
$\kappa_V$ in the 2HDMs with $t_\beta = 1, 2, 3$ (left panel) and  $t_\beta = 10, 15, 20$ (right panel).}
\label{fig2}
\end{center}
\end{figure}

Next, we derive the upper limit on $M_{\text{2nd}}$ in the 2HDM. 
The values of $m_H^{}$, $m_A^{}$, $m_{H^\pm}$ and $M^2$ are scanned so as to 
extract the maximal value of $M_{\text{2nd}}$ for each fixed value 
of $t_\beta$ and $\kappa_V^{}$. We also scan over the sign of $c_{\beta-\alpha}$. 
In Fig.~\ref{fig2}, we show the upper limit on $M_{\text{2nd}}$ as a function of 
$\kappa_V^{}$ obtained by imposing Bound A. 
The left (right) panel shows the result for the case with $\tan\beta=1,2$ and 3 (10, 15 and 20). 
Regardless the value of $t_\beta$, the maximally allowed value of $M_{\text{2nd}}$ is getting small 
as $|\kappa_V^{}-1|$ becomes large. 
For example, for $\kappa_V^{} \leq 0.99$, we find the upper limit on $M_{\text{2nd}}$ to be about 
800, 550 and 500 GeV for $\tan\beta=1,~3$ and 20, respectively. 
Similar to the HSM, the bound disappears in the alignment limit $\kappa_V^{} \to 1$. 
We can see that the extracted limit on $M_{\text{2nd}}$ in the 2HDM (typically less than 1 TeV) 
is much smaller than that given in the HSM (typically a few TeV). 
We note that this result does not depend on the type of Yukawa interaction. 
\begin{figure}[t]
\begin{center}
\includegraphics[scale=0.3]{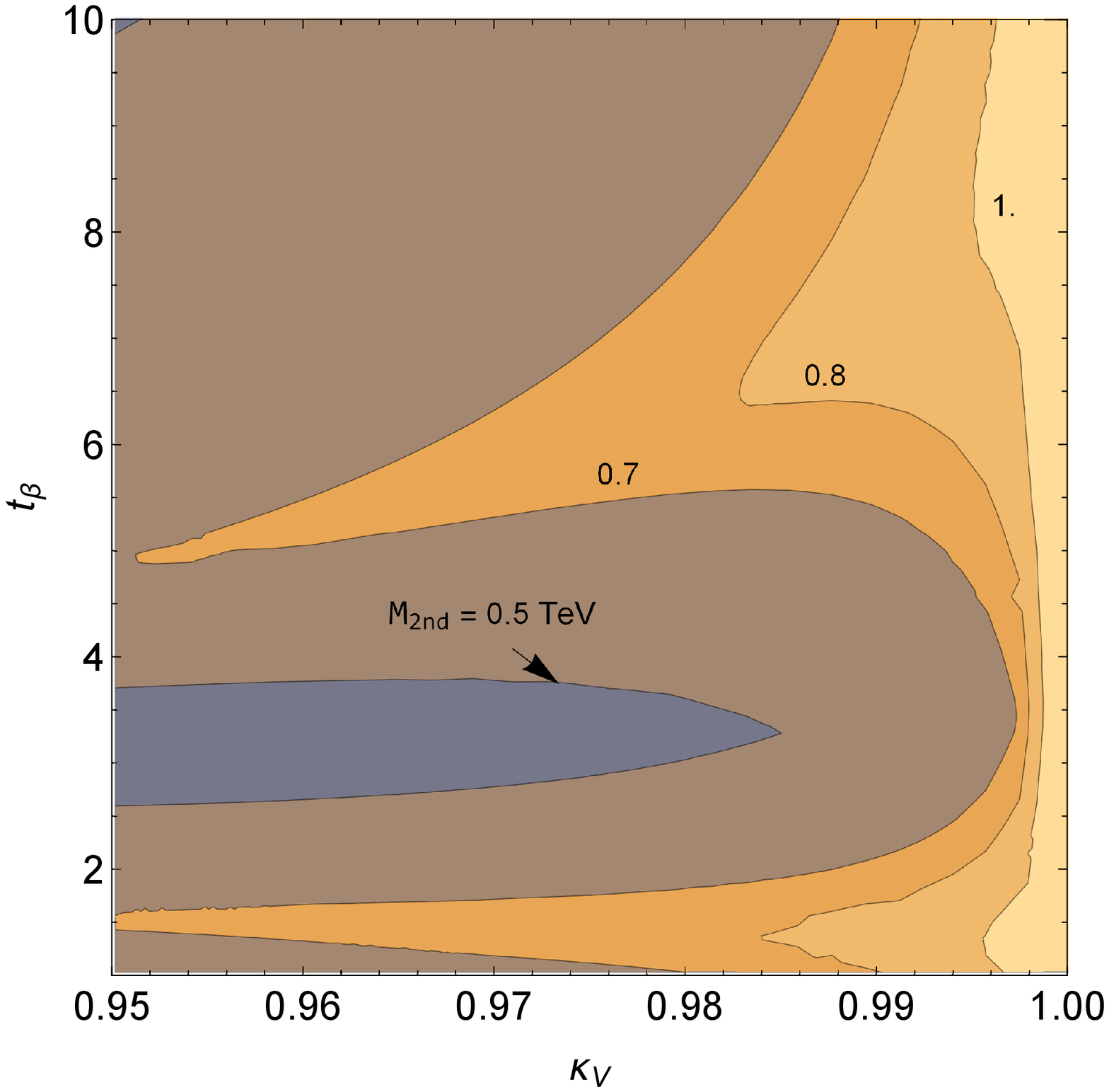}
\includegraphics[scale=0.33]{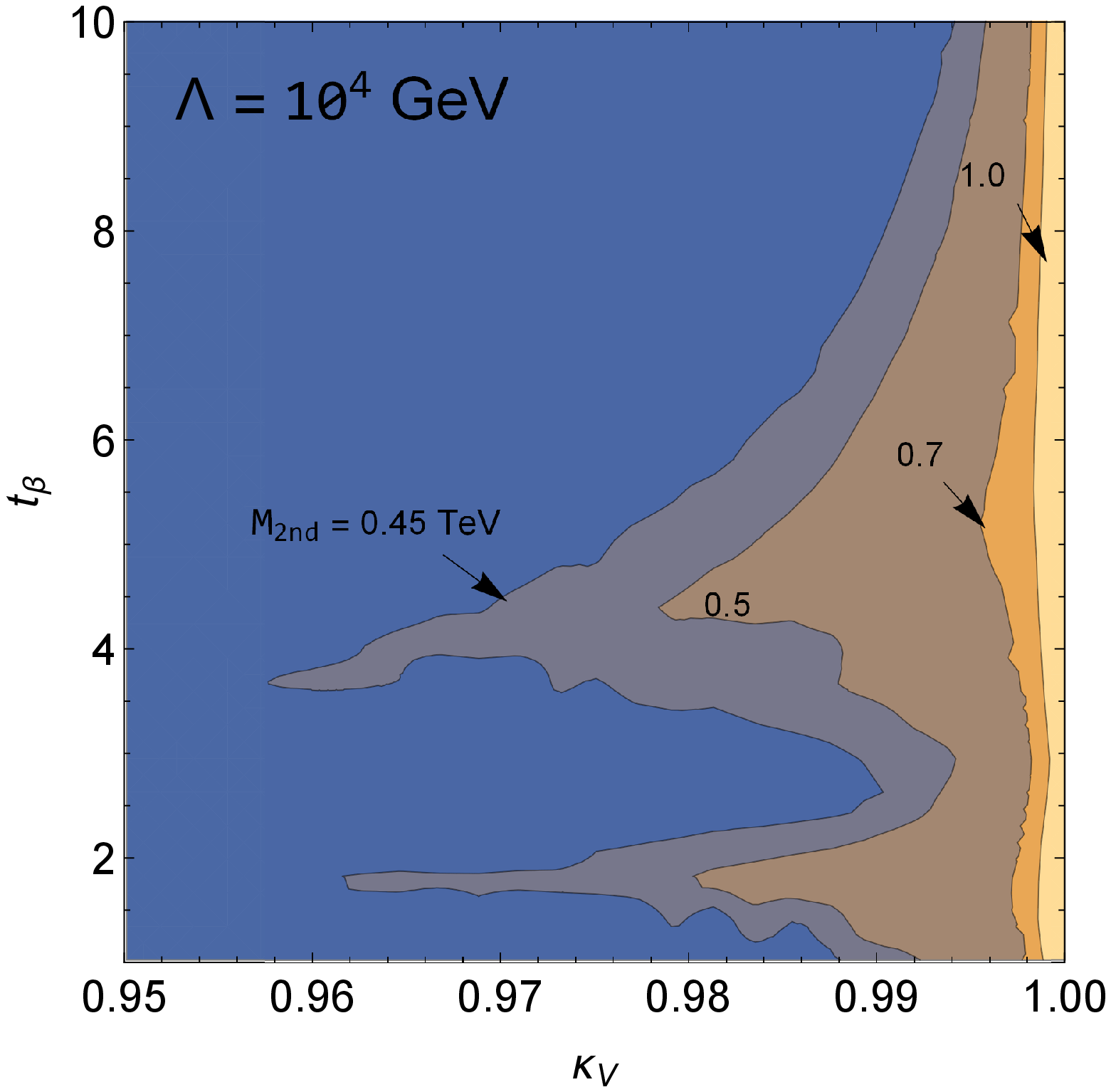}
\includegraphics[scale=0.33]{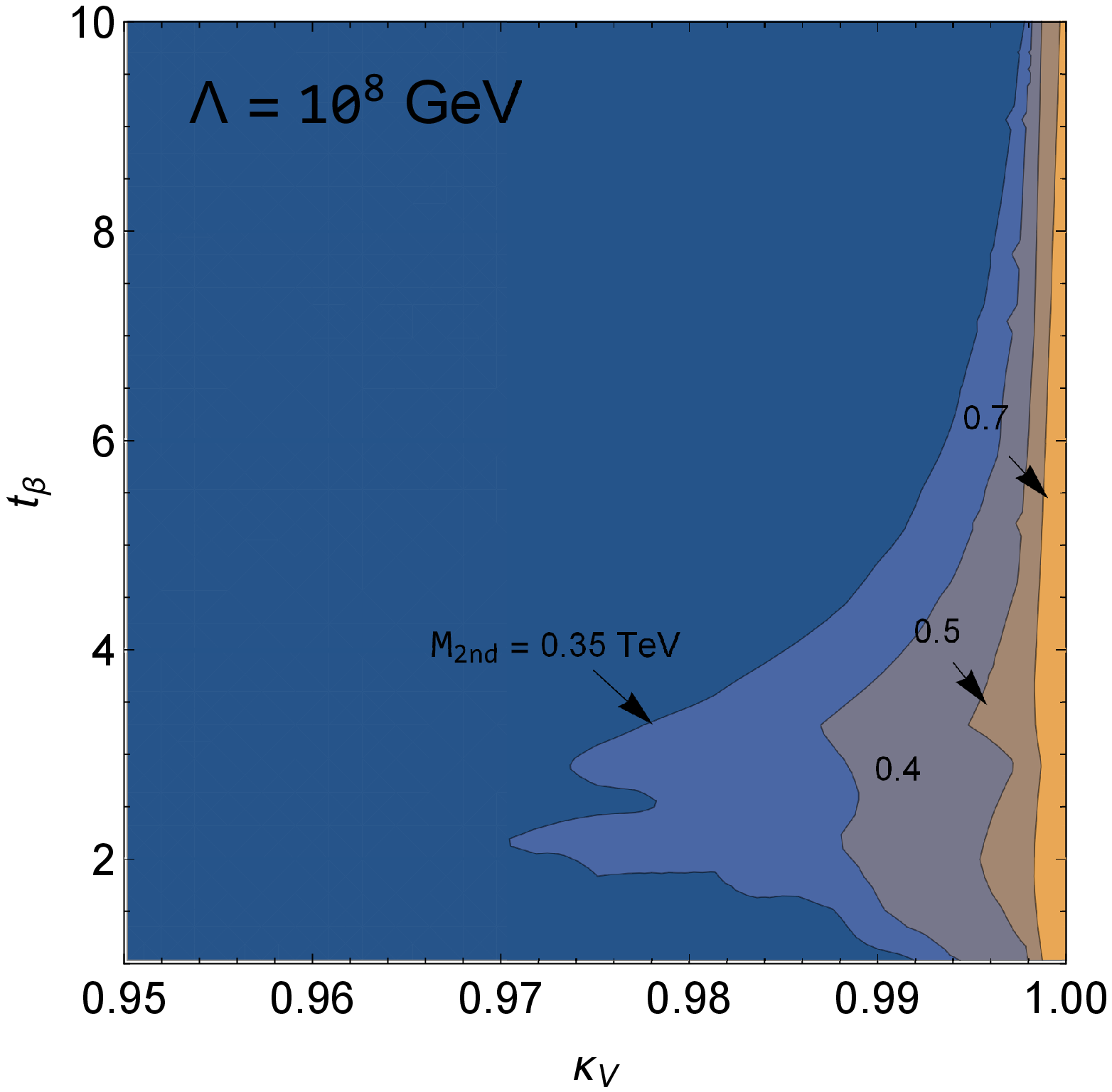}
\end{center}
\caption{Contour plots for the maximally allowed value of $M_{\text{2nd}} = \text{Min}(m_{H^\pm}^{},m_A^{},m_H^{})$ in the 2HDMs
obtained by imposing Bound A (left), Bound B with $\Lambda = 10^4$ GeV (center) 
and Bound B with $\Lambda = 10^8$ GeV (right) on the $\kappa_V^{}$--$t_\beta$ plane. }
\label{fig3}
\end{figure}
In Fig.~\ref{fig3}, we show the contour plot for the maximally allowed value of 
$M_{\text{2nd}}$ in the 2HDM on the $\kappa_V$--$t_\beta$ plane. 
The left panel is obtained by imposing Bound A, while the center and right panels correspond to
Bound B with $\Lambda = 10^4$ GeV (center) and $10^8$ GeV (right). 
Let us first discuss the left panel.
For a fixed value of $\kappa_V^{}$, the maximally allowed value of $M_{\text{2nd}}$ 
strongly depends on $t_\beta$ 
(according to Fig.~\ref{fig2}) and
$M_{\text{2nd}}$ is typically expected to be below 1 TeV,
also for $\kappa_V$ quite close to 1.
If we impose Bound B (center and right panels), the allowed values
of $M_{\text{2nd}}$ become significantly smaller 
than those in the left panel. 
As an example, for $\Lambda = 10^8$ GeV, $M_{\text{2nd}} 
\lesssim 400$ GeV is expected if $\kappa_V^{} \leq 0.99$. 
We note that the dependence on the type of the Yukawa interactions 
only appears through the bottom Yukawa terms in the $\beta$ functions (see Appendix~\ref{sec:rge}), so 
it is negligibly small and Fig.~\ref{fig3} bounds are applicable to all the 2HDM Types.
As shown in Sec.~\ref{sec:compl} for the 2HDM, the LHC bounds obtained after the first run 2 data analysis
are already competitive with the theoretical ones here derived.

\begin{figure}[t]
\begin{center}
\includegraphics[scale=0.7]{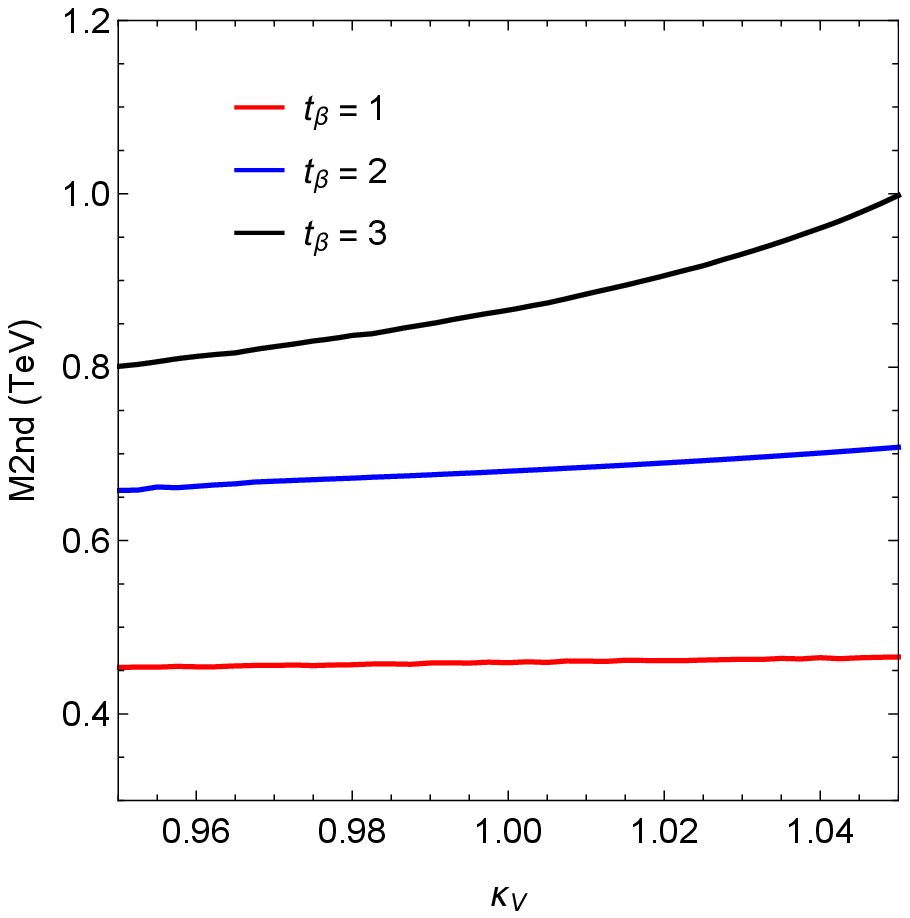}\hspace{5mm}
\includegraphics[scale=0.7]{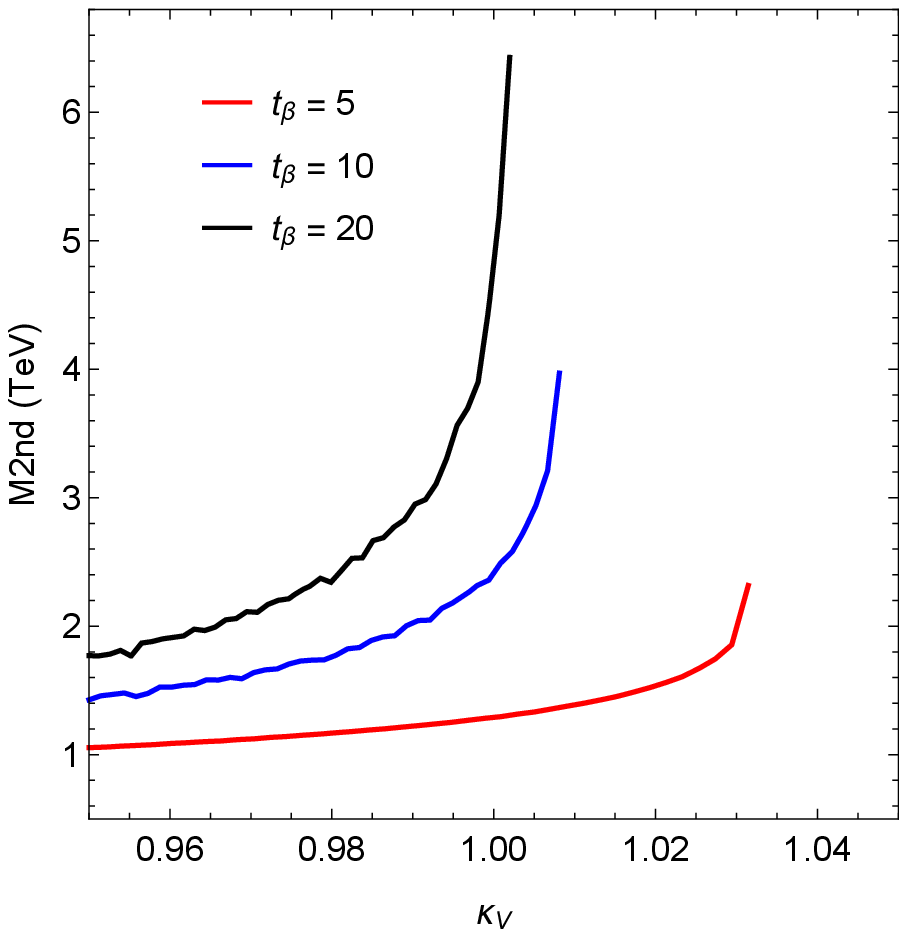}
\caption{Upper limit on $M_{\text{2nd}} = \text{Min}(m_{H_5}^{},m_{H_3},m_{H_1})$ obtained by imposing Bound A as a function of
$\kappa_V$ in the GM model with $t_\beta = 1, 2, 3$ (left panel) and  $t_\beta = 5, 10, 20$ (right panel).}
\label{fig4}
\vspace{5mm}
\includegraphics[scale=0.34]{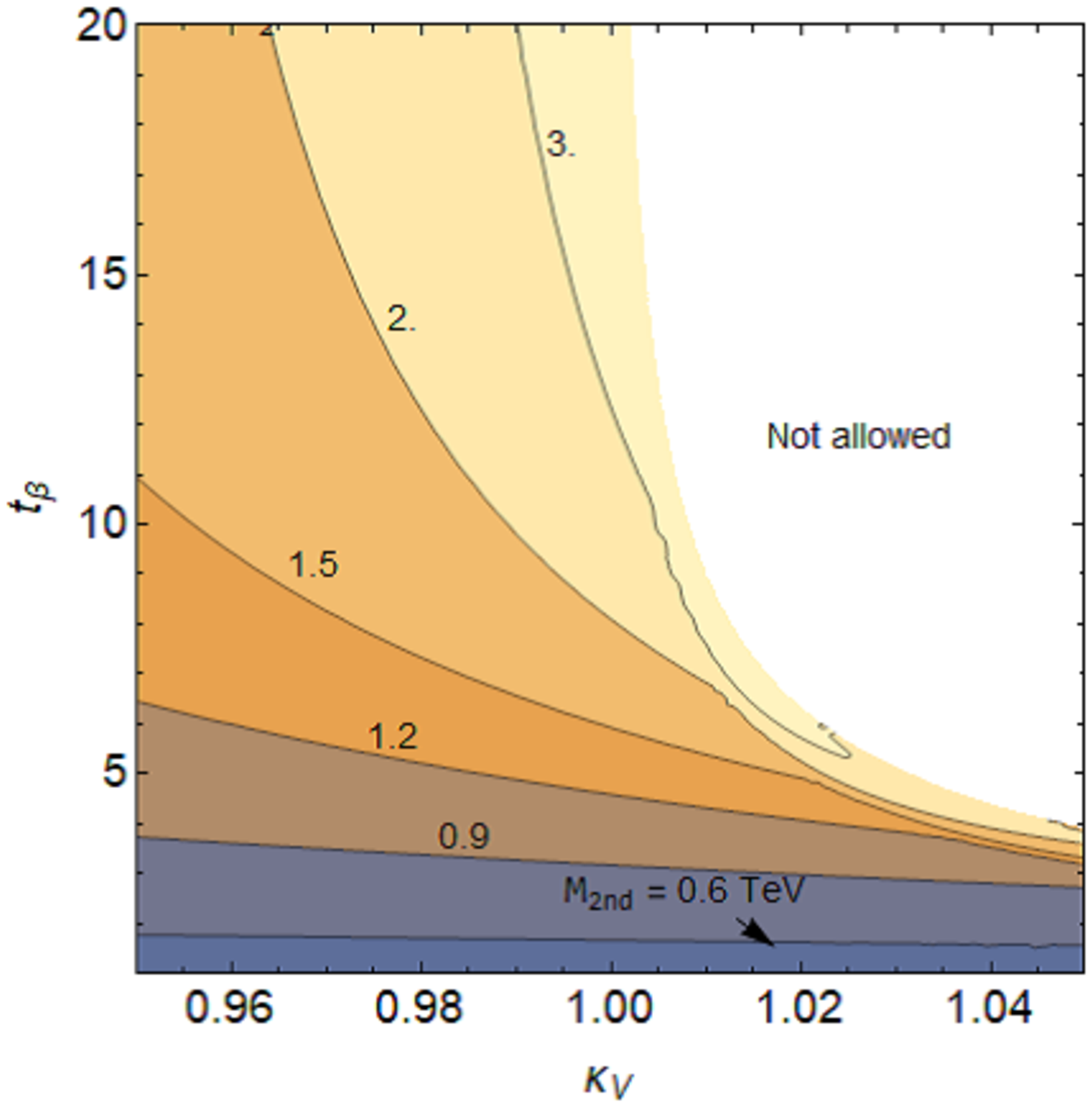}
\includegraphics[scale=0.34]{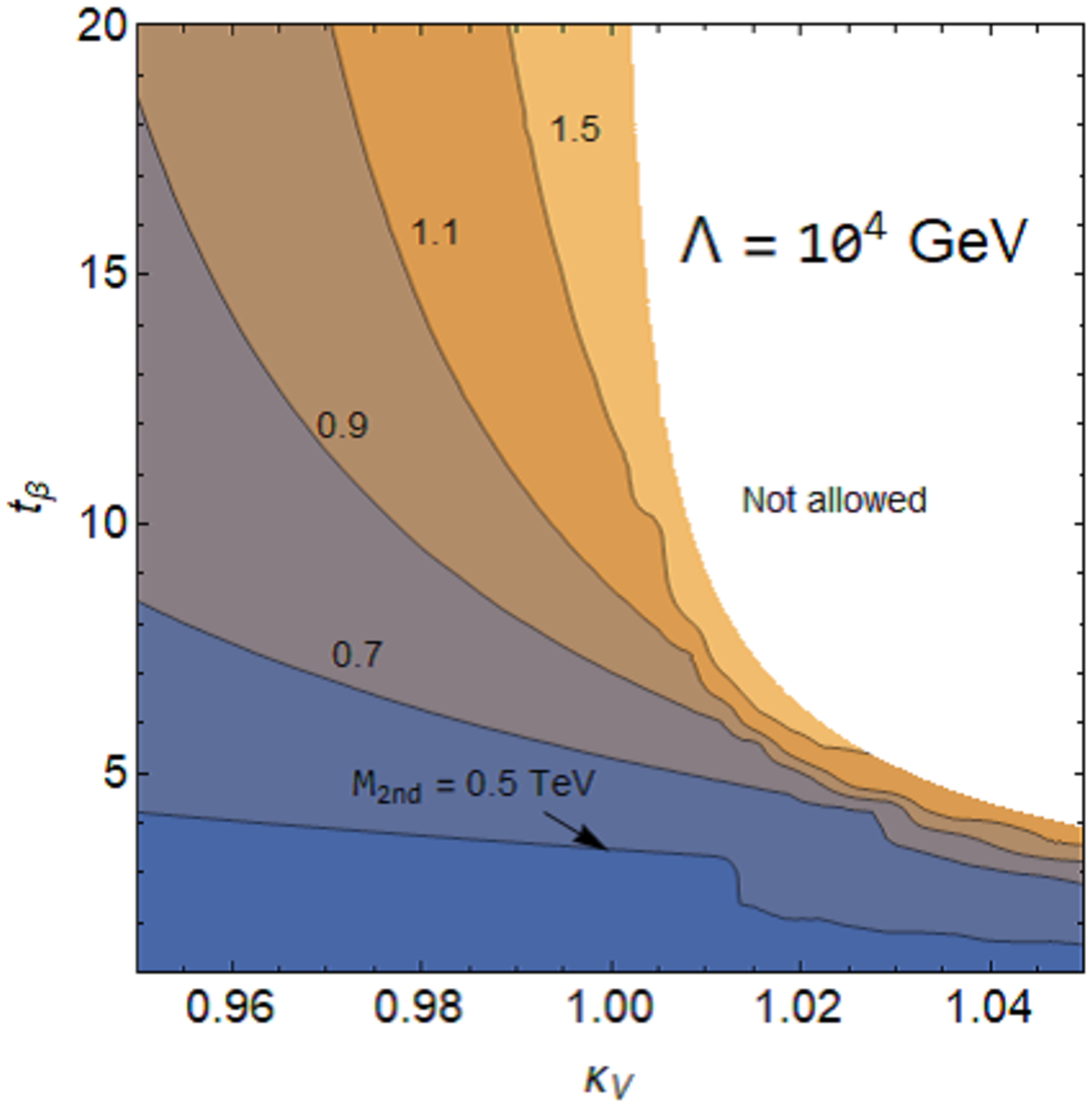}
\includegraphics[scale=0.34]{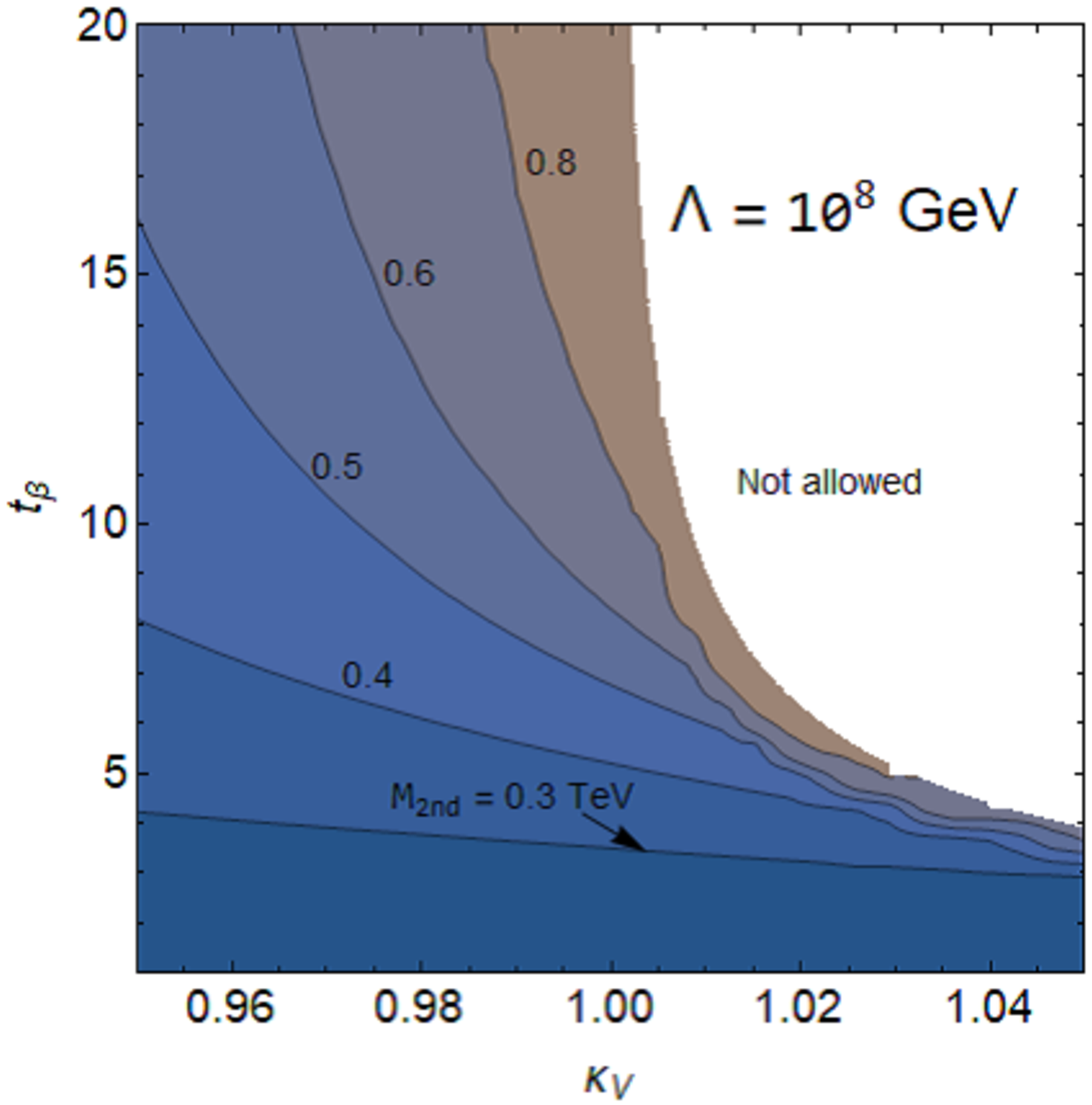}
\caption{Contour plots for the maximally allowed value of $M_{\text{2nd}} = \text{Min}(m_{H_5}^{},m_{H_3},m_{H_1})$ in the GM model
obtained by imposing Bound A (left), Bound B with $\Lambda = 10^4$ GeV (center) 
and Bound B with $\Lambda = 10^8$ GeV (right) on the $\kappa_V^{}$--$t_\beta$ plane. }
\label{fig5}
\end{center}
\end{figure}
Finally, let us discuss the results in the GM model. 
The values of $m_{H_5}$, $m_{H_3}$, $m_{H_1}$, $\mu_1$ and $\mu_2$ are scanned 
with an enough wide range to extract the maximal value of $M_{\text{2nd}}$ for each fixed value 
of $t_\beta$ and $\kappa_V^{}$.
For a fixed value of $t_\beta$, there can be two possible values of $\alpha$ 
giving the same value of $\kappa_V^{}$ (e.g., for $t_\beta = 3$, 
both $\alpha \simeq -0.91$ and $\alpha\simeq -0.087$ give $\kappa_V=0.99$), 
so we here consider both values of $\alpha$. 
In Fig.~\ref{fig4}, we show the maximally allowed value of $M_{\text{2nd}}$ as a function of $\kappa_V^{}$ 
by imposing Bound A for $t_\beta=1,2$ and 3 (left panel) and 5, 10 and 20 (right panel). 
As we mentioned in Sec.~\ref{sec:gm}, $\kappa_V^{}>1$  is allowed and its
maximal value is found at $t_\alpha = -2\sqrt{6}/(3t_\beta)$.
For example, for $t_\beta = 3 (5)$, we obtain Max$(\kappa_V) \simeq 1.08 (1.03)$. 
For this reason, there is an end point in the curves shown on the right panel. 
We find that the maximal allowed value of $M_{\text{2nd}}$ monotonically increases as $\kappa_V$ is getting large. 
However, as long as we take a finite value of $t_\beta$, there is always an upper limit on $M_{\text{2nd}}$, i.e., the decoupling limit
cannot be taken unless
$t_\beta \rightarrow \infty$ (corresponding to the alignment limit). 

In Fig.~\ref{fig5},  we show the contour plots for $M_{\text{2nd}}$ on the $\kappa_V$--$t_\beta$ plane
by imposing Bound A (left) and
by imposing Bound B with $\Lambda = 10^4$ GeV (center) and $10^8$ GeV (right). 
The white regions are not allowed due to the existence of a maximal value for $\kappa_V^{}$. 
We can see that the maximal value of $M_{\text{2nd}}$ is smoothly getting large when $t_\beta$ and/or $\kappa_V^{}$ becomes large. 
As compared to the case of the 2HDMs, $M_{\text{2nd}}$ can be typically above 1 TeV
when we take a large value of $t_\beta$. 
For example, for $\kappa_V<0.99$, 
we obtain the maximal allowed value of $M_{\text{2nd}}$ to be about 1 (2) TeV in the case of $t_\beta = 5 (10)$. 
If we impose Bound B, the maximal $M_{\text{2nd}}$ becomes smaller, 
but still $M_{\text{2nd}}>1$ TeV is allowed for $\kappa_V<0.99$ with $\Lambda = 10^4$ GeV.
 
\section{Complementarity between experimental and theoretical bounds \label{sec:compl}}

In this section, we discuss the complementarity between the bounds discussed in Sec. \ref{sec:bound} and those from the LHC data. 
For the former, we impose Bound A which does not depend on the cutoff $\Lambda$.  
For the latter, we take into account the bounds from direct searches for 
additional Higgs bosons and from the data for the discovered Higgs boson $h(125)$.  
As we saw in the previous section, Bound A provides an upper limit on $M_{\text{2nd}}$ depending on the value of $\kappa_V^{}$. 
Conversely, bounds from LHC data typically provide a lower limit on the mass 
of the extra Higgs boson. 
Therefore, by combining these two types of bounds, we can 
further narrow down the possible allowed region for $M_{\text{2nd}}$ for each extended Higgs model. 

Here, we consider the masses of extra Higgs bosons to be above 350 GeV, i.e., beyond the threshold of the decay of 
neutral Higgs bosons into $t\bar{t}$. 
For the 2HDM and the GM model, we take $t_\beta \geq 1$, because the case with $t_\beta < 1$ is 
highly disfavored by the various $B$ physics experiments~\cite{Bphys1,Bphys2}. 

Let us discuss the bounds from direct searches. 
In the aforementioned scenarios, the main decay modes of the extra Higgs bosons are the following:
\begin{align}
H_S   &\to t\bar{t}/VV/hh~~(\text{in the HSM}), \label{decay_HSM}\\
H     &\to t\bar{t}/VV/hh,~~A \to t\bar{t}/Zh,~~H^\pm \to tb/W^\pm h~~(\text{in the 2HDM}),\label{decay_2HDM} \\
H_1 &\to t\bar{t}/VV/hh,~~H_3^0 \to t\bar{t}/Zh,~~H_3^\pm \to tb/W^\pm h, \notag\\
H_5^0 &\to VV,~~H_5^\pm \to W^\pm Z,~~H_5^{\pm\pm} \to W^\pm W^\pm~~(\text{in the GM model}), \label{decay_GM}
\end{align}
where $V=W,Z$. We note that in the 2HDM with $s_{\beta-\alpha} = 1$, 
the main decay mode of the extra neutral Higgs bosons becomes $H/A \to t\bar{t}$, but it can be replaced by the other 2-fermion final 
states in the Type-II, -X, -Y 2HDMs
if we take a large enough value of $t_\beta$. 
For example, for $\tan\beta \gtrsim 6(10)$, $s_{\beta-\alpha}= 1$, and $m_\Phi^{}(=m_H^{}=m_A^{}=m_{H^\pm}^{}) = M = 500$ GeV,
the channel $H/A \to b\bar{b}$ ($H/A \to \tau\tau$)
can be dominant instead of the $t\bar{t}$ mode in the Type-II (-X) 
2HDM\footnote{In the Type-X 2HDM, $H^\pm \to tb$ can also be replaced by $H^\pm \to \tau^\pm\nu$
for $t_\beta\gtrsim 10$, $s_{\beta-\alpha}= 1$, and $m_\Phi^{} = M = 500$ GeV. This is not the case for the other 2HDM types.}. 
However, as we will show below, in a large $t_\beta$ scenario i.e., $t_\beta = {\cal O}(10)$, 
the production cross section of the extra Higgs bosons is highly suppressed, so that 
the constraint on the mass of the extra Higgs boson becomes weak regardless of the type of 
Yukawa interaction\footnote{In the LHC Run-2 and the High-Luminosity LHC experiments, 
the large $t_\beta$ scenario can also be constrained~\cite{finger,Yokoya} via the pair production 
of the extra Higgs bosons, whose cross section does not depend on $t_\beta$. }. 

First, let us discuss the searches for additional neutral Higgs bosons $({\cal H})$. 
We here consider the following processes:
\begin{align} 
\text{(i)}~gg \to {\cal H} \to t\bar{t},~~
\text{(ii)}~gg \to {\cal H} \to hh, ~~ \text{(iii)}~gg \to {\cal H} \to ZZ,~~ \text{(iv)}~gg \to {\cal H} \to Zh. 
\end{align} 
We note that, although there are other production modes for ${\cal H}$ such as the vector boson fusion
and the vector boson associated processes, 
the cross section of these modes are negligibly small in the scenario with $\kappa_V^{}\sim 1$. 
Thus, we only take into account the gluon fusion production in this analysis. 
Regarding the top quark associated production,
the search for $pp \to t\bar{t} {\cal H}$ followed by ${\cal H} \to t\bar{t}$ 
decay has been performed at the LHC using the data of 13 TeV and 13.2 fb$^{-1}$ 
\cite{ttbar}. 
The current bound assuming 100\% of the $\text{BR}({\cal H}\to t\bar{t})$ is, however, not so stringent. 
For example in the 2HDMs (regardless the type of Yukawa interactions),  
no bound on the mass of $H$ has been taken when $\tan\beta \gtrsim 0.3$. 
Therefore, the top quark associated process will be neglected in this analysis\footnote{In the HSM, 
the production cross section of $pp \to t\bar{t}H_S$ is proportional to $s_\alpha^2$, 
so that the search for $H_S \to t\bar{t}$ is less important with respect to the
2HDM as long as we take $s_\alpha \ll 1$. 
In the GM model, the properties of $H_1$ and $H_3^0$ are quite similar to those of $H$ and $A$ in the Type-I 2HDM, respectively, 
so that we can apply the similar phenomenological analysis of the Type-I 2HDM for these neutral Higgs bosons.}.  
The processes (i)--(iv) have been searched for using the 8 TeV data with the integrated luminosity of 20.3 fb$^{-1}$ 
in Refs.~\cite{ttbar2},~\cite{Hhh}, \cite{HZZ} and \cite{AZh}, respectively. 
Due to the fact that there is no significant excess in the number of signal events with respect to that given in the SM, 
the 95\% CL upper limit on the cross section times the branching ratio has been provided for each process. 
Concerning the process (ii), the $hh \to \gamma\gamma b\bar{b}$, 
$\tau\tau b\bar{b}$, $WW^* b\bar{b}$ and $b\bar{b}b\bar{b}$ decay modes were independently
analysed~\cite{Hhh}, and similarly for the process (iv), for which the
$h \to b\bar{b}$ and $h \to \tau\tau$ modes were analysed~\cite{AZh}. 

In order to compare the bounds driven by the experiments with the corresponding theory predictions, 
let us evaluate the cross section for the gluon fusion production for 
${\cal H}$ ($\sigma_{gg{\cal H}}$) by using the following approximation:
\begin{align}
\sigma_{gg{\cal H}} \simeq \sigma_{ggh_{\text{SM}}} \times \frac{\Gamma({\cal H} \to gg)}{\Gamma(h_{\text{SM}} \to gg)}, 
\end{align}
where $\sigma_{ggh_{\text{SM}}}$ is the gluon fusion cross section in the SM 
and $\Gamma(X \to gg)$ is the decay rate for the $X \to gg$ mode. 
The reference values of $\sigma_{ggh_{\text{SM}}}$ at 8 TeV are given in \cite{ggf}.
 
Second, let us discuss direct searches for singly-charged Higgs bosons.
Typical main decay modes are given in Eqs.~(\ref{decay_2HDM}) and (\ref{decay_GM}) in the 2HDM and the GM model, respectively. 
The search for charged Higgs bosons decaying into the $tb$ mode has been surveyed in Ref.~\cite{ttbar}. 
The current bound is not so stringent. 
For example, in the Type-II 2HDM 
no bound has been taken on the mass of $H^\pm$ when $t_\beta \gtrsim 0.5$, which is also valid for all the other types of Yukawa interaction. 
For the $Wh$ decay mode, there is no available current limit. Detailed simulation studies for the $H^\pm$ search
including the $Wh$ mode at the LHC Run-2 has been done in Ref.~\cite{Moretti:2016jkp} for the Type-II 2HDM. 
Concerning the $WZ$ channel, which is relevant for $H_5^\pm$ in the GM model only, 
it has been searched via the $WZ$ boson fusion process in Ref.~\cite{wz} using the 13 TeV data with the integrated luminosity of 15.2 fb$^{-1}$.
This gives the lower limit on $t_\beta$ (corresponding to the upper limit on $v_\Delta^{}$) for a fixed value of $m_{H_5}^{}$. 
For example, for $m_{H_5}^{} = 500$ GeV, $t_\beta \lesssim 1.65$ $(v_\Delta \gtrsim 45$ GeV) is excluded at 95\% CL. 
Actually, this limit is much weaker than the one given by the search for $H_5^{\pm\pm}\to W^\pm W^\pm$, so that
imposing the latter bound discussed below will be enough.

Third, the search for doubly-charged Higgs bosons in the same-sign diboson decay channel has been performed at the LHC using the
8 TeV data sample with 
the integrated luminosity of 19.4 fb$^{-1}$ in Ref.~\cite{ww}. 
The 95\% CL upper limit on the cross section of the $W^\pm W^\pm$ fusion process times the branching ratio of 
the $H_5^{\pm\pm}\to W^\pm W^\pm$ mode has been provided. 
From this result, we can extract the 95\% CL lower (upper) limit on the value of $t_\beta$ ($v_\Delta^{}$) for a fixed value of $m_{H_5}^{}$. 
Since the $H_5^{\pm\pm}W^\mp W^\mp$ coupling is proportional to $v_\Delta^{}$, 
the vector boson fusion cross section of $H_5^{\pm\pm}$ ($\sigma_{H_5^{\pm\pm}}^{}$) for an arbitrary value of $v_{\Delta}^{}$ can be extracted as follows:
\begin{align}
\sigma_{H_5^{\pm\pm}}^{} = \left(\frac{v_\Delta^{}}{v_\Delta^{\text{Ref.}}} \right)^2\times \sigma_{H_5^{\pm\pm}}\Big|_{v_\Delta^{\text{Ref.}}}, \label{doubly}
\end{align}
where $v_{\Delta}^{\text{Ref.}}$ is a reference value of $v_{\Delta}^{}$ which is set to be 16, 25 and 35 GeV in Ref.~\cite{ww}. 
The values of $ \sigma_{H_5^{\pm\pm}}\big|_{v_\Delta^{\text{Ref.}}}$ are also presented in Ref.~\cite{ww}.  

Apart from the direct searches for extra Higgs bosons, we need to consider also the constraint on the parameter space from the $h(125)$ data at the LHC. 
Here, we take into account the signal strengths $\mu_X^{}$ for $X=\gamma\gamma$, $ZZ$, $WW$ and $\tau\tau$ defined by 
\begin{align}
\mu_X^{}\equiv \sigma_{ggh}\times \text{BR}(h \to XX). 
\end{align}
The measured values of $\mu_X^{}$ are given by 
the combined analysis of the ATLAS and CMS experiments using the LHC Run-1 data~\cite{LHC1} as follows:
\begin{align}
\mu_{\gamma\gamma} = 1.10^{+0.23}_{-0.22},\quad 
\mu_{ZZ}^{} = 1.13^{+0.34}_{-0.31},\quad
\mu_{WW}^{} = 0.84^{+0.17}_{-0.17},\quad
\mu_{\tau\tau} = 1.0^{+0.6}_{-0.6},
\end{align}
and we shall require each prediction for $\mu_X^{}$ to lie within the 95\% CL region. 

 \begin{figure}[t]
  \includegraphics[width=95mm]{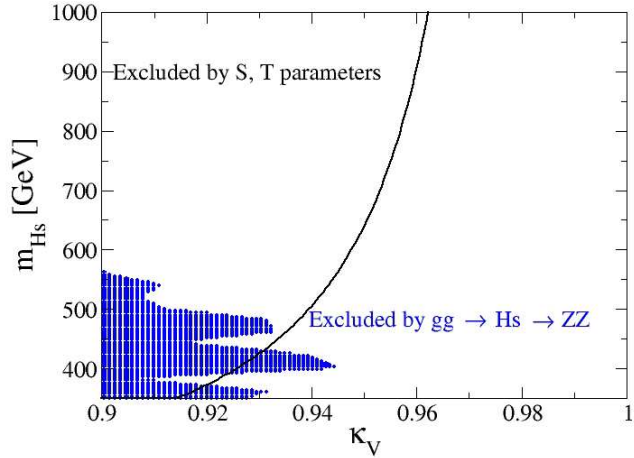} 
 \caption{Allowed parameter space on the $\kappa_V^{}$--$m_{H_S}^{}$ plane in the HSM. 
The blue shaded region is excluded by the search for the $gg \to H_S \to ZZ$ process. 
The region above the black solid curve is excluded by the $S$ and $T$ parameters. 
In this plot, we take $\lambda_{\Phi S} = \mu_S = \lambda_S = 0$.} 
 \label{fig:hsm}
 \end{figure}

In Fig.~\ref{fig:hsm}, we show the allowed parameter space on the $\kappa_V^{}$--$m_{H_S}^{}$ plane in the HSM. 
The region above the black solid curve is excluded at 95\% CL by the $S$ and $T$ parameters, while 
the blue shaded region is excluded at 95\% CL by the direct 
search for the $gg \to H_S \to ZZ$ process, which turns out to set the most
stringent constraint among the direct search and signal strength data. 
However, we see that the region excluded by the direct searches is 
almost ruled out by the constraint from the $S$, $T$ parameters, 
so that the LHC data do not significantly improve our bounds for 
$m_{H_S}^{}$ with respect to Sec. \ref{sec:bound}. 

 \begin{figure}[!t] \hspace{-5mm} 
  \includegraphics[width=60mm]{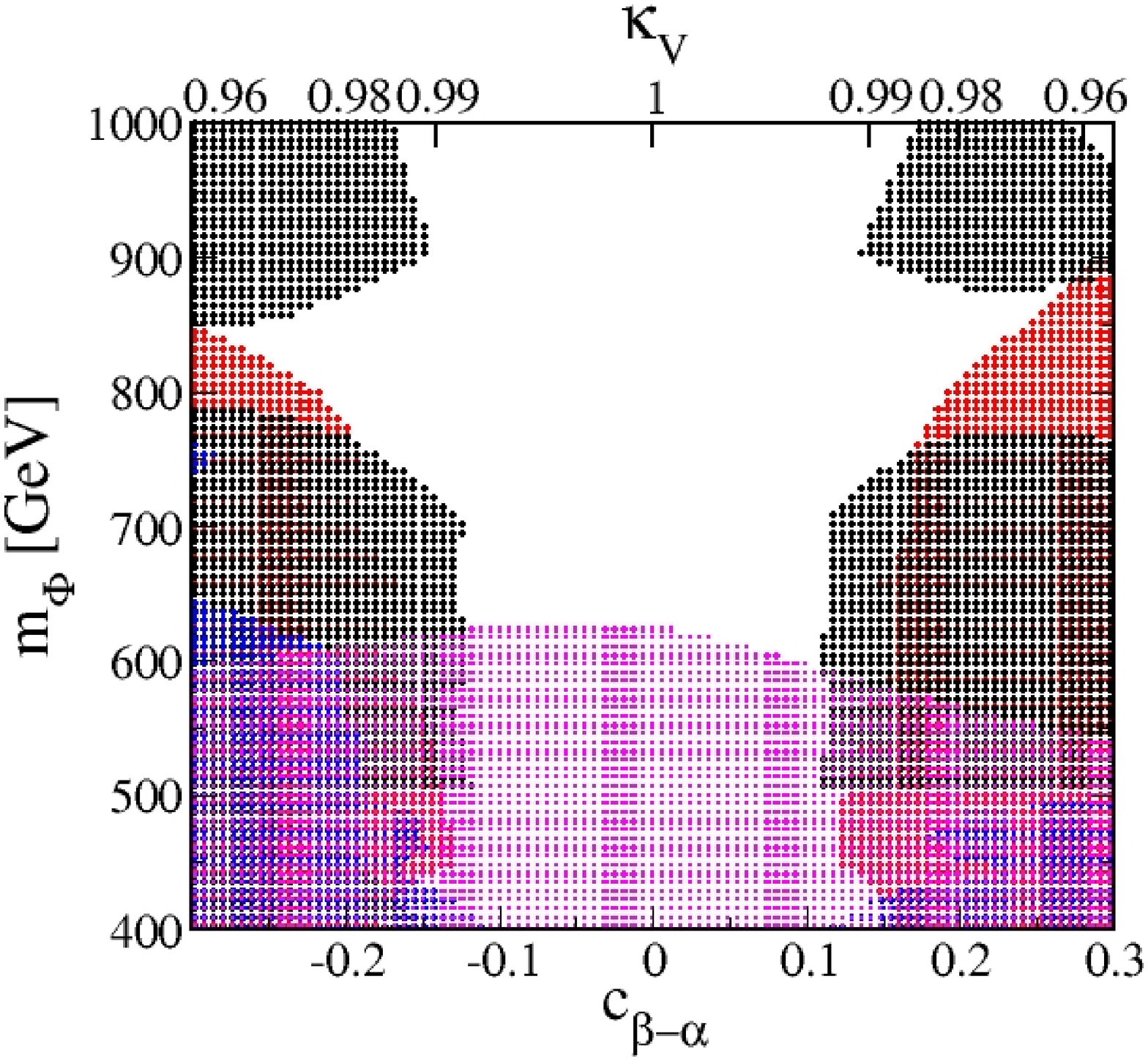}\hspace{-14mm} 
  \includegraphics[width=60mm]{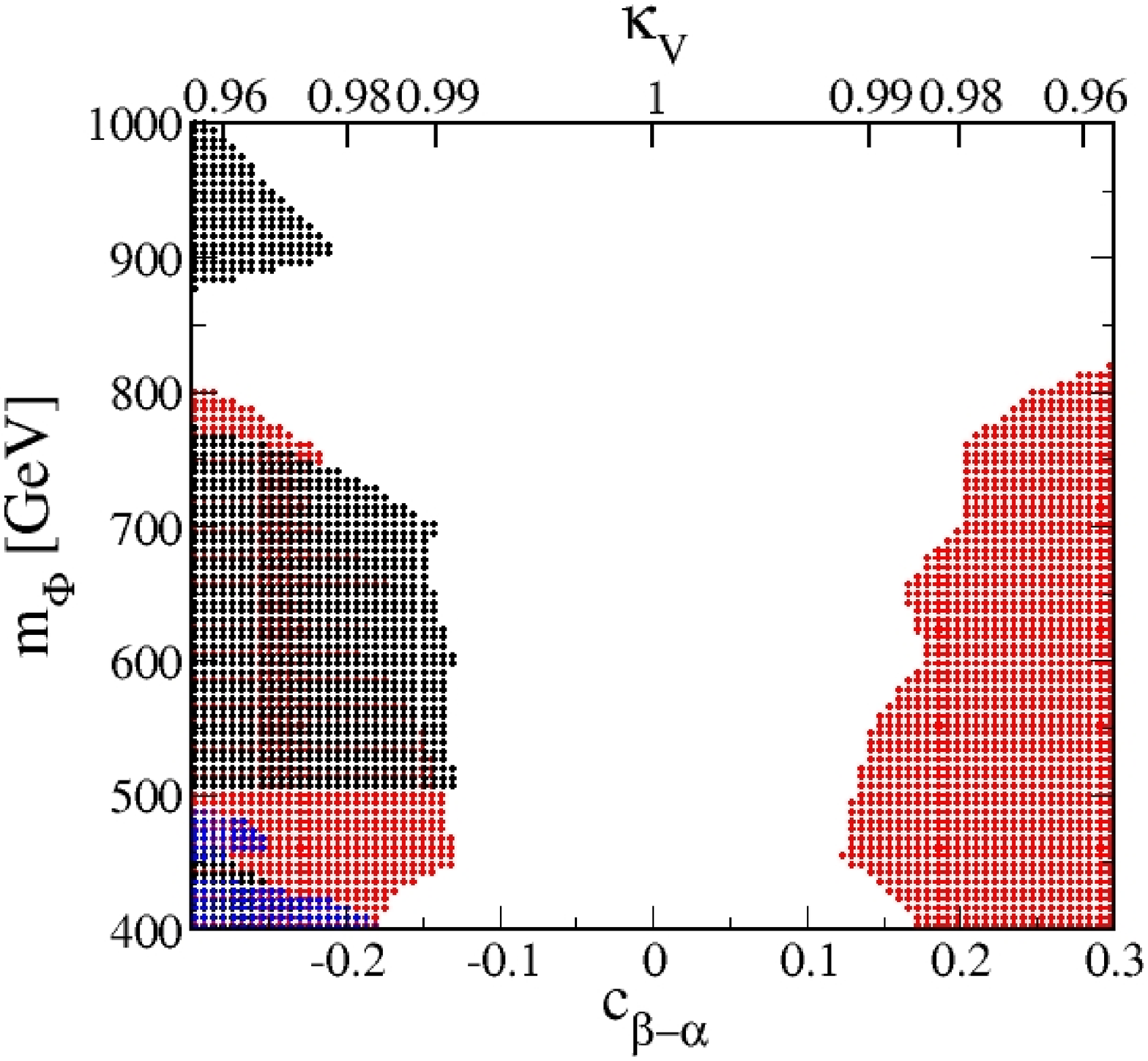}\hspace{-14mm}
  \includegraphics[width=60mm]{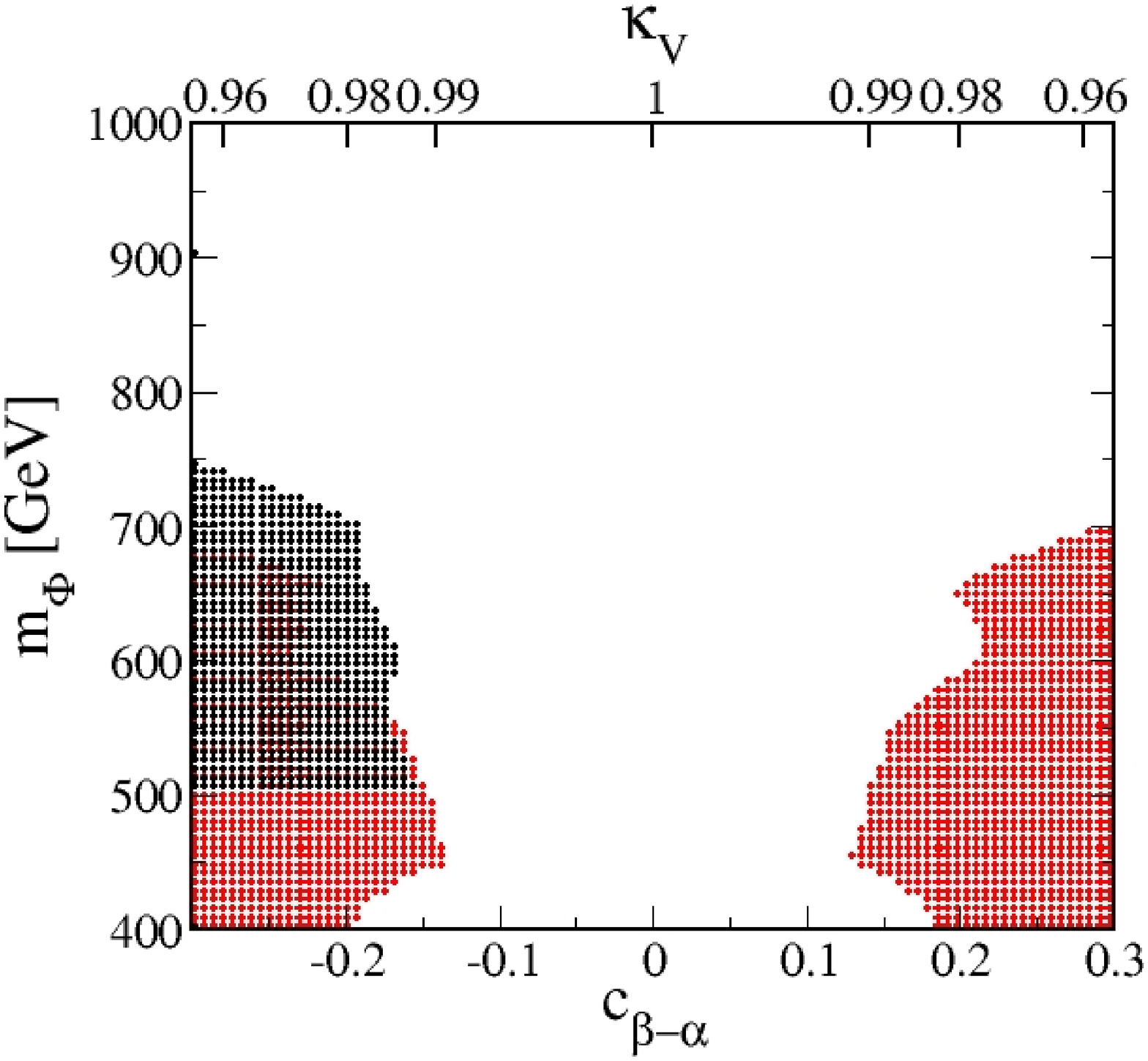}\\ \vspace{-1mm}\hspace{-5mm} 
  \includegraphics[width=60mm]{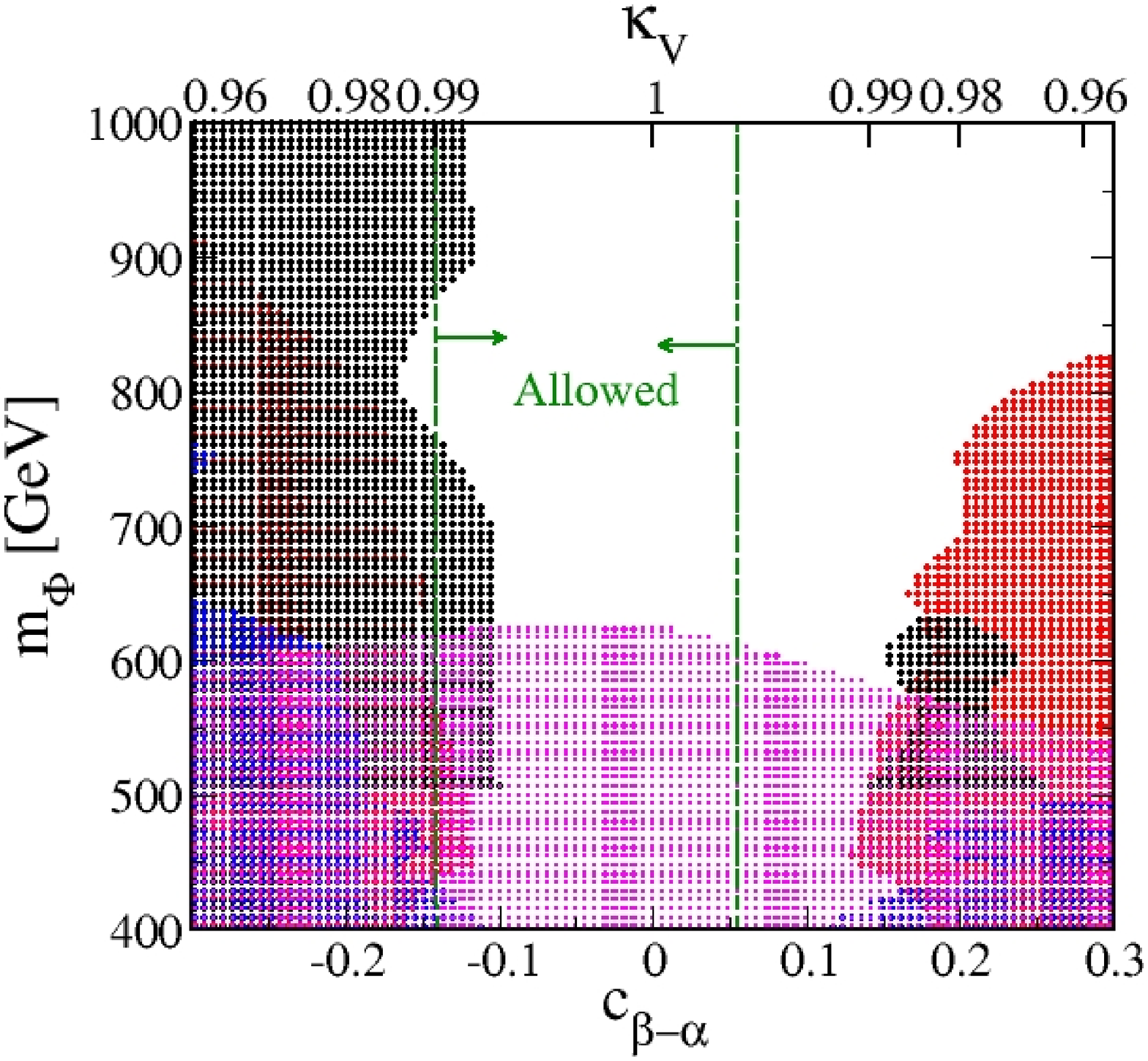}\hspace{-14mm} 
  \includegraphics[width=60mm]{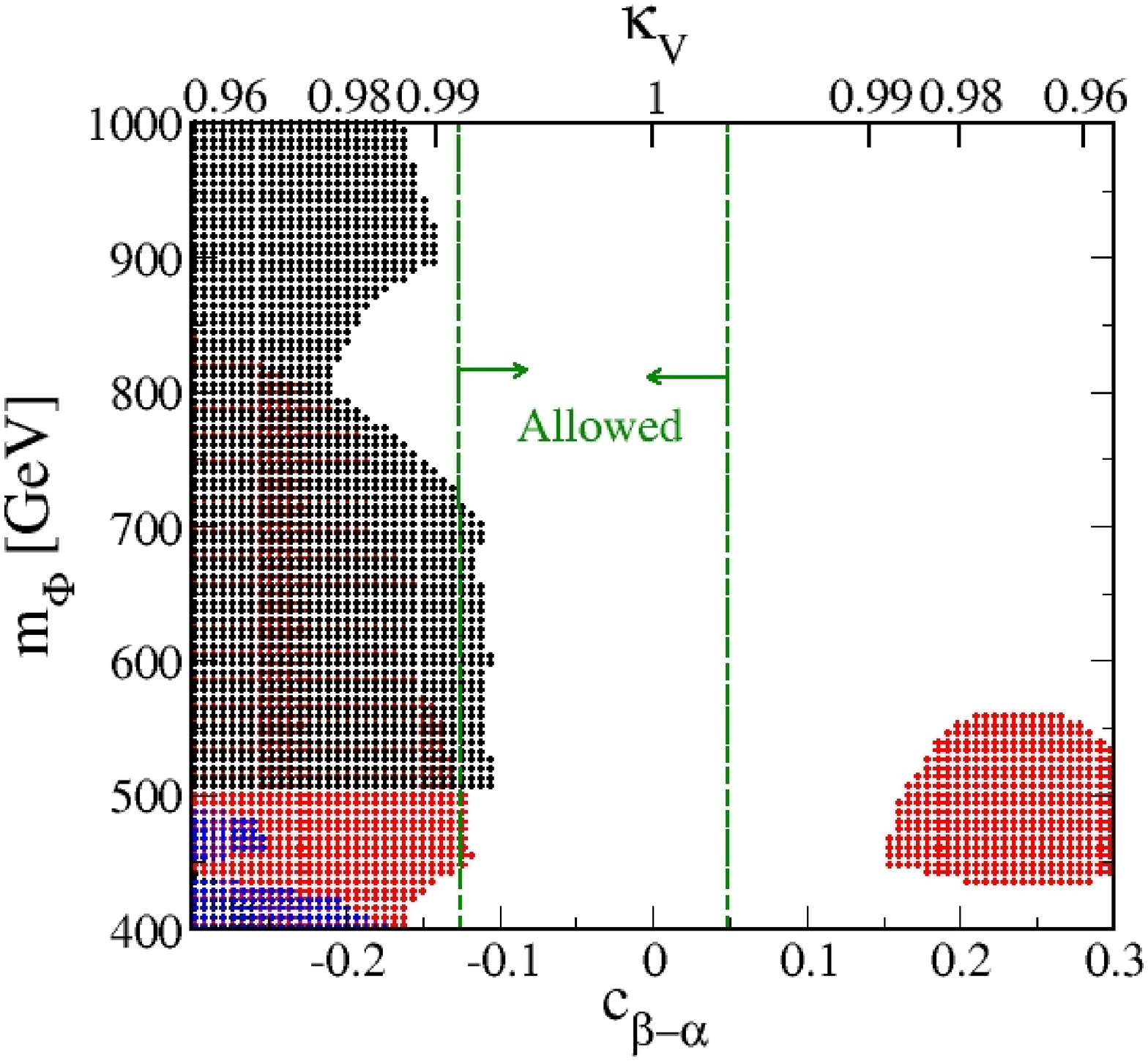}\hspace{-14mm}
  \includegraphics[width=60mm]{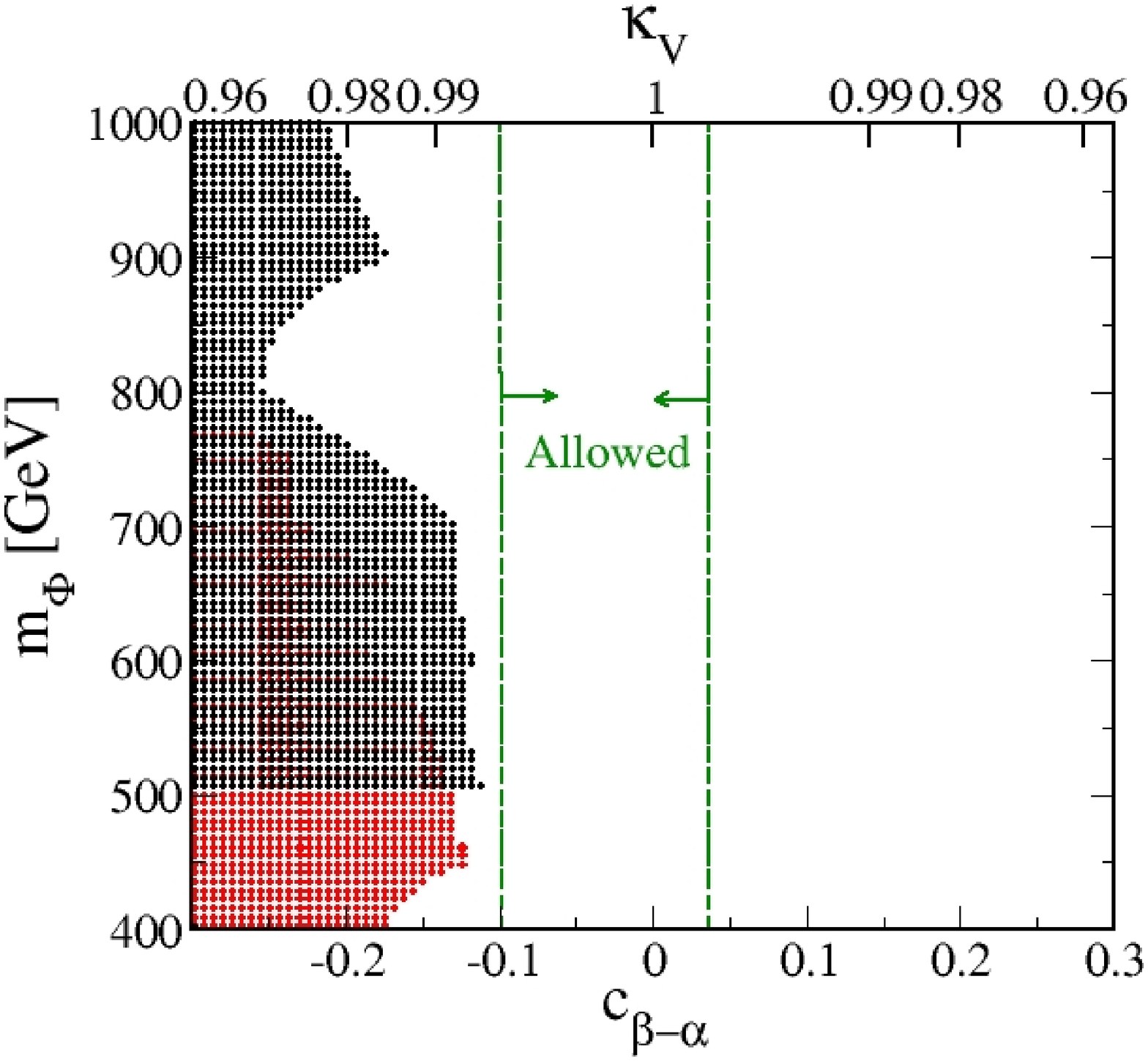}\\ \vspace{-1mm}\hspace{-5mm} 
  \includegraphics[width=60mm]{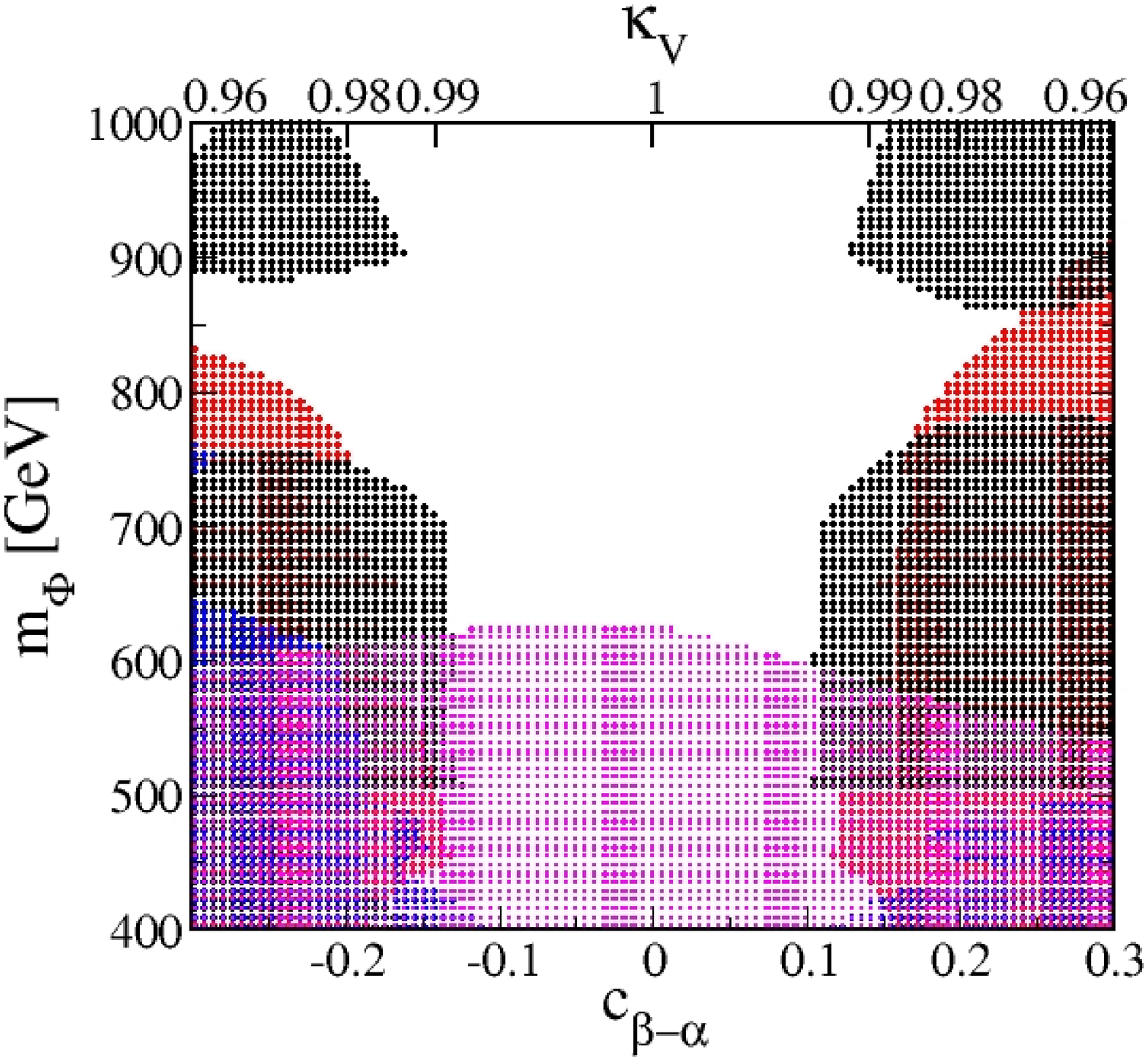}\hspace{-14mm}
  \includegraphics[width=60mm]{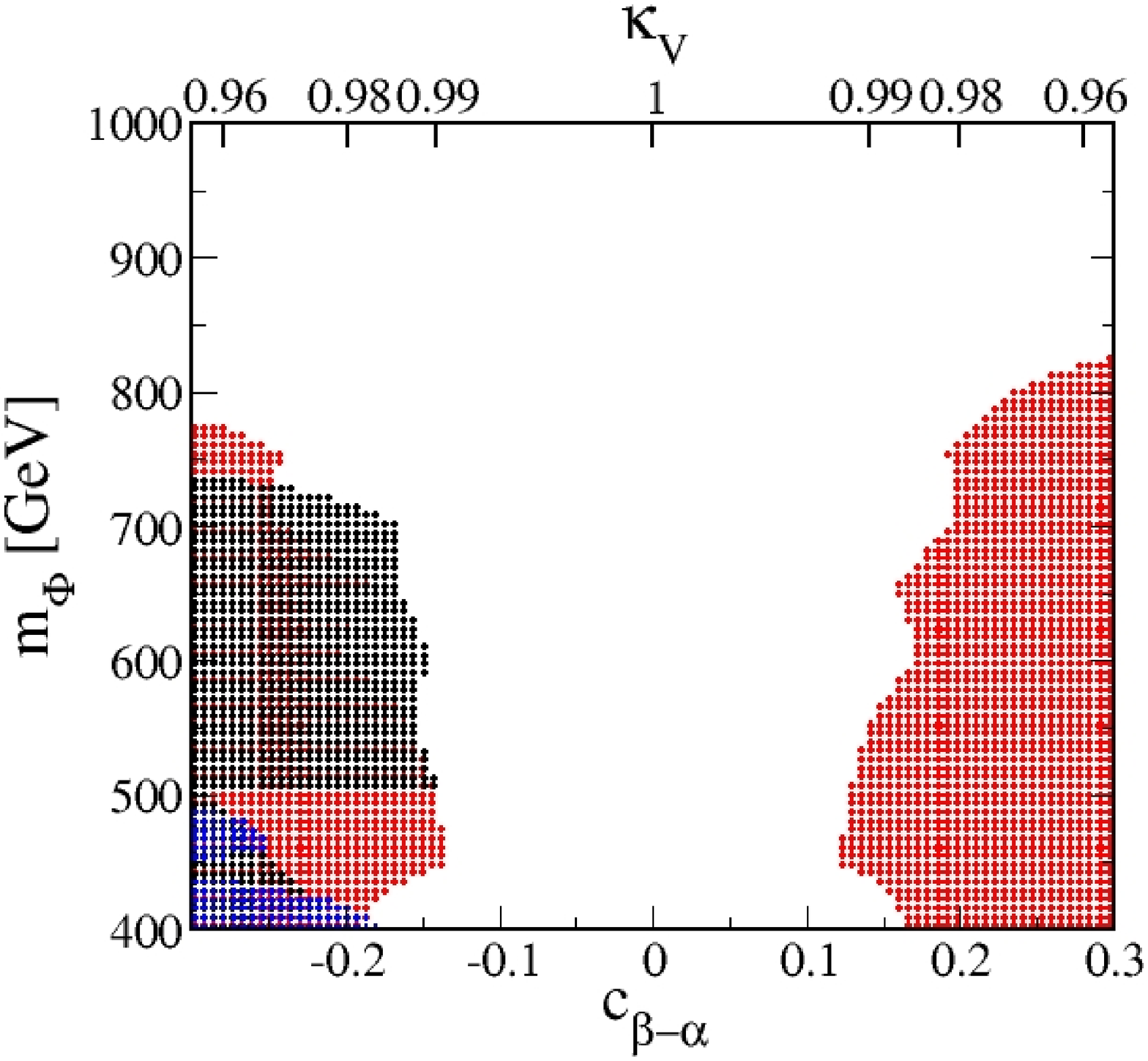}\hspace{-14mm}
  \includegraphics[width=60mm]{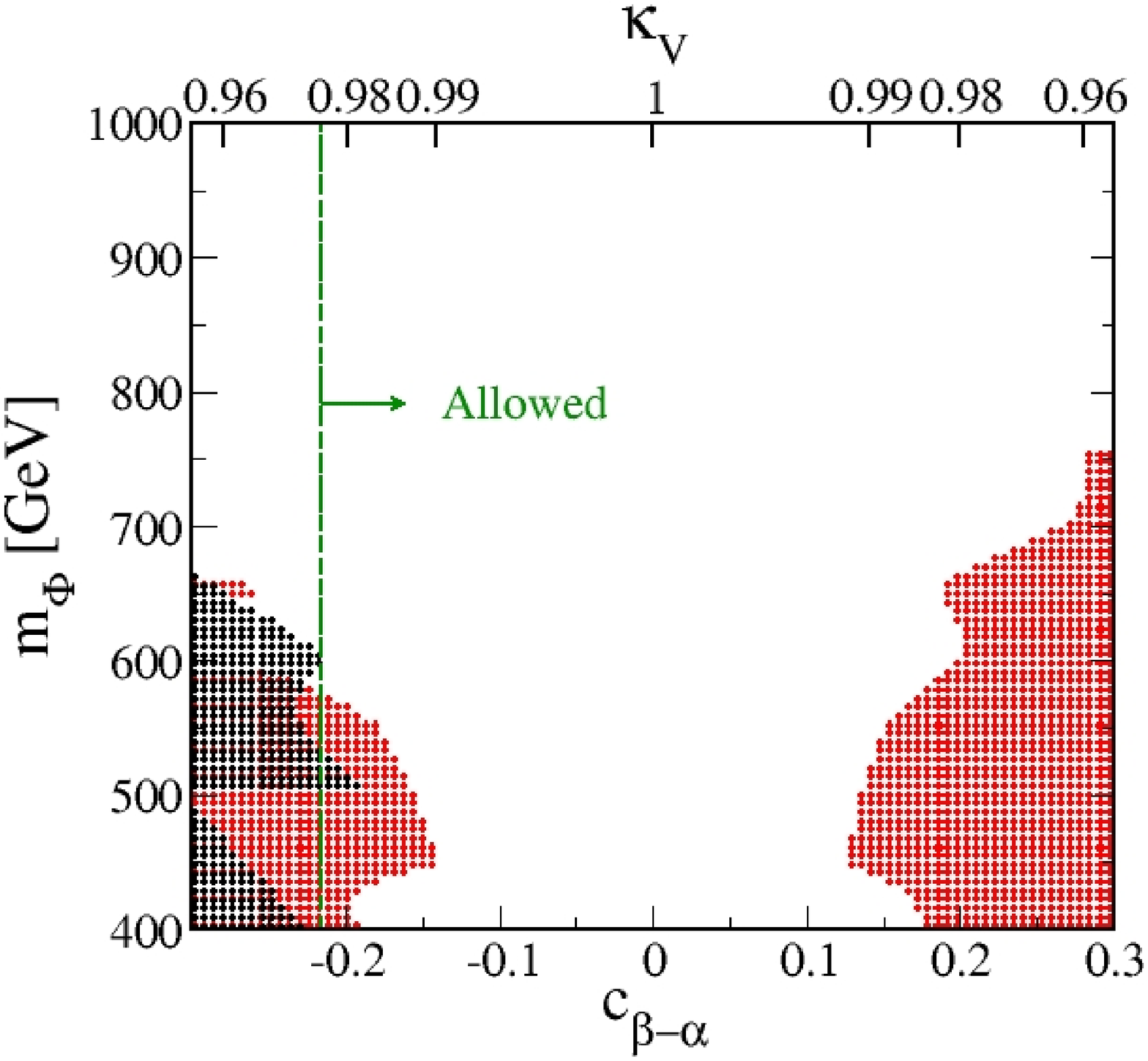}\\ \vspace{-1mm}\hspace{-5mm} 
  \includegraphics[width=60mm]{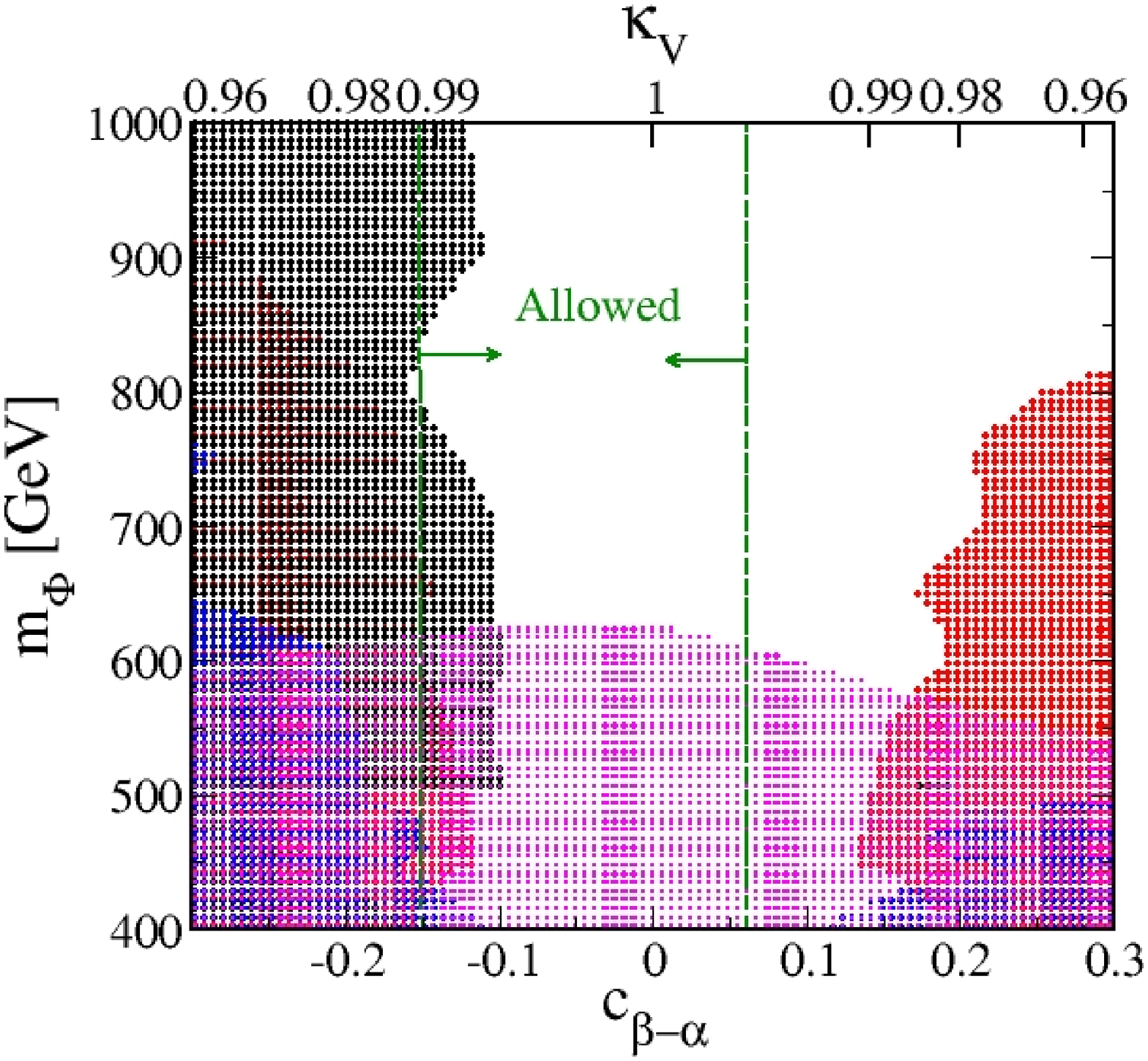}\hspace{-14mm}
  \includegraphics[width=60mm]{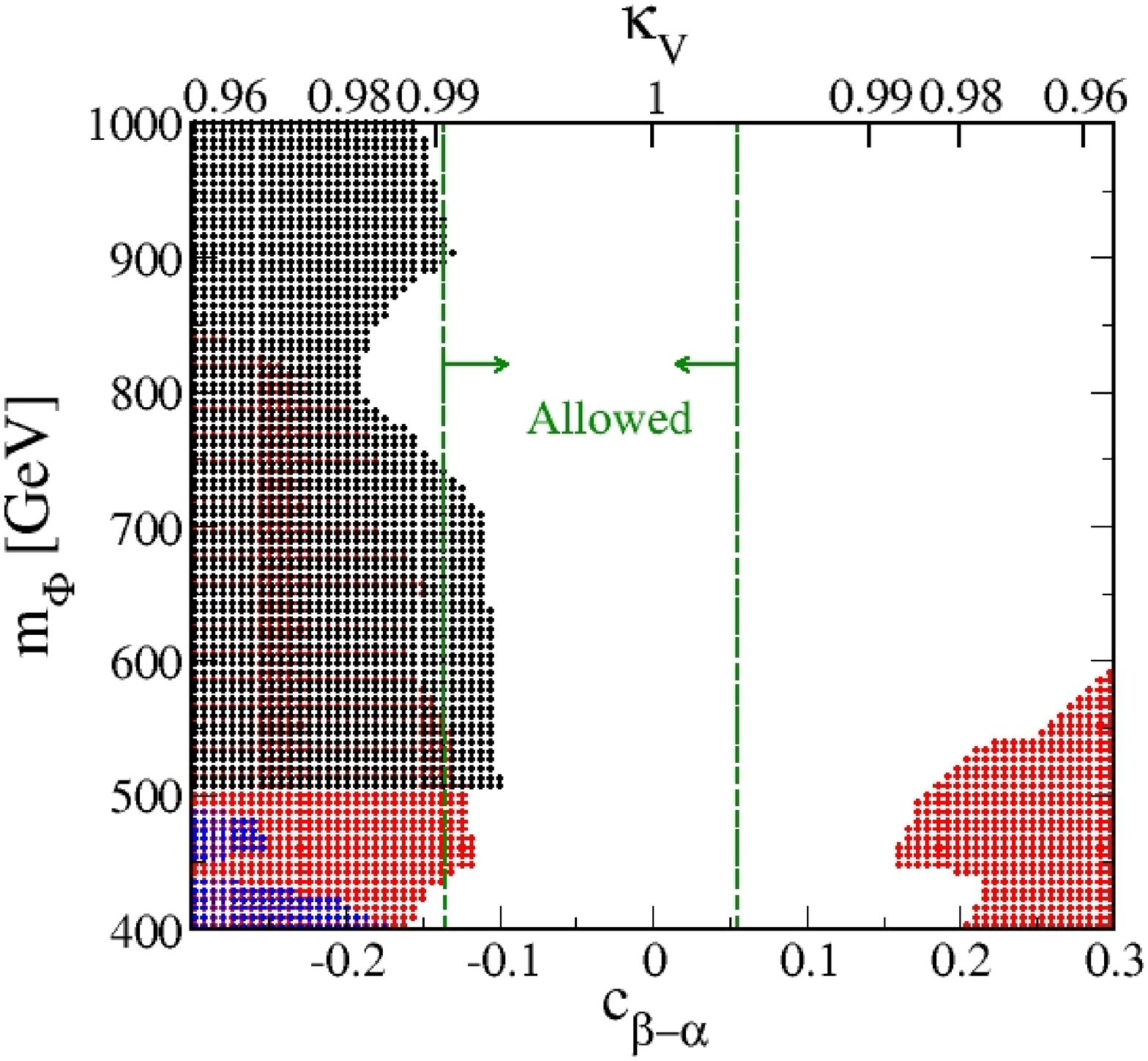}\hspace{-14mm}
  \includegraphics[width=60mm]{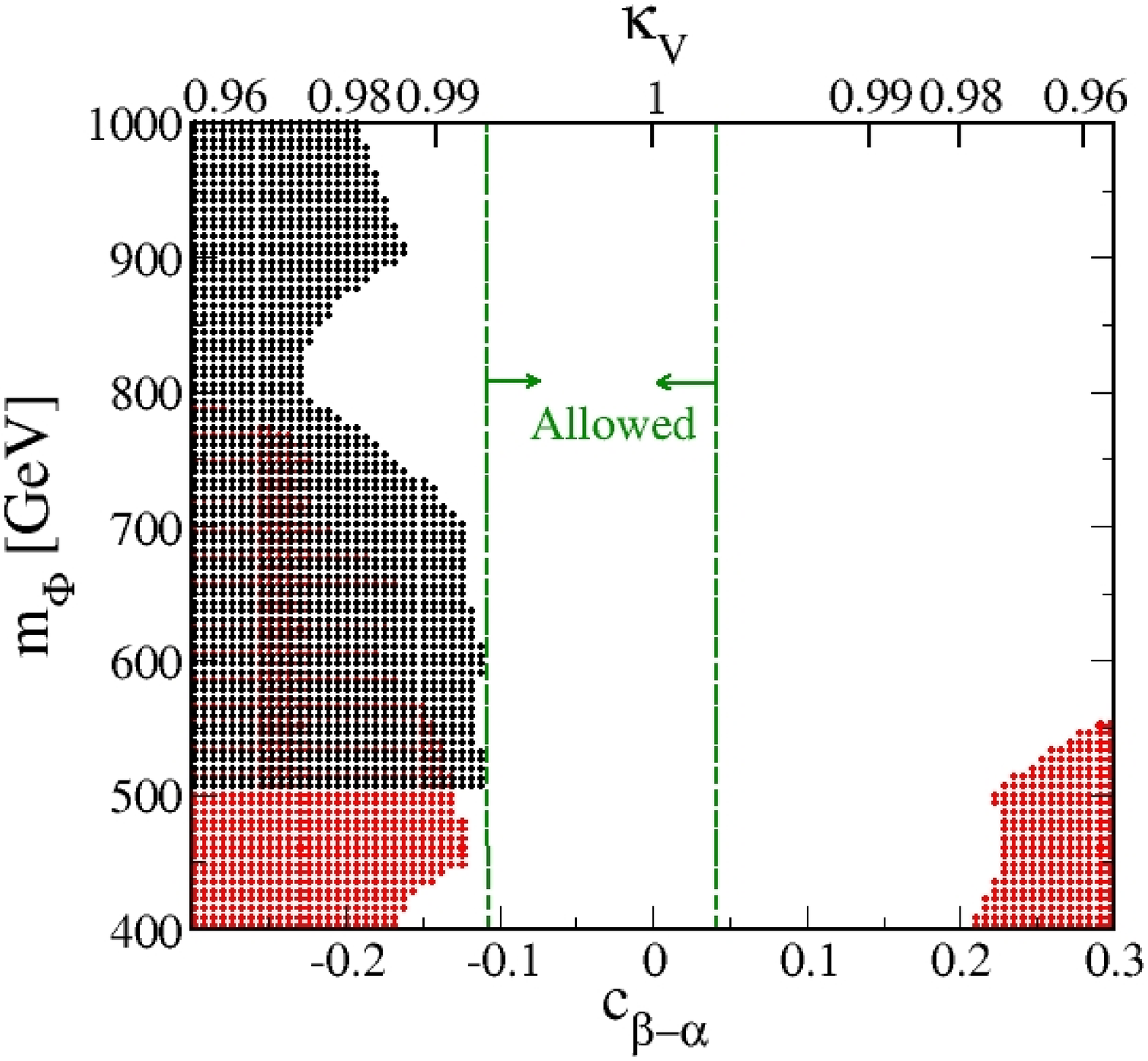}\vspace{-5mm}
 \caption{Constraints on the parameter space with $M/m_\Phi = 1$ in the 2HDMs, where
$\tan\beta$ is taken to be 1 (left), 2 (center) and 3 (right). 
The top to bottom panels show the results in the Type-I, -II, -X and -Y 2HDMs.  
The shaded regions are excluded by the direct searches at the LHC: 
$gg \to H \to hh$ (black), $gg \to A \to Zh$ (red), $gg \to H \to ZZ$ (blue), and $gg \to H/A \to t\bar{t}$ (magenta). 
The region outside the green dashed curves is excluded by the 95\% CL Higgs signal strength constraints.}
 \label{fig:ds_thdm1}
 \end{figure}

 \begin{figure}[!t]
  \includegraphics[width=66mm]{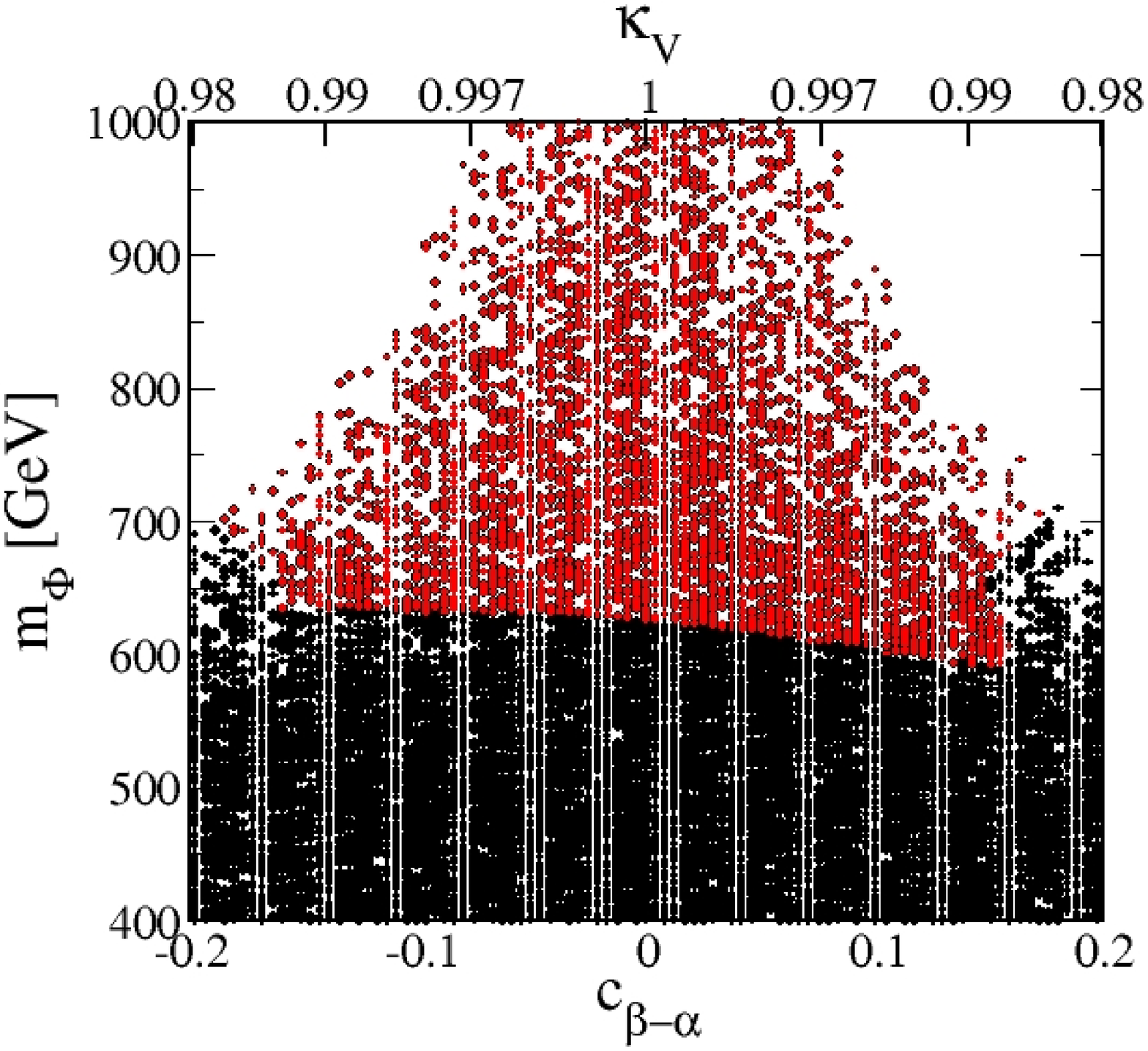}\hspace{-5mm} 
  \includegraphics[width=66mm]{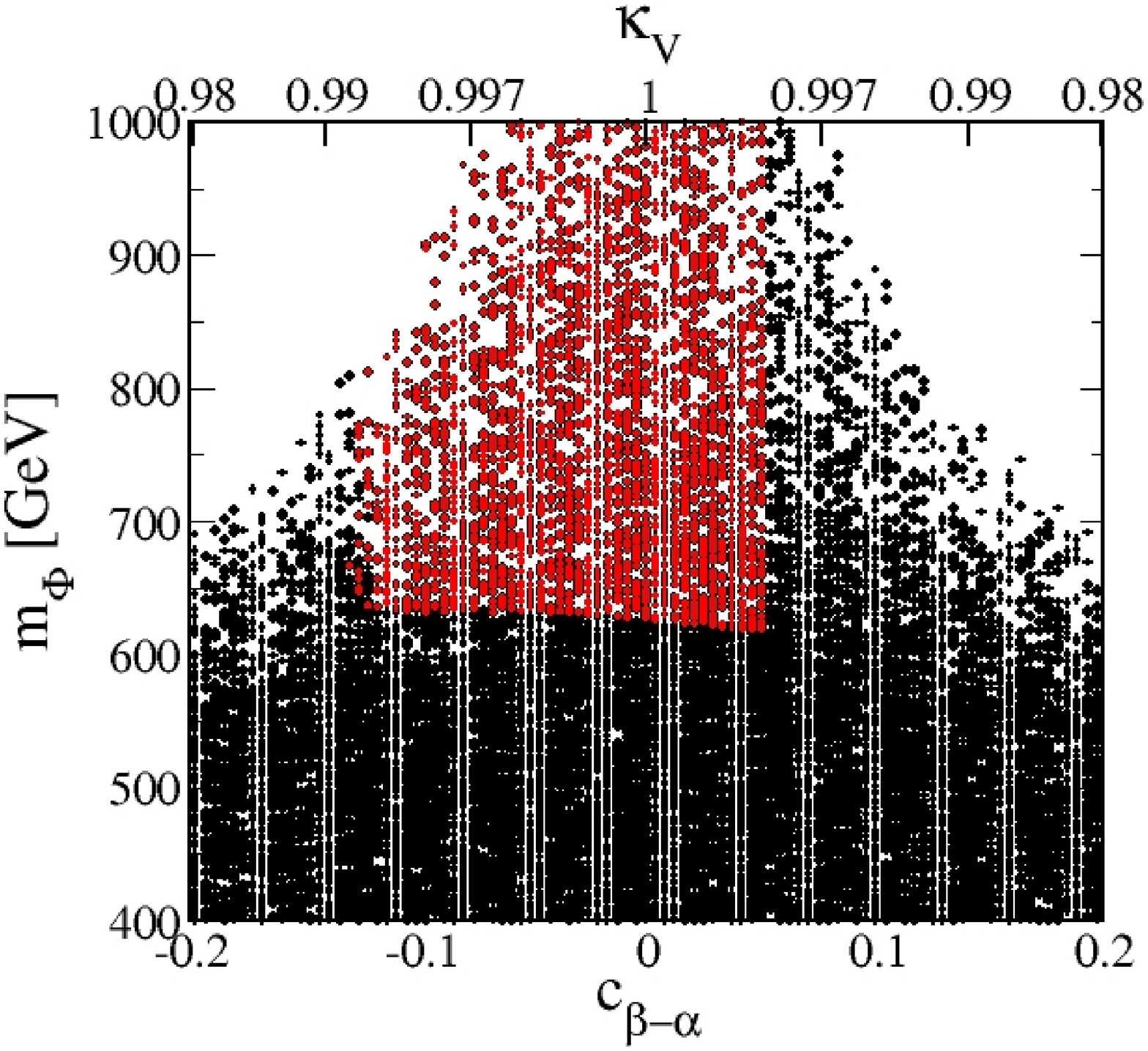}\\\vspace{-2mm}
  \includegraphics[width=66mm]{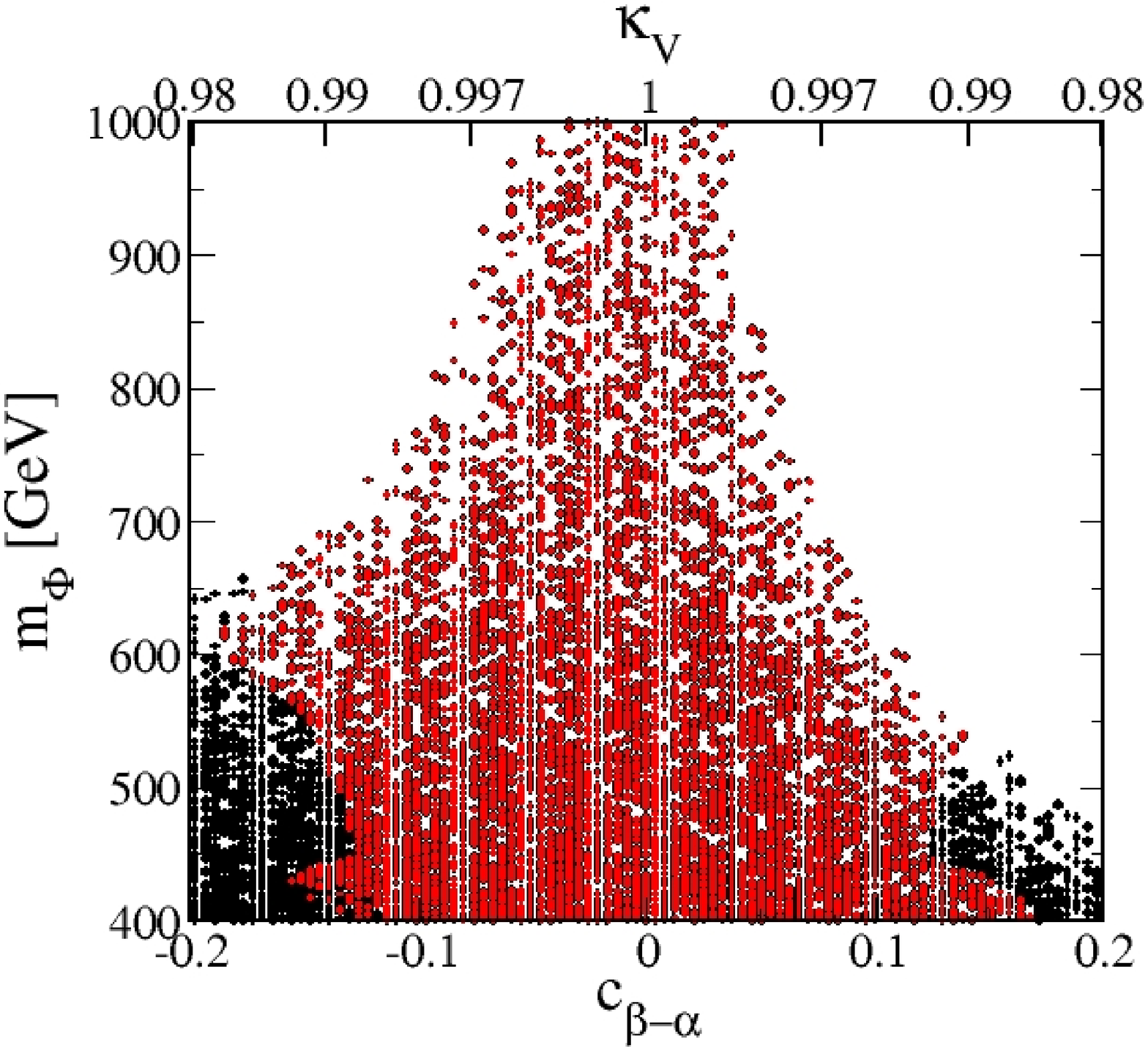}\hspace{-5mm} 
  \includegraphics[width=66mm]{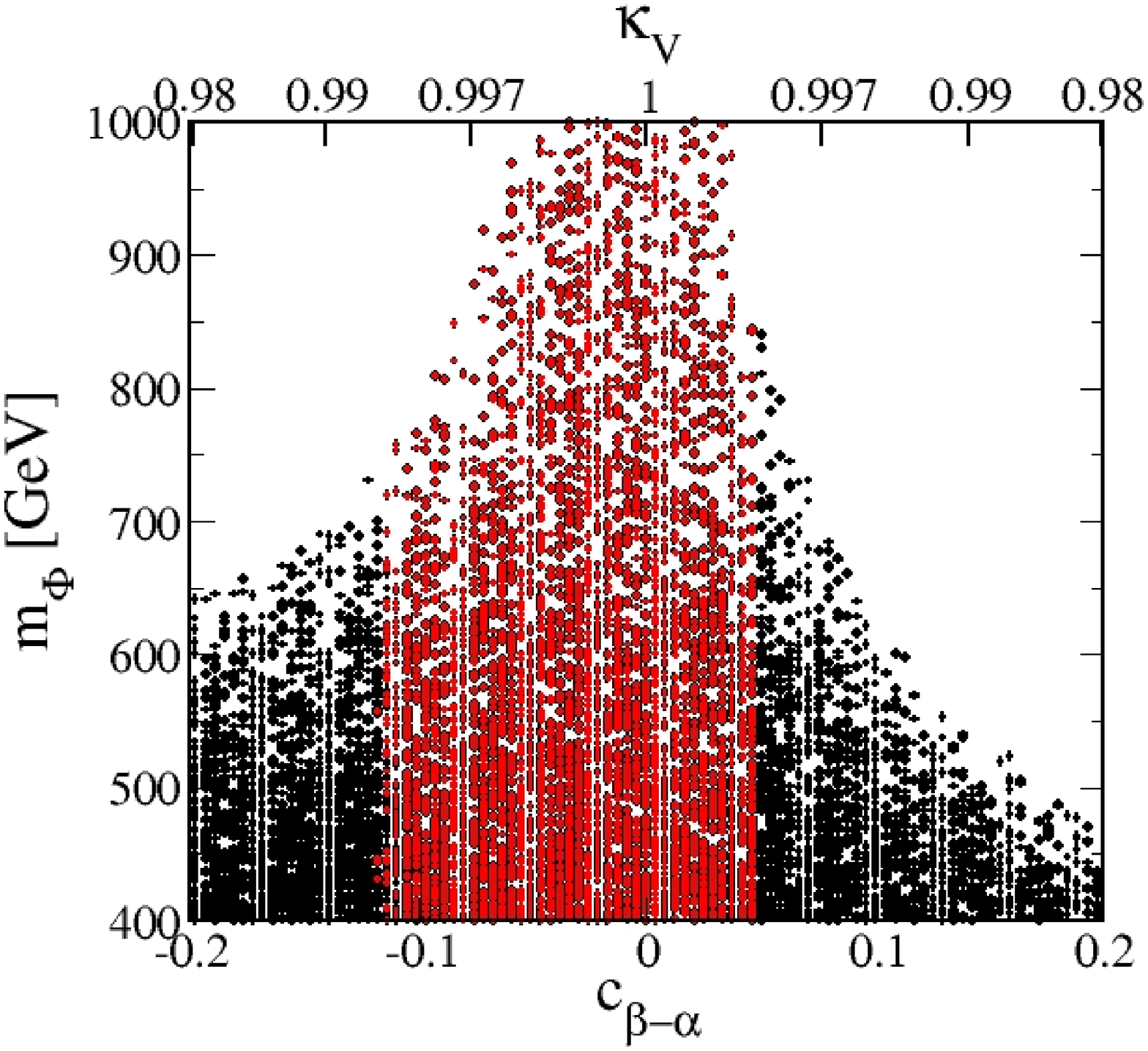}\\\vspace{-2mm}
  \includegraphics[width=66mm]{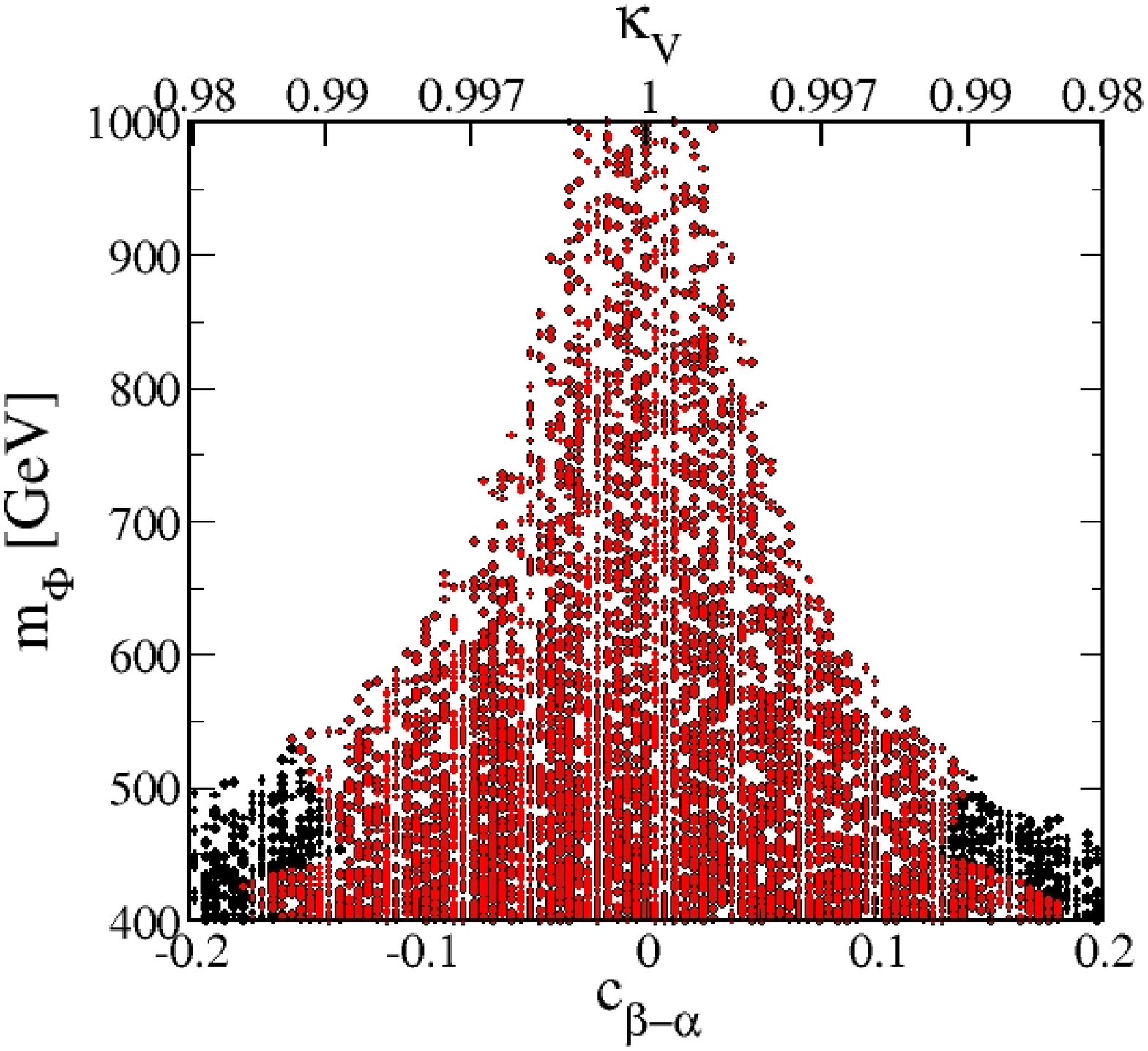}\hspace{-5mm} 
  \includegraphics[width=66mm]{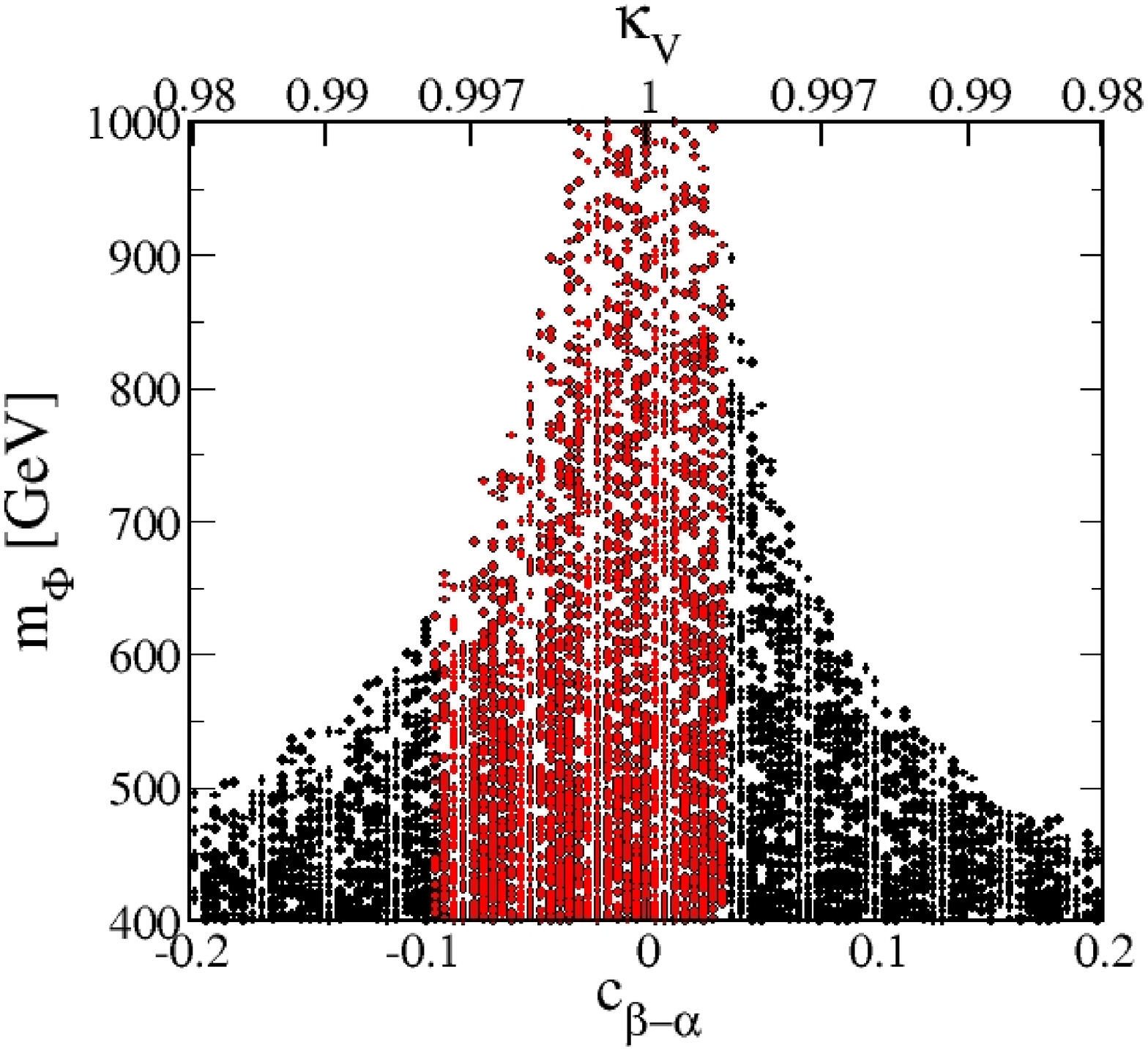}\\\vspace{-2mm}
  \includegraphics[width=66mm]{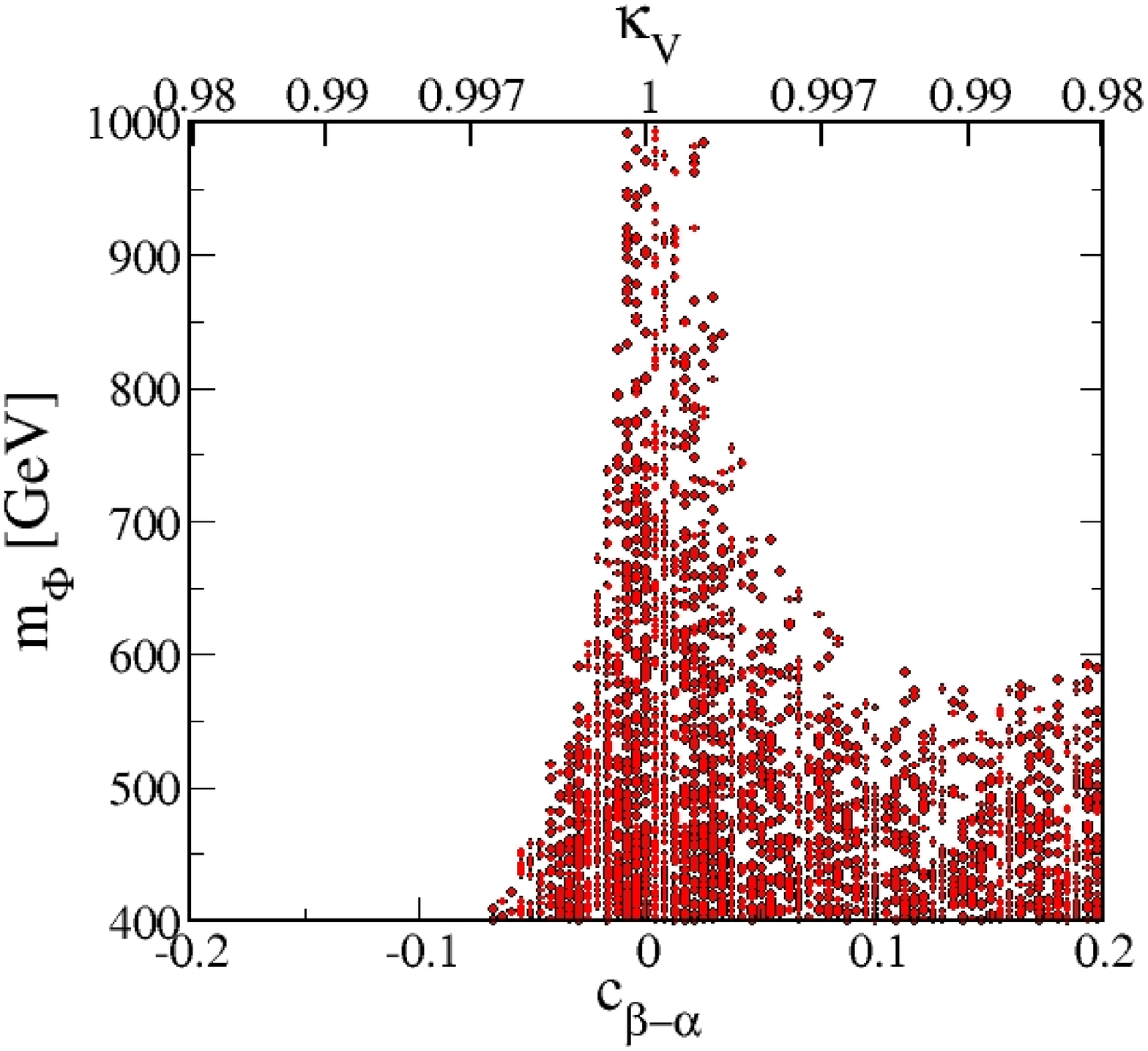}\hspace{-5mm} 
  \includegraphics[width=66mm]{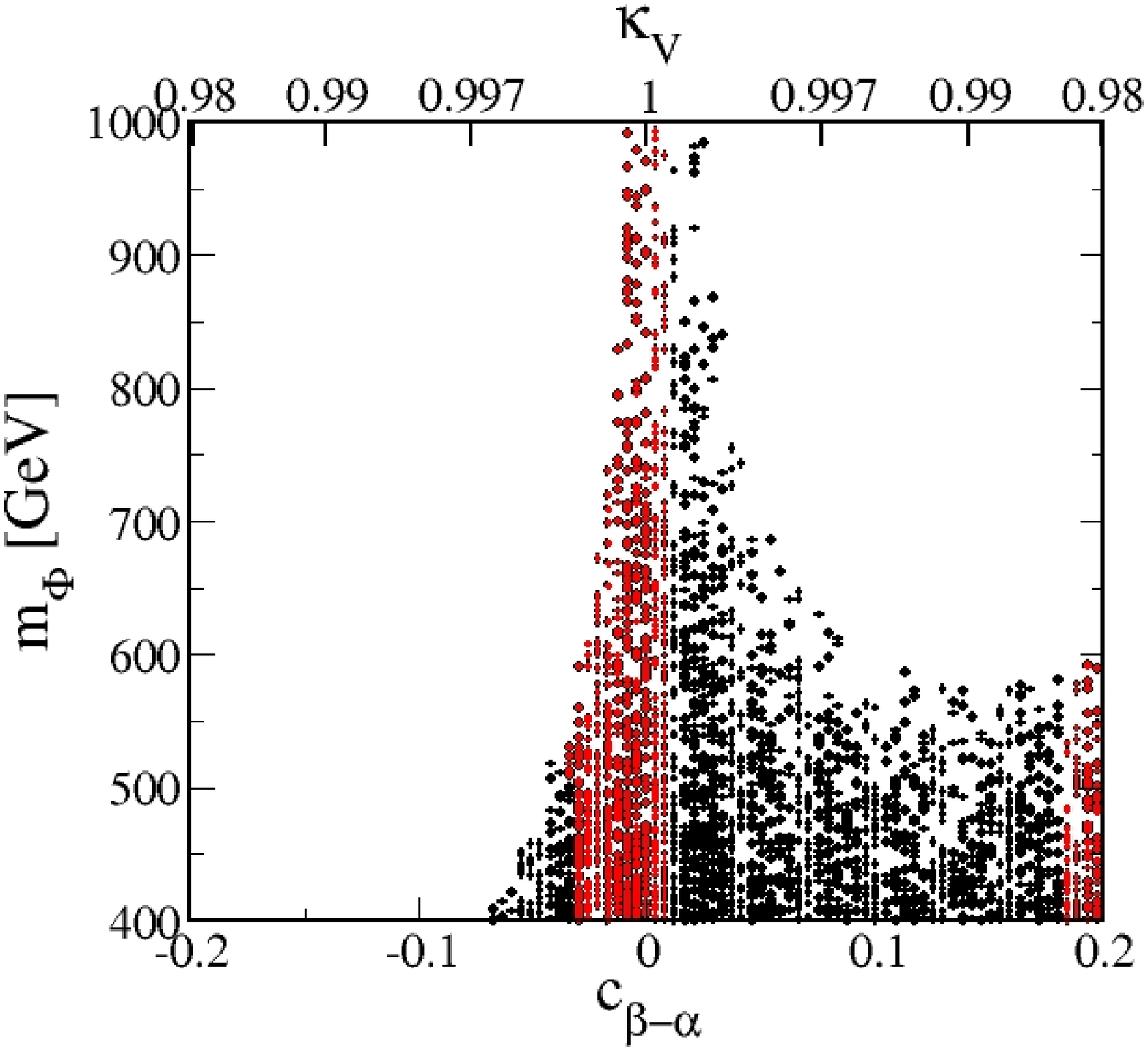}\vspace{-5mm}
\caption{Black (Red) dots are allowed only by Bound A (by both Bound A and the LHC data) 
in the Type-I (left) and Type-II (right) 2HDMs. 
The value of $\tan\beta$ is taken to be  1, 2, 3 and 10 from the top to bottom panels. 
The value of $M$ is scanned to be from 0 to $m_\Phi^{}(=m_{H^\pm}^{}=m_A^{}=m_H^{})$.
}
 \label{fig:2hdm_2}
 \end{figure}

Next, the constraints on the parameter space of the 2HDM are shown in Figs.~\ref{fig:ds_thdm1} and \ref{fig:2hdm_2}. 
In Fig.~\ref{fig:ds_thdm1}, we show the excluded region at 95\% CL by the direct searches (shaded region) and $\mu_X$ (indicated by the green dashed line) on 
the $c_{\beta-\alpha}$--$m_\Phi^{}$ plane, where $m_\Phi^{} = m_A^{} = m_{H^\pm} = m_H^{}$. 
We also display the corresponding value of $\kappa_V^{}$ on the top horizontal axis. 
The results in the Type-I, -II, -X and -Y 2HDMs are shown from the top to bottom panels, while those for $t_\beta=1$, 2 and 3
are shown from the left to right panels. In all the plots, we take $M = m_\Phi^{}$. 
Regardless of the type of Yukawa interactions, the bound from the direct searches 
becomes milder when we take a larger value of $t_\beta$, because the top quark loop contribution to
the gluon fusion cross section is suppressed by a factor $\propto \cot^2\beta$. 
In fact, if we take $t_\beta \gtrsim 10$, almost no region on the $c_{\beta-\alpha}$--$m_\Phi^{}$ plane shown in this figure is excluded by the direct searches. 
Concerning the bound from $\mu_X^{}$, they give a severe constraint on $\kappa_V^{}$ particularly in the Type-II and Type-Y 2HDMs, by which 
$|\kappa_V^{}-1|$ larger than 1\% are not allowed. 
This constraint tends to get stronger when we take a larger value of $t_\beta$ except for the Type-I 2HDM. 
Let us now comment on the case with $M \neq  m_\Phi^{}$.  
For $m_\Phi^{} < M$, the constraint from $gg \to H \to hh$ tends to be milder, because the 
branching ratio of $H \to hh$ becomes small. 
On the contrary, the constraint from $gg \to H \to ZZ$ tends to be slightly stronger 
due to a little enhancement of the branching ratio of $H \to ZZ$. 
In any case, the total excluded region with $M < m_\Phi^{}$ is found not to change so much with respect to 
our initial assumption. 
On the contrary, the case $M > m_\Phi$ is totally disfavored by the vacuum stability bound. 

Let us combine the bounds from the LHC data discussed above and Bound A in the 2HDM. 
In Fig.~\ref{fig:2hdm_2}, the black (red) dots are allowed by only Bound A (both Bound A and the LHC data). 
We here keep the assumption of the degeneracy in mass of the extra Higgs bosons, and we scan over 
the value of $M$.
The results for Type-I and Type-II 2HDMs are shown on the left and right panels, respectively, and the value of 
$t_\beta$ is taken to be 1, 2, 3 and 10 from the top to bottom panels. 
We note that the results for the Type-X and Type-Y 2HDMs are almost the same as those for the Type-I and Type-II 2HDMs, respectively.
We can see that for $t_\beta = 1$ (top panels), $m_\Phi \lesssim 600$ GeV is excluded, because of the $gg \to H/A \to t\bar{t}$ process 
(according to Fig.~\ref{fig:ds_thdm1}). 
For larger $t_\beta$, the region filled by the black and red dots is getting the same in the Type-I 2HDM, namely, the bounds from the direct search and $\mu_X$ become 
less important. In contrast, in the Type-II 2HDM, the region filled by the red dots is much smaller than that filled by the black dots even for the case with large $t_\beta$, 
because of the constraint from $\mu_X^{}$. 
In conclusion, we find that for smaller values of $t_\beta$, $\kappa_V^{} \lesssim 0.98 (0.99)$ is excluded in the Type-I and X (Type-II and Y) 2HDMs. 
For larger values of $t_\beta$, a smaller value of $\kappa_V$, e.g., less than 0.98, is possible in all the four types, but the mass of the extra Higgs bosons 
has to be below $\sim 600$ GeV
due to the theoretical constraints. 

 \begin{figure}[t]
  \includegraphics[width=65mm]{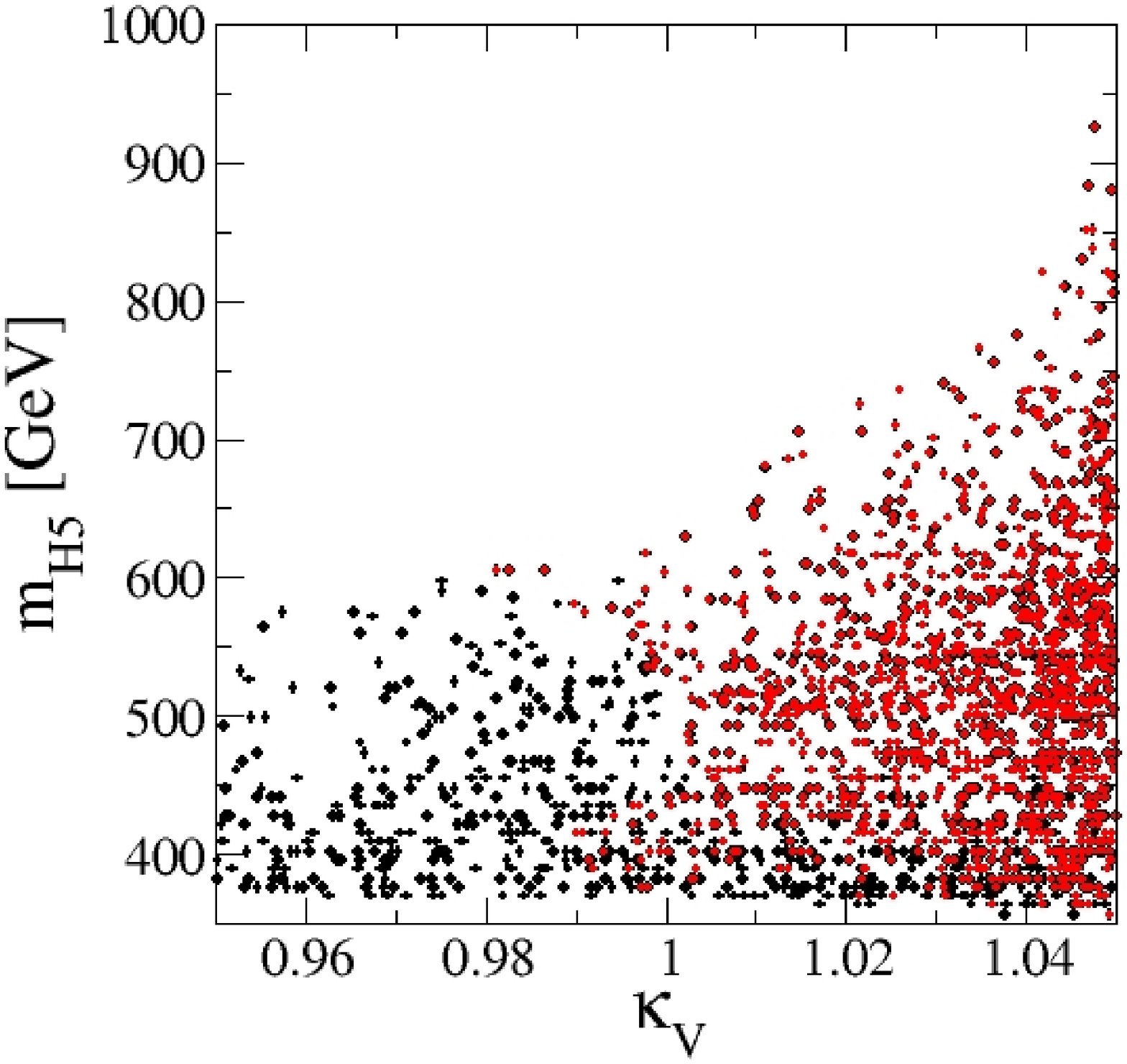}\hspace{-17mm}
  \includegraphics[width=65mm]{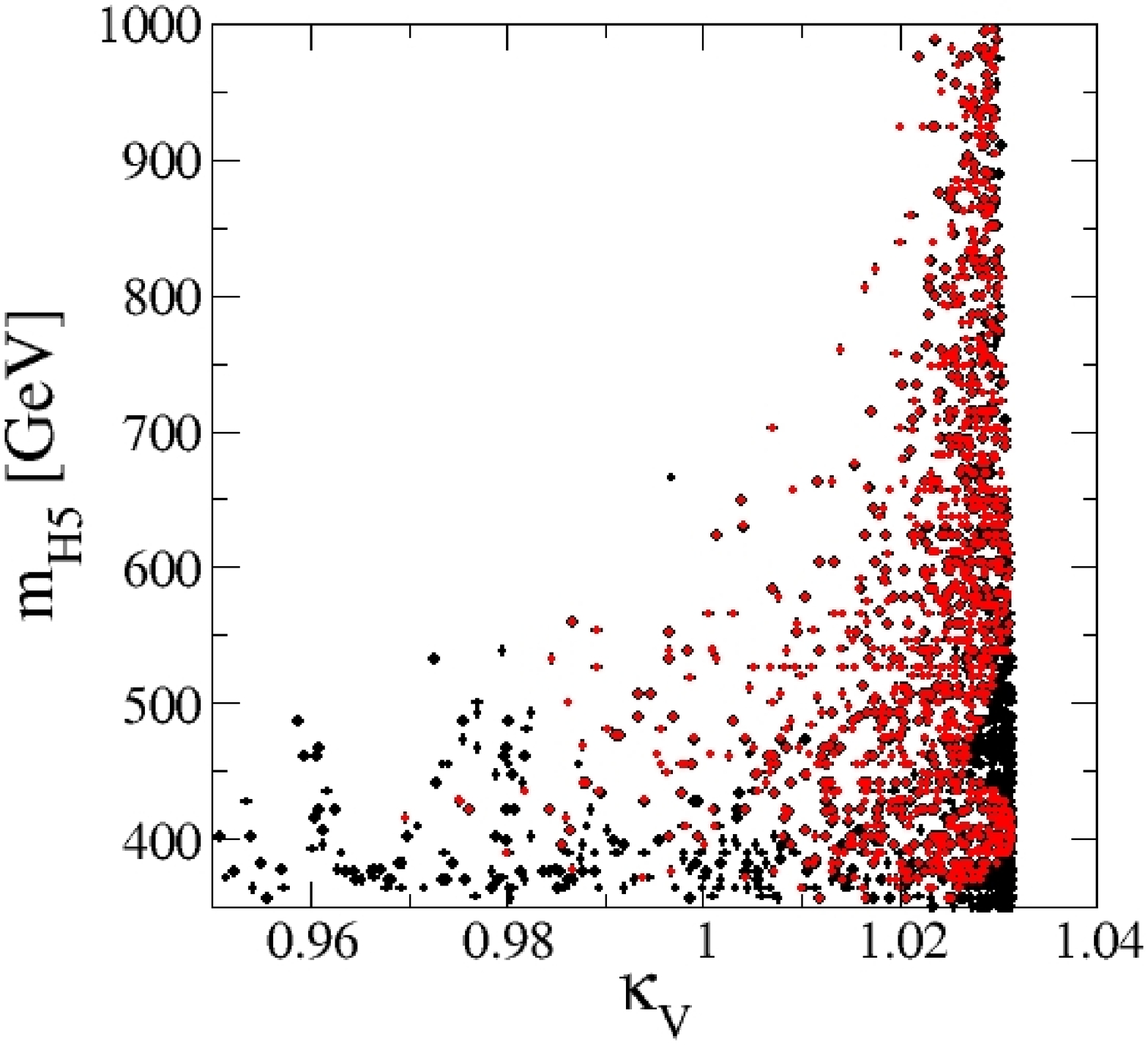}\hspace{-17mm}
  \includegraphics[width=65mm]{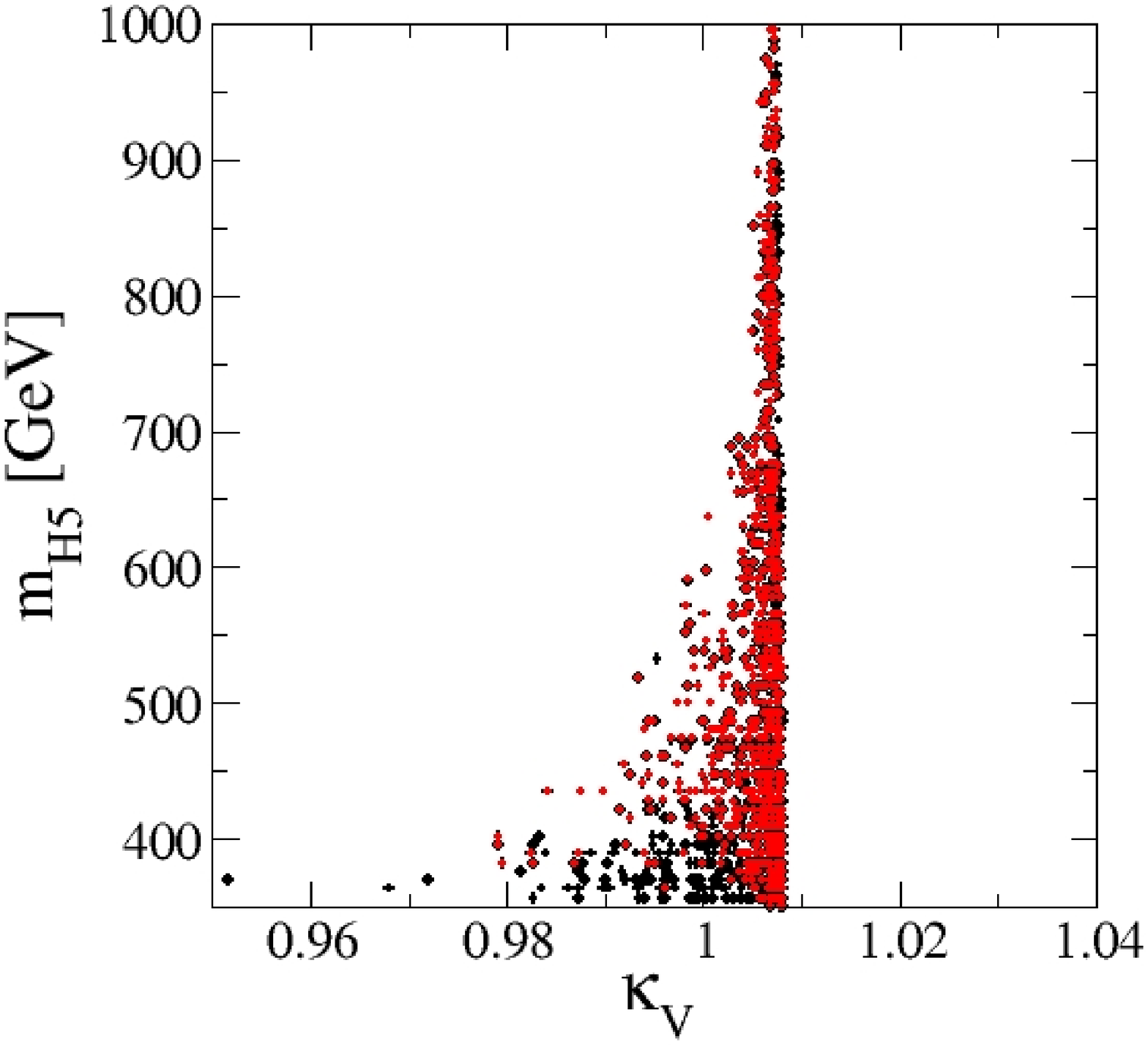}
 \caption{Black (Red) dots are allowed only by Bound A (by both Bound A and the LHC data) in the GM model. 
The value of $\tan\beta$ is taken to be  3, 5 and 10 on the left, center and right panels, respectively. 
We take $m_{H_3}^{} = m_{H_5}^{}$. All the other parameters ($m_{H_1}^{}$, $\mu_1$ and $\mu_2$) are scanned keeping $m_{H_1}^{}\geq 350$ GeV. 
}
 \label{fig:gm}
 \end{figure}

Finally, let us discuss the results in the GM model with $m_{H_5} = m_{H_3}$. 
In Fig.~\ref{fig:gm}, similar to Fig~\ref{fig:2hdm_2}, the black (red) dots are allowed by Bound A (by both Bound A and the LHC data). 
The parameters $m_{H_1}^{}$, $\mu_1$ and $\mu_2$ are scanned ($m_{H_1} \geq 350$ GeV) with a large enough range to maximize the allowed parameter space. 
We note that the case with $t_\beta \lesssim 3$ is excluded by the direct searches for $H_5^{\pm\pm} \to W^\pm W^\pm$ up to $m_{H_5}$ to be 800 GeV~\cite{ww}. 
We see that a larger value of $m_{H_5}$ is allowed in the case with $\kappa_V > 1$ by both the constraints from Bound A and the LHC data. 
For larger $t_\beta$, the allowed range of $\kappa_V^{}$ is getting smaller. 
For example, we obtain $0.96 \lesssim \kappa_V\lesssim 1.03$ and $0.99 \lesssim \kappa_V\lesssim 1.01$ for $t_\beta = 5$ and 10,
respectively, if we require $m_{H_5}\geq 500$ GeV. 
Differently from the cases of the HSM and the 2HDM, 
even when $\kappa_V \simeq 1$ is considered, there is an upper limit on the mass of the extra Higgs bosons, because $\kappa_V^{} \to 1$
does not correspond to the alignment limit in the GM model with any finite value of $t_\beta$. 

In summary, the constraints on the extra Higgs boson masses from the currently available LHC data, 
including both the direct searches for extra Higgs bosons and the $h(125)$ signal strengths,
strongly depend on the model considered. In particular, no further excluded regions
were found by imposing the LHC 
data in the HSM for $\kappa_V \gtrsim 0.95$. In the 2HDMs, it is necessary
to distinguish among the different types of Yukawa interactions and the values of 
$t_\beta$. Assuming mass degeneracy for 
the extra Higgs bosons, in Type-I and -X 2HDMs for, e.g., $\kappa_V = 0.99$ the combined LHC data and theoretical constraints
require $ 600 \, \text{GeV} \lesssim M_{\text{2nd}} \lesssim 800 \, \text{GeV}$ 
for $t_\beta = 1$, while, for $t_\beta = 10$, only the theoretical ones
are relevant and one finds $M_{\text{2nd}} \lesssim 600 \, \text{GeV}$.
Conversely, the Type-II and -Y 2HDMs turn out to be more constrained by the LHC data 
with respect to the Type-I and -X 2HDMs because of the $h(125)$ signal strength bounds, 
the large $t_\beta$ region being especially disfavored (in fact, almost only the alignment limit is allowed).
Regarding the GM model, the LHC data definitely favor the region with $\kappa_V \gtrsim 1$,
which shrinks closer to $\kappa_V = 1$ for larger values of $t_\beta$. For each value
of $\kappa_V$ which fulfills the LHC data constraints, 
the bounds on the extra Higgs boson masses are driven by the theoretical issues:
e.g., for $\kappa_V = 1.01$ and $t_\beta = 5$, one 
finds $m_{H_{3,5}} \lesssim 700 \, \text{GeV}$ (assuming $m_{H_3} = m_{H_5}$).

\section{Conclusions\label{sec:conc}}

We have extracted the mass scale of a possible second Higgs boson $M_{\text{2nd}}$ by imposing theoretical constraints and
by requiring compatibility with the currently available experimental data
in the next--to--minimal Higgs models with $\rho = 1$ at tree level, namely the HSM, the 2HDMs and the GM model.
In particular, we have focused on the correlation between the bound on $M_{\text{2nd}}$ and a possible
deviation in the $hVV$ ($V=W,Z$) couplings from the SM prediction. In doing so, we have assumed
the discovered Higgs boson $h(125)$ to be the lightest state among all the other Higgs bosons. 
As for the theoretical bounds, we took into account perturbative unitarity, vacuum stability and triviality,
while, as for the experimental constraints, we imposed the bounds from the electroweak precision data, 
the direct searches for extra Higgs bosons and the $h(125)$ signal strengths $\mu_X$. 
Assuming a value of $\Delta \kappa_V \equiv \kappa_V -1$, the non-LHC bounds (theoretical
and EWPT constraints) enforce an upper
limit on $M_{\text{2nd}}$, while the LHC data provide a lower bound.
For the former, we applied Bound A and Bound B under the full scan of the model 
parameters. Our results are summarized in Table \ref{tab:concl} for reference
values of $\Delta \kappa_V$, assuming the cutoff to be above 10 TeV under Bound B.
\renewcommand\arraystretch{1.2}
\begin{table}[]
\centering
\begin{tabular}{c|c|c|c|c|c|c|c}
\hline \hline
\multirow{2}{*}{$\,\, \Delta \kappa_V \,\,$} & \multirow{2}{*}{HSM} & \multicolumn{3}{c|}{2HDM}                              & \multicolumn{3}{c}{GM Model}                        \\ \cline{3-8} 
                                   &                      & $t_\beta = 1$    & $t_\beta = 2$    & $t_\beta = 10$   & $t_\beta = 3$    & $t_\beta = 5$   & $t_\beta = 10$  \\ \hline
$-1\%$                             & $4.0 \, (3.5)$       & $0.8 \, (0.5)$ & $0.7 \, (0.5)$ & $0.7 \, (0.4)$ & $0.9\, (0.4)$  & $1.2 \, (0.6)$ & $2.0 \, (1.0)$  \\ \hline
$-2\%$                             & $3.0 \, (2.5)$       & $0.7 \, (0.4)$ & $0.7\, (0.5)$  & $0.6 \, (0.3)$ & $0.8\, (0.4)$  & $1.1 \, (0.6)$ & $1.8 \, (0.9)$ \\ \hline
$-5\%$                             & $0.7 \, (0.7)$     & $0.6 \, (0.3)$ & $0.6\, (0.4)$  & $0.5\, (0.3)$  & $0.8\, (0.4)$  & $1.0 \, (0.6)$ & $1.5 \, (0.8)$ \\ \hline
$+1\%$                             & -                    & -                & -                & -                & $0.9 \, (0.5)$ & $1.4 \, (0.7)$ & $4.8 \, (1.8)$  \\ \hline \hline
\end{tabular}
\caption{Summary of the upper bounds on $M_{\text{2nd}}$ (TeV) by Bound A (by Bound B requiring the cutoff scale to be above 10 TeV) in the models considered
for reference values of $\Delta \kappa_V$.}
\label{tab:concl}
\end{table}
In addition, we have discussed the complementarity between the non-LHC and the LHC bounds, 
i.e., the direct searches and the $h(125)$ signal strengths. 
In the HSM, the LHC data do not improve the non-LHC bounds for $\kappa_V \gtrsim 0.95$. 
On the contrary, in the 2HDMs, the constraint from the LHC data can be important depending on the type of Yukawa interaction and the value of $t_\beta$. 
For the Type-I and Type-X 2HDMs with $t_\beta \sim 1$, the LHC data restrict $\kappa_V \gtrsim 0.98$ and the mass of extra Higgs bosons to be above 
$\sim 600$ GeV. For larger $t_\beta$, the LHC data are less effective and they provide no further constraint with respect to the non-LHC bounds. 
For the Type-II and Type-Y 2HDMs, the $\mu_X$ data constrain $\kappa_V \gtrsim 0.99$ for $t_\beta\sim 1$, and such bound becomes stronger for larger $t_\beta$. 
Finally, in the GM model, we found that the case with $\kappa_V \gtrsim 1$ is favored by both the LHC and non-LHC bounds.
In addition, for a larger value of $t_\beta$, the allowed range of $\kappa_V$ is getting smaller. 
For example, if we require $m_{H_5} (=m_{H_3}) \geq 500$ GeV, we obtain
$0.96 \lesssim \kappa_V\lesssim 1.03$ and $0.99 \lesssim \kappa_V\lesssim 1.01$ for $t_\beta = 5$ and 10, respectively. 

In conclusion, from the analysis performed in this paper, we have clarified 
the connection between the scale of the second Higgs boson mass in the non-minimal Higgs sectors here considered and 
the deviation in the $hVV$ couplings from the SM prediction. 
Since the $hVV$ couplings will be precisely measured 
at future collider experiments such as the High-Luminosity LHC (a few percent level) and $e^+e^-$ colliders (less than 1\% level 
at the International Linear Collider~\cite{Snowmass} with 500 GeV of the collision energy~\cite{ILC}), 
we can obtain precise information of the mass of the next-to-lightest Higgs boson
even without its discovery at the LHC.

\begin{appendix}

\section{Theoretical constraints \label{app1}}

We briefly review the theoretical constraints, i.e., 
the perturbative unitarity, the vacuum stability and the triviality we enforce in the models 
with an extended Higgs sector. 
We then present relevant analytic expressions for these constraints. 

We first discuss the bound from the perturbative unitarity. 
The request of the S-matrix unitarity 
for 2 body to 2 body elastic scattering processes for scalar bosons, 
assuming the validity
of perturbative calculations, leads to the following condition:
\begin{align}
|\textrm{Re}(a_J)|\leq 1/2, 
\label{ll}
\end{align}
where $a_J$ are partial wave amplitudes with total angular momentum $J$.
For the purpose of this paper, we define the perturbative unitarity bound by 
\begin{align}
|x_i|\leq 1/2, 
\end{align}
where $x_i$ are the eigenvalues of the $S$-wave ($J=0$) amplitude matrix, 
because they give the most stringent
constraints.

In the HSM, we obtain 4 independent eigenvalues~\cite{hsm_PU1,hsm_PU2}
\begin{align}
x_1^\pm &= \frac{1}{16\pi}\left[3\lambda +6\lambda_S \pm \sqrt{(6\lambda_S-3\lambda)^2+4\lambda_{\Phi S}^2}\right], \label{eigen1}\\
x_2 & = \frac{1}{8\pi}\lambda,   \label{eigen2}  \\
x_3 & = \frac{1}{8\pi}\lambda_{\Phi S}   \label{eigen3}. 
\end{align}
In the 2HDMs, we obtain 12 independent eigenvalues~\cite{thdm_PU1,thdm_PU2,thdm_PU3,thdm_PU4}
\begin{align}
x_1^\pm &=  \frac{1}{32\pi}
\left[3(\lambda_1+\lambda_2)\pm\sqrt{9(\lambda_1-\lambda_2)^2+4(2\lambda_3+\lambda_4)^2}\right],\\
x_2^\pm &=
\frac{1}{32\pi}\left[(\lambda_1+\lambda_2)\pm\sqrt{(\lambda_1-\lambda_2)^2+4\lambda_4^2}\right],\\
x_3^\pm &= \frac{1}{32\pi}\left[(\lambda_1+\lambda_2)\pm\sqrt{(\lambda_1-\lambda_2)^2+4\lambda_5^2}
\right],\\
x_4^\pm &= \frac{1}{16\pi}(\lambda_3+2\lambda_4\pm 3\lambda_5),\\
x_5^\pm &= \frac{1}{16\pi}(\lambda_3\pm\lambda_4),\\
x_6^\pm &= \frac{1}{16\pi}(\lambda_3\pm\lambda_5).
\end{align}
In the GM model, we obtain 9 independent eigenvalues~\cite{gm_PU1,Logan}
\begin{align}
x_1^\pm &= \frac{1}{16\pi}\left[12\lambda_1+22\lambda_2+14\lambda_3\pm \sqrt{(12\lambda_1-22\lambda_2-14\lambda_3)^2+144\lambda_4^2}\right], \\ 
x_2^\pm &= \frac{1}{16\pi}\left[4\lambda_1+4\lambda_2-2\lambda_3\pm \sqrt{(4\lambda_1-4\lambda_2+2\lambda_3)^2+4\lambda_5^2}\right], \, \\
x_3&=\frac{1}{16\pi}(8\lambda_2+16\lambda_3),  \, \\
x_4&=\frac{1}{16\pi}(8\lambda_2+4\lambda_3), \, \\
x_5&=\frac{1}{16\pi}4(\lambda_4+\lambda_5),  \, \\
x_6&=\frac{1}{16\pi}(4\lambda_4-2\lambda_5), \, \\
x_7&=\frac{1}{16\pi}4(\lambda_4+\lambda_5). 
\end{align}
As stated above, we impose $|x_i| \leq 1/2$ and derive the allowed regions in the parameter space of each model.

Next, let us discuss the vacuum stability bound. We require that the scalar potential is bounded from below in any direction with 
large field values. 
This can be simply expressed by $V^{(4)}\geq 0$, where $V^{(4)}$ is the scalar quartic part of the potential. 
The sufficient and necessary conditions to satisfy the vacuum stability 
constraint in the HSM~\cite{HSM-RGE,HSM-RGE2} are given by
\begin{align}
\lambda \geq 0,\quad \lambda_S \geq 0,\quad 2\sqrt{\lambda \lambda_S} + \lambda_{\Phi S} \geq 0 \label{vs_hsm}.
\end{align}
In the 2HDMs, the sufficient and necessary conditions are expressed by:
\begin{align}
\lambda_1\geq 0,~~ \lambda_2\geq 0,~~ \sqrt{\lambda_1\lambda_2}+\lambda_3+ \text{MIN}(0,\lambda_4+\lambda_5,\lambda_4-\lambda_5) \geq 0. \label{vs_2hdm}
\end{align}
Eqs.~(\ref{vs_hsm}) and (\ref{vs_2hdm}) can be improved by using the running couplings  
evaluated by solving the one-loop RGEs (see App.~\ref{sec:rge}) and requiring that all the above 
inequalities are satisfied at every energy scale $\mu$ with $m_Z^{} \leq \mu \leq \Lambda$, where 
$\Lambda$ is the cutoff of the theory. 

In the GM model, 
it has been clarified in Ref.~\cite{GVW,Simone} that the custodial symmetry in the potential 
is explicitly broken due to the $U(1)_Y$ gauge boson loop effects. 
In order to make the model consistent at high energies, 
we need to use the most general form of the Higgs potential without the custodial symmetry. 
The explicit form of the general potential is given in Eq.~(\ref{pot_gen}) in App.~\ref{sec:gm-gen}. 
In terms of the quartic couplings of the general potential,
the necessary conditions to guarantee the vacuum stability are derived by assuming two non-vanishing complex fields at once.   
Taking into account all the possible two field directions, we obtain the following inequalities:

\begin{align}
\begin{split}
&\lambda \geq 0,~~ \rho_3 \geq 0,~~ \rho_1 + \rho_2 \geq 0,~~ \rho_1 + \frac{\rho_2}{2} \geq 0, \\
& \rho_4 + \frac{\rho_5}{2} +\sqrt{2\rho_3(\rho_1+\rho_2)}\geq 0, \\
& \rho_4  +\sqrt{2\rho_3(\rho_1+\rho_2)}\geq 0, \\
& \rho_4 + 2\sqrt{\rho_3(2\rho_1+\rho_2)}\geq 0, \\
& \rho_4 + \rho_5 + 2\sqrt{\rho_3(2\rho_1+\rho_2)}\geq 0, \\
& \sigma_1 + 2\sqrt{\lambda(\rho_1+\rho_2)}\geq 0, \\
& \sigma_1 + \sigma_2 + 2\sqrt{\lambda(\rho_1+\rho_2)}\geq 0, \\
& \sigma_1 + \frac{\sigma_2}{2} + \sqrt{2\lambda(2\rho_1+\rho_2)}\geq 0, \\
& \sigma_3 +  \sqrt{2\lambda\rho_3}\geq 0. 
\end{split}
\label{stability}
\end{align}

\section{One-loop $\beta$-functions \label{sec:rge}}

We here present the set of $\beta$-functions evaluated at one-loop level 
in the HSM, the 2HDMs and the GM model. 
The $\beta$-function for a coupling constant $y$ is defined by 
\begin{align}
\beta(y) \equiv \frac{d}{d\ln \mu} y. 
\end{align}
The RGE evolution of the gauge, scalar and Yukawa couplings is given by the following
$\beta$-functions in the HSM~\cite{HSM-RGE}: 
\begin{align}
\beta(g_3)&=\frac{g_3^3}{16\pi^2}(-7), \quad \beta(g_2)=\frac{g_2^3}{16\pi^2}\left(-\frac{19}{6}\right), \quad \beta(g_1)=\frac{g_1^3}{16\pi^2}\left(\frac{41}{6}\right), \\
\beta(y_t)&=\frac{1}{16\pi^2}\left[\frac{9}{2}y_t^3+\frac{3}{2}y_b^3 -y_t\left(8g_3^2+\frac{9}{4}g_2^2+\frac{17}{12}g_1^2\right)\right], \\
\beta(y_b)&=\frac{1}{16\pi^2}\left[\frac{9}{2}y_b^3+\frac{3}{2}y_t^3 -y_b\left(8g_3^2+\frac{9}{4}g_2^2+\frac{5}{12}g_1^2\right)\right], \\
\beta(\lambda)&=\frac{1}{16\pi^2}\Big\{24\lambda^2 + 2\lambda_{\Phi S}^2
-6 (y_t^4+y_b^4)+\frac{9}{8}g_2^4+\frac{3}{8}g_1^{4}+\frac{3}{4}g_1^2g_2^2 \notag\\
& \hspace{16mm}-3\lambda[3g_2^2+g_1^2-4(y_t^2+y_b^2) ]\Big\}, \\
\beta(\lambda_{\Phi S}^{})& =
\frac{1}{16\pi^2}
\left[12\lambda \lambda_{\Phi S}^{}+8\lambda_{\Phi S}^{2}+24\lambda_{\Phi S}^{} \lambda_S -\lambda_{\Phi S}^{}\left(\frac{9}{2}g_2^2  +\frac{3}{2}g_1^2 -6y_t^2-6y_b^2\right)\right], \\
\beta(\lambda_{S}^{})& =\frac{1}{16\pi^2}\left(2\lambda_{\Phi S}^2 + 72\lambda_S^2 \right). \label{lams}
\end{align}
In the 2HDMs, we obtain 
\begin{align}
\beta(g_3)&=\frac{g_3^3}{16\pi^2}(-7), \quad \beta(g_2)=\frac{g_2^3}{16\pi^2}\left(-3\right), \quad \beta(g_1)=\frac{g_1^3}{16\pi^2}7, \\
\beta(y_t)&=\frac{1}{16\pi^2}\left[\frac{9}{2}y_t^3+\frac{3}{2}y_b^3\, \Theta -y_t\left(8g_3^2+\frac{9}{4}g_2^2+\frac{17}{12}g_1^2\right)\right], \\
\beta(y_b)&=\frac{1}{16\pi^2}\left[\frac{9}{2}y_b^3+\frac{3}{2}y_t^3   \Theta -y_b\left(8g_3^2+\frac{9}{4}g_2^2+\frac{5}{12}g_1^2\right)\right], \\
\beta(\lambda_1)&=\frac{1}{16\pi^2}\Big[12\lambda_1^2+4\lambda_3^2+2\lambda_4^2+4\lambda_3\lambda_4+2\lambda_5^2 -12y_b^4 \tilde{\Theta}
+\frac{9}{4}g_2^4+\frac{3}{4}g_1^4+\frac{3}{2}g_1^2g_2^2\notag\\
&\quad \quad \quad \quad -3\lambda_1( 3g_2^2+ g_1^2-4y_b^2 \tilde{\Theta}) \Big], \\
\beta(\lambda_2)&=\frac{1}{16\pi^2}\Big[12\lambda_2^2+4\lambda_3^2+4\lambda_3\lambda_4+2\lambda_4^2+2\lambda_5^2 -12(y_t^4+y_b^4 \Theta )
 \notag\\
&\quad \quad \quad \quad +\frac{9}{4}g_2^4+\frac{3}{4}g_1^4+\frac{3}{2}g_1^2g_2^2-3\lambda_2(3 g_2^2+\lambda_2 g_1^2-4y_t^2-4y_b^2\Theta) \Big], \\
\beta(\lambda_3)
&=\frac{1}{16\pi^2}\Big[2(\lambda_1+\lambda_2)(3\lambda_3+\lambda_4)+4\lambda_3^2+2\lambda_4^2+2\lambda_5^2
-12y_t^2y_b^2\tilde{\Theta}\notag\\
&\quad\quad\quad\quad
+\frac{9}{4}g_2^4+\frac{3}{4}g_1^4-\frac{3}{2}g_1^2g_2^2-3\lambda_3(g_2^2+g_1^2-2y_t^2-2y_b^2)\Big], \\
\beta(\lambda_4)&=\frac{1}{16\pi^2}\Big[2\lambda_4(\lambda_1+\lambda_2+4\lambda_3+2\lambda_4)+8\lambda_5^2
+12y_t^2y_b^2\tilde{\Theta}
+3g_2^2g_1^2 \notag\\
&\quad\quad\quad\quad -3\lambda_4\big(3g_2^2+g_1^{2}-2y_t^2-2y_b^2\big)\Big], \\
\beta(\lambda_5)&=\frac{1}{16\pi^2}\Big[2\lambda_5(\lambda_1+\lambda_2+4\lambda_3+6\lambda_4)-3\lambda_5(3g_2^2+g_1^2-2y_t^2-2y_b^2)\Big], 
\end{align}
where 
\begin{align}
\Theta &= 1 (0)~~\text{for}~~\text{Type-I,~X~(Type-II,~Y)}, \\
\tilde{\Theta} &= 0(1)~~\text{for}~~\text{Type-I,~X~(Type-II,~Y)}.
\end{align}
In the GM model, we obtain~\cite{Simone} 
\begin{align}
\beta(g_3)&=\frac{g_3^3}{16\pi^2}(-7), \quad \beta(g_2)=\frac{g_2^3}{16\pi^2}\left(-\frac{11}{6}\right), \quad \beta(g_1)=\frac{g_1^3}{16\pi^2}\frac{47}{6}, \\
\beta(y_t)&=\frac{1}{16\pi^2}\left[\frac{9}{2}y_t^3+\frac{3}{2}y_b^3 -y_t\left(8g_3^2+\frac{9}{4}g_2^2+\frac{17}{12}g_1^2\right)\right], \\
\beta(y_b)&=\frac{1}{16\pi^2}\left[\frac{9}{2}y_b^3+\frac{3}{2}y_t^3    -y_b\left(8g_3^2+\frac{9}{4}g_2^2+\frac{5}{12}g_1^2\right)\right], 
\end{align}
\begin{align}
\beta(\lambda) & =  \frac{1}{16\pi^2}\Big[\frac{3}{8} \left(3 g_2^4+2g_2^2g_1^2+g_1^4 \right)   +24\lambda^2-6 (y_t^4+y_b^4)
+3 \sigma_1^2+3 \sigma_1 \sigma_2+\frac{5 \sigma_2^2}{4}+6 \sigma_3^2+2 \sigma_4^2 \notag\\
&-3\lambda(g_1^2  +3 g_2^2-4y_t^2-4y_b^2) \Big], \\
\beta(\rho_1)& = \frac{1}{16\pi^2}\Big[15 g_2^4-12 g_1^2 g_2^2+6 g_1^4
+28 \rho_1^2+24 \rho_1 \rho_2+6 \rho_2^2+6 \rho_4^2+4 \rho_4 \rho_5+3\rho_5^2\notag\\
&+2 \sigma_1^2+2 \sigma_1 \sigma_2-12\rho_1( g_1^2 + 2g_2^2) \Big], \\
\beta(\rho_2)& =  \frac{1}{16\pi^2}\Big[24 g_1^2 g_2^2-6 g_2^4+24\rho_1 \rho_2+18 \rho_2^2-2
\rho_5^2+\sigma_2^2-12 \rho_2 (2g_2^2 + g_1^2) \Big], \\
\beta(\rho_3)& =  \frac{1}{16\pi^2}2\left(3 g_2^4+22 \rho_3^2+3 \rho_4^2+2\rho_4 \rho_5+\rho_5^2+2 \sigma_3^2-12 g_2^2 \rho_3\right), \\
\beta( \rho_4)& =  \frac{1}{16\pi^2}2 \Big[3 g_2^4+\rho_4 \left(8 \rho_1+6 \rho_2+10 \rho_3+4\rho_4 \right)+2 \rho_5 (\rho_1+\rho_2+\rho_3)\notag\\
&+\rho_5^2+2 \sigma_1 \sigma_3+\sigma_2 \sigma_3+\sigma_4^2-3\rho_4( g_1^2 +4 g_2^2) \Big], \\
\beta(\rho_5)& =  \frac{1}{16\pi^2}2 \Big[3 g_2^4+\rho_5 \left(2 \rho_1+4 \rho_3+8 \rho_4+5\rho_5\right)-\sigma_4^2 -3\rho_5(4 g_2^2 +g_1^2)\Big], \\
\beta(\sigma_1)& =  \frac{1}{16\pi^2}\Big[3 g_1^4-6g_1^2g_2^2+6 g_2^4 +2 \sigma_1 \left(6 \lambda +8 \rho_1+6 \rho_2+2\sigma_1\right) 
+2 \sigma_2 (2 \lambda +3 \rho_1+\rho_2) \notag\\
&+2\left(6 \rho_4 \sigma_3+2\rho_5 \sigma_3+\sigma_4^2\right)+\sigma_2^2
-\frac{3}{2}\sigma_1(5g_1^2 +11g_2^2-4y_t^2-4y_b^2)\Big], \\
\beta(\sigma_2)& = \frac{1}{16\pi^2}\Big[12g_1^2g_2^2
+4\sigma_2 [\lambda +\rho_1+2 (\rho_2+\sigma_1)+\sigma_2] +4 \sigma_4^2\notag\\
& -\frac{3}{2}\sigma_2(5g_1^2+11g_2^2 -4 y_t^2-4y_b^2)\Big],\\
\beta(\sigma_3)& = \frac{1}{16\pi^2}\Big[3 g_2^4 + 2\sigma_3\left(6 \lambda +10 \rho_3+4 \sigma_3\right) + (3 \rho_4+\rho_5) (2\sigma_1+\sigma_2)+4 \sigma_4^2\notag\\
& -\frac{3}{2}\sigma_3( g_1^2 +11 g_2^2-4y_t^2-4y_b^2)\Big] , \\
\beta(\sigma_4)& =  \frac{1}{16\pi^2}\frac{\sigma_4}{2}  \left[4 \left(2 \lambda +2\rho_4-\rho_5
+ 2\sigma_1+2\sigma_2+4\sigma_3\right)-3(3g_1^2+11 g_2^2-4y_t^2-4y_b^2)\right]. 
\end{align}

\section{Contributions to the $S$, $T$ parameters}\label{appST}

In this section, we present the analytic expressions for the oblique $S$ and $T$ parameters
in the extended Higgs sector models considered. 
Because we are interested in the NP contribution to these parameters, we define the differences as 
$\Delta S = S_{\text{NP}}-S_{\text{SM}}$ and $\Delta T = T_{\text{NP}}-T_{\text{SM}}$. 

Let us start with the 2HDM within which the $\Delta S$ and $\Delta T$ parameters have the following expressions:
\begin{align}
\Delta S & = \frac{1}{4\pi} \Bigg\{ -F'(m_Z^2;m_{H^\pm},m_{H^\pm})+s_{\beta-\alpha}^2F'(m_Z^2;m_H,m_A)\notag\\
&+c_{\beta-\alpha}^2\left[F'(m_Z^2;m_h,m_A)-G'(m_Z^2;m_h,m_Z)+G'(m_Z^2;m_H,m_Z)\right] \Bigg\}, \\
\Delta T & = 
\frac{1}{4\pi e^2v^2} \Bigg\{
F(0;m_{H^\pm},m_A)
+s_{\beta-\alpha}^2\Big[ F\left(0;m_{H^\pm},m_H\right)-F\left(0;m_A,m_H\right)\Big] \notag\\
&+c_{\beta-\alpha}^2\Big[ F\left(0;m_{H^\pm},m_h\right)-F\left(0;m_A,m_h\right)\notag\\
&-G(0;m_h,m_W)+G(0;m_H,m_W) +G(0;m_h,m_Z)-G(0;m_H,m_Z)\Big] \Bigg\}. 
\end{align}
The loop functions are given by  
\begin{align}
F(p^2,m_1,m_2)&= \int_0^1 dx \left[(2x-1)(m_1^2-m_2^2)+p^2(2x-1)^2\right]\ln\Delta_B,\\ 
G(p^2,m_1,m_2)&= \int_0^1 dx \left[(2x-1)m_1^2 - (2x-5)m_2^2 +p^2(2x-1)^2\right]\ln\Delta_B,
\end{align} 
where $e^2 = 4 \pi \alpha_{\text{em}}$ and
\begin{align}
\Delta_B  &= -x(1-x)p^2 + xm_1^2 + (1-x)m_2^2, \\
F'(m_V^2;m_1,m_2)&=[F(m_V^2;m_1,m_2)-F(0;m_1,m_2)]/m_V^2, \\
G'(m_V^2;m_1,m_2)&=[G(m_V^2;m_1,m_2)-G(0;m_1,m_2)]/m_V^2. 
\end{align}
In the HSM, the expressions are simpler: 
\begin{align}
\Delta S & = \frac{s_{\alpha}^2}{4\pi} \left[G'(m_Z^2;m_H,m_Z)-G'(m_Z^2;m_h,m_Z)\right], \\
\Delta T & = 
\frac{s_{\alpha}^2}{4\pi e^2v^2} \Big[G(0;m_H,m_W)-G(0;m_h,m_W)+G(0;m_h,m_Z)-G(0;m_H,m_Z)\Big].
\end{align}
In the GM model, we need a special treatment for the calculation of the $T$ parameter.
In fact, in this model an additional counter term $\delta T$ appears
due to the fact that the kinetic term is 
described by four independent quantities, namely $g_1,~g_2,~v$ and $\rho$. We are here 
imposing $\rho = 1$ at the tree level by taking the two triplet VEVs to be the same. 
This means that the $T$ parameter is not predictable in the GM model.
We can indeed take any value of $\delta T$ by setting a suitable renormalization condition. 
In our numerical evaluation, we simply set $\delta T$ so as to satisfy $\Delta T = 0$~\cite{GVW,Chiang:2017vvo}. 

The $\Delta S$ parameter in the GM model is given by: 
\begin{align}
\Delta S & = \frac{1}{4\pi} \Bigg\{
-5F'(m_Z^2;m_{H_5},m_{H_5})
-F'(m_Z^2;m_{H_3},m_{H_3}) \\
&+\frac{10s_\beta^2}{3}F(m_Z^2,m_{H_5},m_{H_3})
+2c_\beta^2G'(m_Z^2,m_{H_5},m_W) + \frac{4c_\beta^2}{3}G'(m_Z^2,m_{H_5},m_{Z})\notag\\
&+\left(s_\alpha c_\beta-\frac{2}{3}\sqrt{6}c_\alpha s_\beta\right)^2F'(m_Z^2,m_{H_3},m_{H_1})
+\left(c_\alpha c_\beta+\frac{2}{3}\sqrt{6}s_\alpha s_\beta\right)^2F'(m_Z^2,m_{H_3},m_h)\notag\\
&+\left(s_\alpha s_\beta+\frac{2}{3}\sqrt{6}c_\alpha c_\beta\right)^2G'(m_Z^2,m_{H_1},m_Z^{})
+\left[\left(s_\beta c_\alpha - \frac{2}{3}\sqrt{6} c_\beta s_\alpha \right)^2-1\right]G'(m_Z^2,m_h,m_Z^{}) \Bigg\}.
\end{align}

\section{General potential in the GM model \label{sec:gm-gen}}

The most general scalar potential, not custodial symmetric, in the GM model is given by~\cite{Simone}: 
\begin{align}
V(\Phi,\chi,\xi)_{\text{GM}}&=m_\Phi^2(\Phi^\dagger \Phi)+m_\chi^2\text{tr}(\chi^\dagger\chi)+m_\xi^2\text{tr}(\xi^2)\notag\\
&+\bar{\mu}_1\Phi^\dagger \xi\Phi +\bar{\mu}_2 [\Phi^T(i\tau_2) \chi^\dagger \Phi+\text{h.c.}] +\bar{\mu}_3\text{tr}(\chi^\dagger \chi \xi)+\lambda (\Phi^\dagger \Phi)^2 \notag\\
&
+\rho_1[\text{tr}(\chi^\dagger\chi)]^2+\rho_2\text{tr}(\chi^\dagger \chi\chi^\dagger \chi)
+\rho_3\text{tr}(\xi^4)
+\rho_4 \text{tr}(\chi^\dagger\chi)\text{tr}(\xi^2)
+\rho_5\text{tr}(\chi^\dagger \xi)\text{tr}(\xi \chi)\notag\\
&+\sigma_1\text{tr}(\chi^\dagger \chi)\Phi^\dagger \Phi+\sigma_2 \Phi^\dagger \chi\chi^\dagger \Phi
+\sigma_3\text{tr}(\xi^2)\Phi^\dagger \Phi 
+\sigma_4 (\Phi^\dagger \chi\xi \Phi^c + \text{h.c.}), \label{pot_gen}
\end{align}
where $\bar{\mu}_2$ and $\sigma_4$ are complex in general, but we here assume them real for simplicity. 
The doublet $\Phi$ and the triplet $\chi$ and $\xi$ fields are parameterized as 
\begin{align}
\Phi = \left(
\begin{array}{c}
\phi^+ \\
\phi^0
\end{array}\right),~
\chi = \left(
\begin{array}{cc}
\frac{\chi^+}{\sqrt{2}} & -\chi^{++}\\
\chi^0 & -\frac{\chi^+}{\sqrt{2}}
\end{array}\right),~
\xi = \left(
\begin{array}{cc}
\frac{\xi^0}{\sqrt{2}} & -\xi^+\\
-\xi^- & -\frac{\xi^0}{\sqrt{2}}
\end{array}\right), \label{par}
\end{align}
where the neutral components are expressed by
\begin{align}
\phi^0 =\frac{1}{\sqrt{2}}(\phi_r+v_\phi +i\phi_i),\quad 
\chi^0 =\frac{1}{\sqrt{2}}(\chi_r+i\chi_i^0)+v_\chi,\quad 
\xi^0 = \xi_r+v_\xi.    \label{neut}
\end{align}
By comparing the custodial symmetric potential given in Eq.~(\ref{eq:pot}) with the general one in Eq.~(\ref{pot_gen}), we find the following relations:
\begin{align}
\begin{split}
&m_\phi^2=2m_\Phi^2,~ m_\chi^2=2m_\Delta^2,~ m_\xi^2= m_\Delta^2,~
\bar{\mu}_1=-\frac{\mu_1}{\sqrt{2}},~ \bar{\mu}_2=-\frac{\mu_1}{2},~
\bar{\mu}_3=6\sqrt{2}\mu_2, \\
&\lambda=4\lambda_1,~ 
\rho_1=4\lambda_2+6\lambda_3,~ 
\rho_2=-4\lambda_3,~ 
\rho_3=2(\lambda_2+\lambda_3),~
\rho_4=4\lambda_2,~
\rho_5=4\lambda_3, \\
& \sigma_1 =4\lambda_4-\lambda_5,~ 
 \sigma_2 =2\lambda_5,~ 
 \sigma_3 =2\lambda_4,~
 \sigma_4 =\sqrt{2}\lambda_5,
\end{split}
\label{rel} 
\end{align}
which express the 16 parameters of the most general potential in
terms of the 9 parameters defined in the custodial symmetric one.
\end{appendix}


\vspace*{-4mm}

\begin{thebibliography}{1}

\bibitem{LHC1}

  G.~Aad {\it et al.} [ATLAS and CMS Collaborations],
  JHEP {\bf 1608}, 045 (2016)
  [arXiv:1606.02266 [hep-ex]].
  
\bibitem{review_thdm}
G. C. Branco, P. M. Ferreira, L. Lavoura, M. N. Rebelo, M. Sher and J. P. Silva, Phys. Rept.
\textbf{516}, 1 (2012), arXiv:1106.0034 [hep-ph].

 
\bibitem{GM1}

  H.~Georgi and M.~Machacek,
  Nucl.\ Phys.\ B {\bf 262}, 463 (1985).

\bibitem{GM2} 
  M.~S.~Chanowitz and M.~Golden,
  Phys.\ Lett.\ B {\bf 165}, 105 (1985).

    
\bibitem{HSM-RGE} 
  M.~Gonderinger, Y.~Li, H.~Patel and M.~J.~Ramsey-Musolf,
  JHEP {\bf 1001}, 053 (2010)
  [arXiv:0910.3167 [hep-ph]].
 
\bibitem{type2}

 T.~P.~Cheng and L.~F.~Li,
 Phys.\ Rev.\  D {\bf 22}, 2860 (1980);

 J.~Schechter and J.~W.~F.~Valle,
 Phys.\ Rev.\  D {\bf 22}, 2227 (1980);

  G.~Lazarides, Q.~Shafi and C.~Wetterich,
  Nucl.\ Phys.\  B {\bf 181}, 287 (1981);

  R.~N.~Mohapatra and G.~Senjanovic,
  Phys.\ Rev.\  D {\bf 23}, 165 (1981);

  M.~Magg and C.~Wetterich,
  Phys.\ Lett.\  B {\bf 94}, 61 (1980).
  
\bibitem{HLLHC_ATLAS} 
  [ATLAS Collaboration],
  arXiv:1307.7292 [hep-ex].

\bibitem{HLLHC_CMS} 
  [CMS Collaboration],
  arXiv:1307.7135.
  
\bibitem{ILC} 
  K.~Fujii {\it et al.},
  arXiv:1506.05992 [hep-ex].
  
  
\bibitem{LQT} 
  B.~W.~Lee, C.~Quigg and H.~B.~Thacker,
  Phys.\ Rev.\ D {\bf 16}, 1519 (1977).
  
\bibitem{thdm_PU3}  
  I.~F.~Ginzburg and I.~P.~Ivanov,
  Phys.\ Rev.\ D {\bf 72}, 115010 (2005).

\bibitem{thdm_PU4}  
  S.~Kanemura and K.~Yagyu,
  Phys.\ Lett.\ B {\bf 751}, 289 (2015)
  [arXiv:1509.06060 [hep-ph]].


\bibitem{thdm_PU1}

  S.~Kanemura, T.~Kubota and E.~Takasugi,
  Phys.\ Lett.\  B {\bf 313}, 155 (1993). 

\bibitem{thdm_PU2}  
A.~G.~Akeroyd, A.~Arhrib and E.~M.~Naimi,
  Phys.\ Lett.\  B {\bf 490}, 119 (2000). 
 [arXiv:hep-ph/0006035].
  
\bibitem{Dawson} 
  C.~Y.~Chen, S.~Dawson and I.~M.~Lewis,
  Phys.\ Rev.\ D {\bf 91}, no. 3, 035015 (2015), [arXiv:1410.5488 [hep-ph]].

\bibitem{HSM} 
  S.~Kanemura, M.~Kikuchi and K.~Yagyu,
  Nucl.\ Phys.\ B {\bf 917}, 154 (2017)
  [arXiv:1608.01582 [hep-ph]].

\bibitem{GW}
  S.~L.~Glashow and S.~Weinberg,
  Phys.\ Rev.\  D {\bf 15}, 1958 (1977).

\bibitem{HBasis} 
  S.~Davidson and H.~E.~Haber,
  Phys.\ Rev.\ D {\bf 72}, 035004 (2005)
  [Phys.\ Rev.\ D {\bf 72}, 099902 (2005)]
  [hep-ph/0504050].

\bibitem{Barger}
  V.~D.~Barger, J.~L.~Hewett and R.~J.~N.~Phillips,
  Phys.\ Rev.\ D {\bf 41}, 3421 (1990). 

\bibitem{Grossman}
  Y.~Grossman,
  Nucl.\ Phys.\ B {\bf 426}, 355 (1994).
  
\bibitem{typeX}

  M.~Aoki, S.~Kanemura, K.~Tsumura and K.~Yagyu,
  Phys.\ Rev.\ D {\bf 80}, 015017 (2009)
  [arXiv:0902.4665 [hep-ph]].
  
\bibitem{Simone}
  S.~Blasi, S.~De Curtis and K.~Yagyu,
  Phys.\ Rev.\ D {\bf 96} (2017) no.1,  015001
  [arXiv:1704.08512 [hep-ph]].


\bibitem{Peskin} 
 M.~E.~Peskin and T.~Takeuchi,
  Phys.\ Rev.\ Lett.\  {\bf 65}, 964 (1990); 
  Phys.\ Rev.\  D {\bf 46}, 381 (1992). 

\bibitem{st} 
  M.~Baak {\it et al.},
  Eur.\ Phys.\ J.\ C {\bf 72}, 2205 (2012)
  [arXiv:1209.2716 [hep-ph]].

\bibitem{Bphys1}

  F.~Mahmoudi and O.~Stal,
  Phys.\ Rev.\ D {\bf 81}, 035016 (2010)
  [arXiv:0907.1791 [hep-ph]]. 

\bibitem{Bphys2}

  T.~Enomoto and R.~Watanabe,
  JHEP {\bf 1605}, 002 (2016)
  [arXiv:1511.05066 [hep-ph]].

\bibitem{finger}
  S.~Kanemura, K.~Tsumura, K.~Yagyu and H.~Yokoya,
  Phys.\ Rev.\ D {\bf 90}, no. 7, 075001 (2014)
 [arXiv:1406.3294 [hep-ph]].
\bibitem{Yokoya} 
  S.~Kanemura, H.~Yokoya and Y.~J.~Zheng,
  Nucl.\ Phys.\ B {\bf 886}, 524 (2014)
  [arXiv:1404.5835 [hep-ph]].
  
\bibitem{ttbar}
  The ATLAS collaboration [ATLAS Collaboration],
  ATLAS-CONF-2016-089.

\bibitem{ttbar2} 
  G.~Aad {\it et al.} [ATLAS Collaboration],
  JHEP {\bf 1508}, 148 (2015)
  [arXiv:1505.07018 [hep-ex]].
\bibitem{Hhh} 
  G.~Aad {\it et al.} [ATLAS Collaboration],
  Phys.\ Rev.\ D {\bf 92}, 092004 (2015)
   [arXiv:1509.04670 [hep-ex]].

\bibitem{HZZ} 
  G.~Aad {\it et al.} [ATLAS Collaboration],
  Eur.\ Phys.\ J.\ C {\bf 76}, no. 1, 45 (2016)
  [arXiv:1507.05930 [hep-ex]].

\bibitem{AZh} 
  G.~Aad {\it et al.} [ATLAS Collaboration],
  Phys.\ Lett.\ B {\bf 744}, 163 (2015)
  [arXiv:1502.04478 [hep-ex]].
\bibitem{ggf} 
  G.~Aad {\it et al.} [ATLAS Collaboration],
  Eur.\ Phys.\ J.\ C {\bf 76} (2016) no.1,  6
  [arXiv:1507.04548 [hep-ex]].

\bibitem{Moretti:2016jkp} 
  S.~Moretti, R.~Santos and P.~Sharma,
  Phys.\ Lett.\ B {\bf 760}, 697 (2016)
  [arXiv:1604.04965 [hep-ph]].


\bibitem{wz} 
  A.~M.~Sirunyan {\it et al.} [CMS Collaboration],
  arXiv:1705.02942 [hep-ex].


\bibitem{ww} 

  V.~Khachatryan {\it et al.} [CMS Collaboration],
  Phys.\ Rev.\ Lett.\  {\bf 114}, no. 5, 051801 (2015)
  [arXiv:1410.6315 [hep-ex]].



 \bibitem{Snowmass} 
   S.~Dawson, A.~Gritsan, H.~Logan, J.~Qian, C.~Tully, R.~Van Kooten, A.~Ajaib and A.~Anastassov {\it et al.},
   ``Working Group Report: Higgs Boson,''
   arXiv:1310.8361 [hep-ex].

\bibitem{hsm_PU1} 
  G.~Cynolter, E.~Lendvai and G.~Pocsik,
  Acta Phys.\ Polon.\ B {\bf 36}, 827 (2005)
  [hep-ph/0410102].


\bibitem{hsm_PU2} 
  S.~Kanemura, M.~Kikuchi and K.~Yagyu,
  Nucl.\ Phys.\ B {\bf 907}, 286 (2016)
  [arXiv:1511.06211 [hep-ph]].

\bibitem{gm_PU1}
  M.~Aoki and S.~Kanemura,
  Phys.\ Rev.\ D {\bf 77}, no. 9, 095009 (2008)
  Erratum: [Phys.\ Rev.\ D {\bf 89}, no. 5, 059902 (2014)]
  [arXiv:0712.4053 [hep-ph]].

\bibitem{Logan} 
  K.~Hartling, K.~Kumar and H.~E.~Logan,
  Phys.\ Rev.\ D {\bf 90}, no. 1, 015007 (2014)
  [arXiv:1404.2640 [hep-ph]].

\bibitem{HSM-RGE2} 
  L.~Basso, O.~Fischer and J.~J.~van Der Bij,
  Phys.\ Lett.\ B {\bf 730}, 326 (2014)
  [arXiv:1309.6086 [hep-ph]].

\bibitem{GVW}

  J.~F.~Gunion, R.~Vega and J.~Wudka,
  Phys.\ Rev.\ D {\bf 43}, 2322 (1991).

\bibitem{Chiang:2017vvo}
  C.~W.~Chiang, A.~L.~Kuo and K.~Yagyu,
  arXiv:1707.04176 [hep-ph].
  
  
  
\end{thebibliography}
\end{document}